\documentclass[prx,reprint,twocolumn,10pt,superscriptaddress,english,showpacs,longbibliography,nofootinbib]{revtex4-1}
\usepackage[T1]{fontenc}
\usepackage[utf8]{inputenc}
\usepackage{amsmath}
\usepackage{color}
\usepackage[table]{xcolor}
\usepackage{amsmath}
\usepackage{amssymb}
\usepackage{longtable}
\usepackage{multirow}
\usepackage{dcolumn}% Align table columns on decimal point
\usepackage{bm}
\usepackage[none]{hyphenat} % avoid the hyphen "-" that splits a word in the end of a line.
\usepackage{bbm}
\usepackage{braket}
\usepackage{tabularx}
\usepackage{blindtext}
\usepackage{chemformula}

\usepackage{cleveref}

\def\ZZ{\mathbb{Z}}
\def\ie{{\it i.e.},\ }

\input{epsf}
\renewcommand{\paragraph}[1]{\vspace{0.2cm}{\bf \textit{#1}}}

\definecolor{orange}{rgb}{1,0.5,0}
\definecolor{Red}{RGB}{255,204,204}
\definecolor{Blue}{RGB}{153,204,255}
\definecolor{Green}{RGB}{153,255,153}
\definecolor{Yellow}{RGB}{255,255,153}
\definecolor{Black}{RGB}{224,224,224}

\newcommand{\PreserveBackslash}[1]{\let\temp=\\#1\let\\=\temp}
\newcolumntype{C}[1]{>{\PreserveBackslash\centering}p{#1}}
\newcolumntype{R}[1]{>{\PreserveBackslash\raggedleft}p{#1}}
\newcolumntype{L}[1]{>{\PreserveBackslash\raggedright}p{#1}}

\crefname{equation}{Eq.}{Eqs.}
\crefname{figure}{Fig.}{Figs.}
\crefname{table}{Table}{Tables}

\renewcommand{\paragraph}[1]{\vspace{0.2cm}{\bf \textit{#1}}}
\def\ie{{\it i.e.},\ }

\newcommand{\mrm}{\mathrm}

\def\beq#1\eeq{\begin{equation}#1\end{equation}}
\def\beqs#1\eeqs{\begin{align}#1\end{align}}
\def\kk{\mathbf{k}}

\def\pare#1{\left( #1 \right)}

\def\ket#1{| #1 \rangle}

\def\nono{\nonumber}

\def\pr{\prime}

\def\ZZ{\mathbb{Z}}

\begin{document}

\title{High-Throughput Calculations of Magnetic Topological Materials}

\author{Yuanfeng Xu}
\affiliation{Max Planck Institute of Microstructure Physics, 06120 Halle, Germany}

\author{Luis Elcoro}
\affiliation{Department of Condensed Matter Physics, University of the Basque Country UPV/EHU, Apartado 644, 48080 Bilbao, Spain}

\author{Zhida Song}
\affiliation{Department of Physics, Princeton University, Princeton, New Jersey 08544, USA}

\author{Benjamin J. Wieder}
\affiliation{Department of Physics, Massachusetts Institute of Technology, Cambridge, MA 02139, USA}
\affiliation{Department of Physics, Northeastern University, Boston, MA 02115, USA}
\affiliation{Department of Physics, Princeton University, Princeton, New Jersey 08544, USA}

\author{M. G. Vergniory}
\affiliation{Donostia International Physics Center, P. Manuel de Lardizabal 4, 20018 Donostia-San Sebastian, Spain}
\affiliation{IKERBASQUE, Basque Foundation for Science, Bilbao, Spain}

\author{Nicolas Regnault}
\affiliation{Laboratoire de Physique de l'Ecole normale sup\'{e}rieure, ENS, Universit\'{e} PSL, CNRS, Sorbonne Universit\'{e}, Universit\'{e} Paris-Diderot, Sorbonne Paris Cit\'{e}, Paris, France}
\affiliation{Department of Physics, Princeton University, Princeton, New Jersey 08544, USA}

\author{Yulin Chen}
\affiliation{School of Physical Science and Technology, ShanghaiTech University, Shanghai 201210, China}
\affiliation{ShanghaiTech Laboratory for Topological Physics, Shanghai 200031, China}
\affiliation{Clarendon Laboratory, Department of Physics, University of Oxford, Oxford OX1 3PU, UK}
\affiliation{State Key Laboratory of Low Dimensional Quantum Physics, Department of Physics and Collaborative Innovation Center of Quantum Matter, Tsinghua University, Beijing 100084, China}

\author{Claudia Felser}
\affiliation{Max Planck Institute for Chemical Physics of Solids, Dresden D-01187, Germany}
\affiliation{Center for Nanoscale Systems, Faculty of Arts and Science, Harvard University, 11 Oxford Street, LISE 308Cambridge, MA 021138, USA}

\author{B. Andrei Bernevig }
\email{bernevig@princeton.edu}
\affiliation{Department of Physics, Princeton University, Princeton, New Jersey 08544, USA}
\affiliation{Max Planck Institute of Microstructure Physics, 06120 Halle, Germany}
\affiliation{Physics Department, Freie Universitat Berlin, Arnimallee 14, 14195 Berlin, Germany}

\date{\today}

\begin{abstract}
The discoveries of intrinsically magnetic topological materials, including semimetals with a large anomalous Hall effect and axion insulators \cite{wang2018large,liu2018giant,otrokov2019prediction}, have directed fundamental research in solid-state materials.
Topological Quantum Chemistry \cite{bradlyn_topological_2017} has enabled the understanding of and the search for paramagnetic topological materials \cite{vergniory_complete_2019, zhang2019catalogue}. 
Using magnetic topological indices obtained from magnetic topological quantum chemistry (MTQC)~\cite{MTQC}, here we perform the first high-throughput search for magnetic topological materials. 
here we perform a high-throughput search for magnetic topological materials based on first-principles calculations. We use as our starting point the Magnetic Materials Database on the Bilbao Crystallographic Server, which contains more than 549 magnetic compounds with magnetic structures deduced from neutron-scattering experiments, and identify 130 enforced semimetals (for which the band crossings are implied by symmetry eigenvalues), and topological insulators. For each compound, we perform complete electronic structure calculations, which include complete topological phase diagrams using different values of the Hubbard potential. Using a custom code to find the magnetic co-representations of all bands in all magnetic space groups, we generate data to be fed into the algorithm of MTQC to determine the topology of each magnetic material. Several of these materials display previously unknown topological phases, including symmetry-indicated magnetic semimetals, three-dimensional anomalous Hall insulators and higher-order magnetic semimetals. We analyse topological trends in the materials under varying interactions: 60 per cent of the 130 topological materials have topologies sensitive to interactions, and the others have stable topologies under varying interactions. We provide a materials database for future experimental studies and open-source code for diagnosing topologies of magnetic materials.
\end{abstract}

\maketitle
\tableofcontents

\section{Introduction}
Non-magnetic topological materials have dominated the landscape of topological physics for the past two decades. 
Research in this field has led to a rapid succession of theoretical and experimental discoveries; notable examples include the theoretical prediction of the first topological insulators (TIs) in two~\cite{PhysRevLett.95.146802,bernevig2006quantum} and three spatial dimensions~\cite{zhang2009topological}, topological crystalline insulators~\cite{hsieh2012topological}, Dirac and Weyl semimetals~\cite{burkov_weyl_2011,wan2011topological,PhysRevLett.107.186806, wang2012dirac, yang_classification_2014,weng2015weyl,huang2015weyl}, and non-symmorphic topological insulators and semimetals~\cite{PhysRevLett.108.140405,slager_space_2013,Liu2014,wang2016hourglass,wieder2018wallpaper}. 
Though topological materials were once believed to be rare and esoteric, recent advances in nonmagnetic topological materials have found that TIs and enforced semimetals (ESs) are much more prevalent than initially thought. In 2017 {\it Topological Quantum Chemistry} (TQC)  and the equivalent method of symmetry-based indicators provided a description of the universal global properties of all possible atomic limit band structures in all non-magnetic symmetry groups, in real and momentum space \cite{bradlyn_topological_2017, po2017symmetry, kruthoff_topological_2017, song_quantitative_2018,khalaf_symmetry_2017}. 
This allowed for a classification of the non-magnetic, non-trivial (topological) band structures through high-throughput methods that have changed our understanding of the number of topological materials existent in nature. About 40\%$-$50\% of all non-magnetic materials can be classified as topological at the Fermi level \cite{vergniory_complete_2019, zhang2019catalogue, tang2019comprehensive}, leading to a ``periodic table'' of topological materials.

These breakthroughs in non-magnetic materials have not yet been matched by similar advances in magnetic compounds, due to a multitude of challenges.  First, although a method for classifying band topology in the 1651 magnetic and nonmagnetic space groups (MSGs and SGs, respectively) was recently introduced \cite{watanabe2018structure}, there still does not exist a theory similar to TQC or equivalent methods \cite{bradlyn_topological_2017, po2017symmetry,kruthoff_topological_2017,song_quantitative_2018,khalaf_symmetry_2017} by which the indicator groups in Ref.~\cite{watanabe2018structure} can be linked to topological (anomalous) surface (and hinge) states.  Second, a full classification of the magnetic co-representations and compatibility relations has not yet been tabulated.  Third, code to compute the magnetic co-representations from {\it ab initio} band structures does not exist.  Fourth, and finally, even if all the above were available, 
the {\it ab initio} calculation of magnetic compounds is notoriously inaccurate for complicated magnetic structures beyond ferromagnets. 
Specifically, unless the magnetic structure of a material is known a priori, then the {\it ab initio} calculation will likely converge to a misleading ground state.
This has rendered the number of accurately predicted magnetic topological materials to be less than 10 \cite{hirschberger2016chiral,yang2017topological,PhysRevX.9.041039,wang2018large,liu2018giant,otrokov2019prediction,liu2019magnetic,belopolski2019discovery,PhysRevLett.122.206401,PhysRevB.100.201102,nie2017topological,zou2019study,PhysRevB.98.201116}. 

In the present work and in ref.~\cite{MTQC}, we present substantial advances towards solving all of the above challenges—which we have made freely available to the public on internet repositories (https://www.cryst.ehu.es/cryst/checktopologicalmagmat) -covering four years of our work on the subject, and over 70 years \cite{shubnikov1964colored} of research on the group theory, symmetry, and topology of magnetic materials. 
We present a full theory of magnetic indices, co-representations, compatibility relations, code with which to compute the magnetic co-representations directly from ab initio calculations, and we perform full local density approximation (LDA) + Hubbard U calculations on 549 magnetic structures, which have been accurately tabulated through the careful analysis of neutron-scattering data. We predict several novel magnetic topological phases in real materials, including higher-order magnetic Dirac semimetals with hinge arcs~\cite{wieder2019strong}, magnetic chiral crystals with long Fermi arcs, Dirac semimetals with nodes not related by time-reversal symmetry, Weyl points and nodal lines in non-collinear antiferrimagnets, and ideal axion insulators with gapped surface states and chiral hinge modes \cite{PhysRevLett.122.256402,wieder2018axion}. 

\section{Workflow}
Starting from the material database MAGNDATA \href{http://webbdcrista1.ehu.es/magndata/}{MAGNDATA} on the Bilbao Crystallographic Server (BCS) (the BCSMD), which contains portable magnetic structure files determined by neutron scattering experiments of more than 707 magnetic structures, we select 549 high-quality magnetic structures for the ab initio calculations. We take the magnetic configurations provided by BCSMD as the initial inputs and then perform ab initio calculations incorporating spin–orbit coupling. LDA + U are applied for each material with different Hubbard U parameters to obtain a full phase diagram. Then, we calculate the character tables of the valence bands of each material using the MagVasp2trace package. By feeding the character tables into the machinery of MTQC, that is, the Check Topological Magnetic Mat. (https://www.cryst.ehu.es/cryst/checktopologicalmagmat) function on BCS \cite{MTQC}, we identify the corresponding magnetic co-representations (irreps) and classify the materials into different topological categories. Here we define six topological categories:
\begin{enumerate}
\item  Band representation. Insulating phase consistent with atomic insulators.
\item Enforced semimetal with Fermi degeneracy (ESFD). Semimetal phase with a partially filled degenerate level at a high symmetry point in the Brillouin zone.
\item Enforced semimetal. Semimetal phase with un-avoidable level crossings along high symmetry lines or in high-symmetry planes in the Brillouin zone.
\item Smith-index semimetal (SISM). Semimetal phase with un-avoidable level crossings at generic points (away from high symmetry points/lines/planes) in the Brillouin zone.
\item Stable topological insulator. Insulating phase inconsistent with atomic insulators. The topology (inconsistency with atomic insulators) is stable against being coupled to atomic insulators.
\item Fragile topological insulator. Insulating phase inconsistent with atomic insulators. The topology is unstable against being coupled to certain atomic insulators.
\end{enumerate}

Further details about BCSMD, the calculation methods, the MagVasp2trace package, the identification of magnetic irreps, and definitions of the topological categories are given in the Methods.

\section{Topological phase diagrams}
With the irreps successfully identified, we classify 403 magnetic structures with convergent ground states into the six topological categories. We find that there are 130 materials (about 32\% of the total) that exhibit nontrivial topology for at least one of the U values in the phase diagram. We sort these materials into four groups based on their U-dependence: (1) 50 materials belong to the same topological categories for all values of U. These are the most robust topological materials. (2) 49 materials, on the other hand, are nontrivial at U = 0 but become trivial when U is larger than a critical value. (3) 20 materials have non-monotonous dependence on U: they belong to one topologically nontrivial categories at U = 0 and change to a different topologically nontrivial category at a larger value of U. (4) Six materials are trivial at U = 0 but become nontrivial after a critical value of U. The topology of these six interesting materials is thus driven by electron–electron interactions. The materials in this category are: CaCo$_2$P$_2$, YbCo$_2$Si$_2$, Ba$_5$Co$_5$ClO$_{13}$, U$_2$Ni$_2$Sn, CeCoGe$_3$, and CeMnAsO. The self-consistent calculations of the remaining 5 materials do not converge for at least one value of $U$, and hence the phase diagrams are not complete.
Complete classifications of the converged materials are tabulated in Appendix \ref{app:E}; the corresponding band structures are given in Appendix \ref{app:I}. 
In Table \cref{topopermsg}, we summarize the total number of topological materials in each magnetic space group (MSG) at different values of U. We have also provided full topological classifications and band structures of each material on the Topological Magnetic Materials Database (https://www.topologicalquantumchemistry.fr/magnetic).

In the scheme of MTQC, the stable magnetic topological insulators and SISMs are characterized by non-zero stable indices. These indices can be understood as generalizations of the Fu-Kane parity criterion for a three-dimensional topological insulator~\cite{FuKaneInversion}.
A complete table of the stable indices and the index-implied topological invariants, which include (weak) Chern numbers, the axion $\theta$ angle, and magnetic higher-order topological insulator (HOTI) indices, is given in ref. \cite{MTQC}.
In Appendix \ref{app:F}, we present examples of stable indices relevant to the present work, as well as their physical interpretations. Alhough there are many (1,651) magnetic and nonmagnetic space groups, we find in ref.~\cite{MTQC} that the stable indices of all of the MSGs are dependent on minimal indices in the set of the so-called minimal groups. Thus, to determine the stable indices of a material, we first subduce the representation of the MSG formed by the material to a representation of the corresponding minimal group—a subgroup of the MSG on which the indices are dependent. We then calculate the indices in the minimal group. Using this method, we find a tremendous variety of topological phases among the magnetic materials studied in this work, including axion insulators~\cite{wilczek1987two,qi2008topological,PhysRevB.81.245209,PhysRevLett.102.146805}, mirror topological crystalline insulators ~\cite{hsieh2012topological}, three-dimensional quantum anomalous Hall insulators, and SISMs. A complete table of the topology of all of the magnetic materials studied in this work is provided in Appendix \ref{app:E}.

We additionally discover many ESFD and enforced semimetal materials, in which unavoidable electronic band crossings respectively occur at high-symmetry ${\bf k}$ points or on high-symmetry lines or planes in the Brillouin zone. For each of the ESFD and enforced semimetal magnetic materials, we tabulate the ${\bf k}$ points where unavoidable crossings occur (see Appendix \ref{app:G}).

We did not discover any examples of magnetic materials for which the entire valence manifold is fragile topological. However, as will be discussed below, we discovered many magnetic materials with well isolated fragile bands in their valence manifolds, thereby providing examples of magnetic fragile bands in real materials.

\begin{table*}[t]
\caption{Topological categories vary with $U$. Shown are the number of magnetic topological insulators/SISMs and enforced semimetals/ESFDs in each MSG for different values of the Hubbard interaction $U =$ 0 eV, 2 eV and 4 eV. For $U = 0$, there are 38 topological insulators/SISMs and 73 enforced semimetals/ESFDs in total. For $U = 2$ eV, the numbers of topological insulators/SISMs and enforced semimetals/ESFDs decrease to 27 and 58, respectively. For $U = 4$ eV, the numbers of topological insulators/SISMs and enforced semimetals/ESFDs decrease to 24 and 57, respectively. Choosing the value of $U$ for each material for which the magnetic moments calculated ab initio lie closest to their experimentally measured values, there are 29 topological insulators/SISMs and 62 enforced semimetals/ESFDs.
}\label{topopermsg}
\begin{centering}
\begin{tabular}{c|ccc|ccc|c|ccc|ccc|c|ccc|ccc}
\hline \hline
\multirow{2}{*}{MSG} & \multicolumn{3}{c|}{TIs/SISMs} & \multicolumn{3}{c|}{ESs/ESFDs} & 
\multirow{2}{*}{MSG} & \multicolumn{3}{c|}{TIs/SISMs} & \multicolumn{3}{c|}{ESs/ESFDs} &
\multirow{2}{*}{MSG} & \multicolumn{3}{c|}{TIs/SISMs} & \multicolumn{3}{c}{ESs/ESFDs}  \\
& U=0 & U=2 & U=4 & U=0 & U=2 & U=4 & & U=0 & U=2 & U=4 & U=0 & U=2 & U=4 & & U=0 & U=2 & U=4  & U=0 & U=2 & U=4 \\
\hline
 2.7 & 1 & 0 & 1 & 0 & 1 & 0 & 62.447 & 0 & 1 & 0 & 1 & 0 & 0 & 129.416 & 0 & 0 & 1 & 0 & 1 & 0 \\
 4.7 & 0 & 1 & 0 & 0 & 0 & 0 & 62.450 & 2 & 2 & 1 & 2 & 1 & 2 & 130.432 & 0 & 1 & 0 & 1 & 0 & 1 \\
 11.54 & 1 & 0 & 1 & 0 & 0 & 0 & 63.462 & 0 & 2 & 0 & 2 & 0 & 2 & 132.456 & 0 & 1 & 0 & 1 & 0 & 1 \\
 11.57 & 1 & 0 & 0 & 0 & 0 & 0 & 63.463 & 0 & 1 & 0 & 1 & 0 & 1 & 134.481 & 3 & 0 & 1 & 1 & 2 & 0 \\
 12.62 & 2 & 0 & 1 & 0 & 1 & 0 & 63.464 & 1 & 1 & 1 & 1 & 0 & 1 & 135.492 & 0 & 2 & 0 & 2 & 0 & 2 \\
 12.63 & 0 & 0 & 0 & 0 & 0 & 0 & 63.466 & 0 & 0 & 2 & 0 & 1 & 1 & 138.528 & 1 & 0 & 1 & 0 & 1 & 0 \\
 13.73 & 2 & 0 & 2 & 0 & 1 & 0 & 63.467 & 0 & 0 & 0 & 0 & 1 & 0 & 139.536 & 0 & 1 & 0 & 1 & 0 & 1 \\
 14.75 & 0 & 1 & 0 & 0 & 0 & 0 & 64.480 & 3 & 0 & 2 & 0 & 0 & 1 & 139.537 & 0 & 1 & 0 & 1 & 0 & 1 \\
 14.80 & 1 & 0 & 0 & 0 & 0 & 0 & 65.486 & 0 & 0 & 0 & 1 & 0 & 1 & 140.550 & 0 & 2 & 1 & 2 & 1 & 2 \\
 15.89 & 2 & 1 & 3 & 0 & 3 & 0 & 65.489 & 1 & 0 & 0 & 1 & 0 & 1 & 141.556 & 0 & 1 & 0 & 0 & 1 & 0 \\
 15.90 & 4 & 0 & 0 & 0 & 0 & 0 & 67.510 & 0 & 0 & 1 & 0 & 0 & 0 & 141.557 & 1 & 4 & 0 & 1 & 0 & 1 \\
 18.22 & 0 & 1 & 0 & 0 & 0 & 0 & 70.530 & 0 & 1 & 0 & 0 & 0 & 0 & 148.19 & 0 & 1 & 0 & 1 & 0 & 1 \\
 33.154 & 0 & 1 & 0 & 0 & 0 & 0 & 71.536 & 0 & 1 & 0 & 1 & 0 & 1 & 155.48 & 0 & 0 & 0 & 0 & 0 & 1 \\
 36.178 & 0 & 0 & 0 & 0 & 0 & 1 & 73.553 & 1 & 0 & 1 & 0 & 1 & 0 & 161.69 & 0 & 1 & 0 & 1 & 0 & 0 \\
 38.191 & 0 & 1 & 0 & 0 & 0 & 0 & 74.559 & 0 & 1 & 0 & 1 & 0 & 1 & 161.71 & 0 & 2 & 0 & 2 & 0 & 0 \\
 49.270 & 0 & 1 & 0 & 1 & 0 & 1 & 85.59 & 0 & 1 & 0 & 1 & 0 & 1 & 165.95 & 0 & 1 & 0 & 0 & 0 & 0 \\
 49.273 & 0 & 1 & 1 & 0 & 1 & 0 & 88.81 & 0 & 1 & 0 & 0 & 0 & 0 & 166.101 & 1 & 3 & 0 & 3 & 1 & 2 \\
 51.295 & 0 & 1 & 0 & 1 & 0 & 1 & 92.114 & 0 & 1 & 0 & 1 & 0 & 1 & 166.97 & 0 & 1 & 0 & 1 & 0 & 1 \\
 51.298 & 0 & 1 & 0 & 1 & 0 & 1 & 107.231 & 0 & 0 & 0 & 1 & 0 & 1 & 167.108 & 0 & 1 & 0 & 0 & 0 & 0 \\
 53.334 & 0 & 0 & 0 & 0 & 1 & 0 & 114.282 & 0 & 1 & 0 & 0 & 0 & 0 & 185.201 & 1 & 0 & 0 & 0 & 0 & 0 \\
 57.391 & 1 & 0 & 1 & 0 & 1 & 0 & 123.345 & 0 & 1 & 0 & 1 & 0 & 1 & 192.252 & 0 & 2 & 0 & 2 & 0 & 2 \\
 58.398 & 0 & 1 & 0 & 0 & 0 & 0 & 124.360 & 1 & 3 & 0 & 4 & 0 & 4 & 194.268 & 0 & 0 & 0 & 0 & 0 & 1 \\
 58.399 & 0 & 2 & 0 & 2 & 0 & 2 & 125.373 & 0 & 1 & 0 & 1 & 0 & 1 & 205.33 & 1 & 0 & 0 & 0 & 0 & 0 \\
 59.407 & 0 & 1 & 0 & 0 & 0 & 0 & 126.386 & 0 & 1 & 0 & 1 & 0 & 1 & 222.103 & 0 & 1 & 0 & 1 & 0 & 1 \\
 59.416 & 0 & 0 & 1 & 0 & 1 & 0 & 127.394 & 1 & 1 & 1 & 1 & 1 & 1 & 224.113 & 2 & 1 & 2 & 0 & 2 & 0 \\
 60.431 & 1 & 0 & 0 & 0 & 0 & 0 & 127.397 & 0 & 1 & 0 & 1 & 0 & 1 & 227.131 & 0 & 1 & 0 & 0 & 0 & 0 \\
 61.439 & 0 & 0 & 1 & 0 & 0 & 0 & 128.408 & 0 & 1 & 0 & 1 & 0 & 1 & 228.139 & 2 & 1 & 0 & 3 & 0 & 3 \\
 62.441 & 0 & 3 & 0 & 0 & 0 & 0 & 128.410 & 0 & 4 & 0 & 4 & 0 & 4 & Total & 38 & 27 & 24 & 73 & 58 & 57 \\
\hline
\end{tabular}
\end{centering}
\end{table*}

\section{High-quality topological materials }
\begin{figure*}[htbp] 
\centering\includegraphics[width=7.1in]{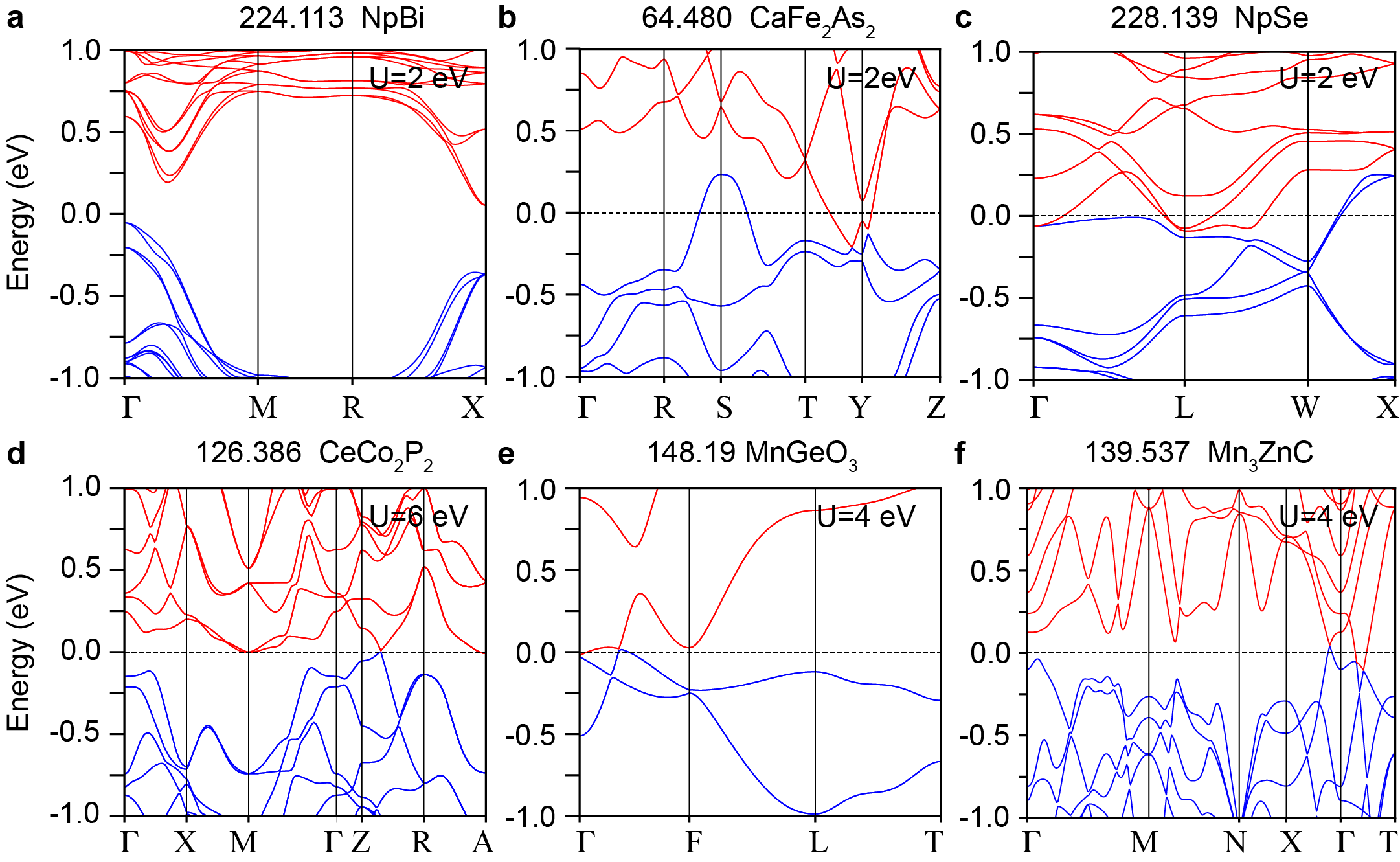}
\caption{Band structures of the ‘high-quality’ magnetic topological materials predicted by MTQC.
(a, b) The antiferromagnetic axion topological insulators, NpBi and CaFe$_2$As$_2$. Although there are Fermi pockets around $S$ and $Y$ in CaFe$_2$As$_2$, the insulating compatibility relations are fully satisfied. We note that there is a small gap (about 5 meV) along the path T–Y; this indicates that the valence bands are well separated from the conduction bands, and thus have a well defined topology. (c) The antiferromagnetic ESFD NpSe, which has a partially filled fourfold degeneracy at $\Gamma$. (d) The antiferromagnetic nodal-line semimetal CeCo$_2$P$_2$. A gapless nodal ring protected by mirror symmetry lies in the Z–R–A plane. (e) The antiferromagnetic Dirac semimetal MnGeO$_3$. One of the two Dirac nodes protected by the $C_3$-rotation symmetry lies along the high-symmetry line $\Gamma$–F. Note that there is a small bandgap at the $\Gamma$ point. (f) The non-collinear ferrimagnetic Weyl semimetal Mn$_3$ZnC. Two Weyl points are pinned to the rotation-invariant line $\Gamma$–T by $C_4$-rotation symmetry. Mn$_3$ZnC also hosts nodal lines at the Fermi level $E_F$; we specifically observe five nodal rings protected by the mirror symmetry ($M_z$) in the plane $k_z = 0$. The sequential number of each MSG in the BNS setting and the chemical formula of each material are provided on the top of each panel.
}\label{fig1} 
\end{figure*}

We here select several representative ``high-quality'' topological materials with clean band structures at the Fermi level:
NpBi in MSG 224.113 ($Pn\bar3m'$) (antiferromagnetic stable topological insulator), 
CaFe$_2$As$_2$ in MSG 64.480 ($C_Amca$) (antiferromagnetic stable topological insulator), 
NpSe in MSG 228.139 ($F_Sd\bar3c$) (antiferromagnetic ESFD), 
CeCo$_2$P$_2$ in MSG 126.386 ($P_I4/nnc$) (antiferromagnetic enforced semimetal), 
MnGeO$_3$ in MSG 148.19 ($R\bar3'$) (antiferromagnetic enforced semimetal), 
Mn$_3$ZnC in MSG 139.537 ($I4/mm'm'$) (non-collinear ferrimagnetic enforced semimetal), 
as shown in \cref{fig1}.

To identify the stable topologies of the antiferrimagnets NpBi and CaFe$_2$As$_2$, we calculate the stable indices subduced onto MSG 2.4 ($P\bar1$), a subgroup of MSG 224.113 ($Pn\bar3m'$) and 64.480 ($C_Amca$).

\begin{figure*}[htbp] 
\centering\includegraphics[width=7.1in]{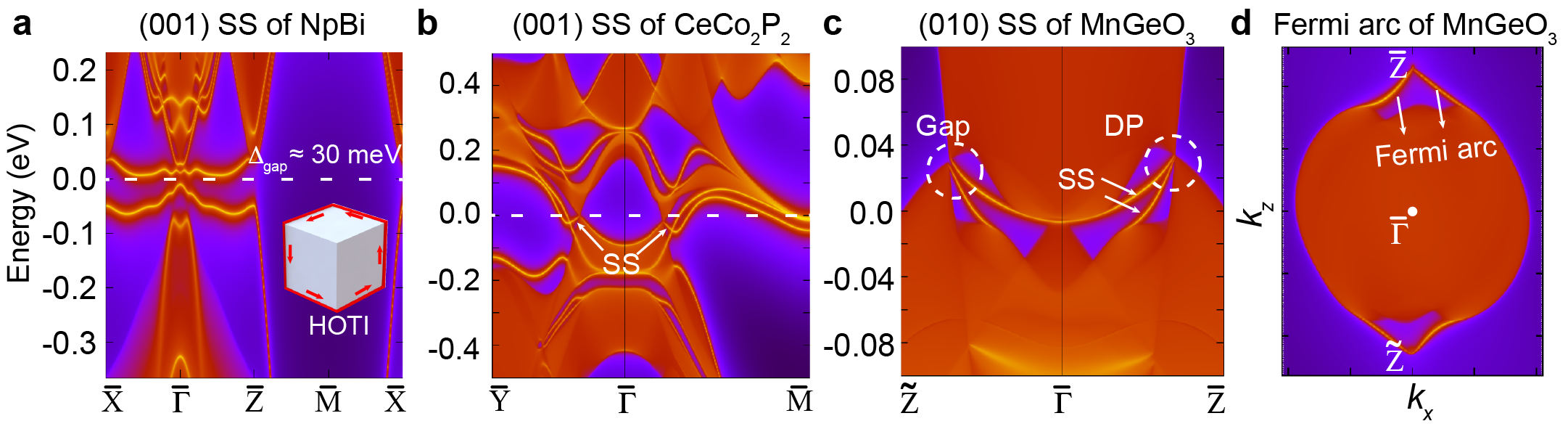}
\caption{Topological surface states of representative magnetic topological insulator and enforced semimetal phases.
(a) The (001) surface state of the axion insulator NpBi, which has an energy gap of 30 meV. The inset shows a schematic of the chiral hinge states on a cubic sample. (b) The (001) surface state of the enforced semimetal CeCo$_2$P$_2$. The drumhead-like topological surface states connect the projections of the bulk nodal rings. (c) The (010) surface state of the enforced semimetal MnGeO$_3$. The bulk Dirac point along the $\bar\Gamma-\bar{Z}$ line is protected by $C_3$ symmetry. However, because time-reversal symmetry is broken, the projected band crossing on $\bar\Gamma-\tilde{Z}$ (along -$k_z$) is no longer protected, and is instead weakly gapped. The coordinates of $\bar{Z}$ and $\tilde{Z}$ on the (010) surface are (0, $k_z=\pi/c$) and  (0, $k_z=-\pi/c$), respectively. (d) The surface Fermi arcs connecting the Dirac points on the (010) surface of MnGeO$_3$.
}
\label{fig2} 
\end{figure*}

The stable indices in MSG 2.4 ($P\bar{1}$) are defined using only parity (inversion) eigenvalues~\cite{MTQC, PhysRevB.83.245132,PhysRevB.85.165120, watanabe2018structure,PhysRevB.97.205136,PhysRevLett.122.256402}: \begin{align}
\eta_{4I}=\sum_{K} n_{K}^- \mod 4, \label{eq:z4-main}\\
z_{2I,i}=\sum_{K, K_i=\pi} n_{K}^- \mod 2
\end{align}
where $K$ sums over the eight inversion-invariant momenta, and $n_K^-$ is the number of occupied states with odd parity eigenvalues at momentum $K$. 
$z_{2I,i}$ is the parity of the Chern number of the Bloch states in the plane $k_i=\pi$. 
As explained in Appendix \ref{app:F1}, $\eta_{4I}=1,3$ correspond to Weyl semimetal (WSM) phases with odd-numbered Weyl points in each half of the BZ; $\eta_{4I}=2$ indicates an axion insulator phase provided that the band structure is fully gapped and the weak Chern numbers in all directions are zero.
The inversion eigenvalues of NpBi with $U$=2 eV and CaFe$_2$As$_2$ with $U$=2 eV are tabulated in \cref{tablez4}.
The corresponding band structures are shown in Fig. \ref{fig2}a and Fig. \ref{fig2}b, respectively.
Both NpBi and  CaFe$_2$As$_2$ have the indices ($\eta_{4I},z_{2I,1},z_{2I,2},z_{2I,3}$)=(2,0,0,0).
As shown in Appendix \ref{app:F6}, the $\eta_{4I}$ index of NpBi is the same for $U$=0, 2, 4, 6 eV; whereas the $\eta_{4I}$ index of CaFe$_2$As$_2$ is 2 for $U$=0, 1, 2 eV and 0 for $U$=3, 4 eV. 
We have confirmed that both NpBi and CaFe$_2$As$_2$ have vanishing weak Chern numbers, implying that the $\eta_{4I}=2$ phases must be axion insulators.

An axion insulator is defined by a nontrivial $\theta$ angle, which necessitates a quantized magneto-electric response in the bulk and chiral hinge modes on the boundary~\cite{wieder2018axion,wilczek1987two,qi2008topological,PhysRevB.81.245209,PhysRevLett.102.146805}.
We have calculated the surface states of NpBi and find that, as expected, the (001) surface is fully gapped (\cref{fig2}a).
Owing to the $C_3$-rotation symmetry of the MSG 224.113, the (100) and (010) surfaces are also gapped. 
Therefore, a cubic sample with terminating surfaces in the (100), (010) and (001) directions, as shown in \cref{fig2}a, exhibits completely gapped surfaces. 
However, as an axion insulator, it must exhibit chiral hinge modes when terminated in an inversion-symmetric geometry \cite{wieder2018axion, PhysRevLett.122.256402}. 
We predict that the chiral hinge modes exist on the edges shown in \cref{fig2}a.
More details about NpBi are provided in Appendix \ref{app:H1}.

\begin{table*}[t]
\caption{Parities and topological indices of two magnetic topological insulators. 
Shown are the numbers of occupied bands with odd/even parity eigenvalues at the eight inversion-invariant points ($\eta_{\alpha}$) for the magnetic topological insulators NpX (where X = Sb, Bi) and XFe$_2$As$_2$ (where X = Ca, Ba) with $U =$ 2 eV. The $\eta_{4I} = 2$ phase corresponds to an axion insulator or Weyl semimetal phase with even pairs of Weyl points at generic locations in the Brillouin zone interior (given vanishing weak Chern numbers), and $\eta_{4I} = 1, 3$ corresponds to a Weyl semimetal phase with odd number of Weyl points at generic locations within each half of the Brillouin zone. We have confirmed that the weak Chern numbers vanish in NpX and XFe$_2$As$_2$, implying that both materials are axion insulators.}
\label{tablez4}
\begin{centering}
\begin{tabular}{c|p{1.4cm}p{1.4cm}p{1.4cm}p{1.4cm}p{1.4cm}p{1.4cm}p{1.4cm}p{1.4cm}|c}
\hline \hline
$\Lambda_\alpha$ & (0,0,0) & ($\pi$,0,0) & (0,$\pi$,0) & ($\pi$,$\pi$,0) & (0,0,$\pi$) & ($\pi$,0,$\pi$) & (0,$\pi$,$\pi$) & ($\pi$,$\pi$,$\pi$) &  ($\eta_{4I},z_{2I,1},z_{2I,2},z_{2I,3}$) 
\tabularnewline
\hline
NpX & 58/22  & 40/40  & 40/40 & 40/40  & 40/40  & 40/40 & 40/40  & 48/32 & (2,0,0,0)  \tabularnewline
 \hline
XFe$_2$As$_2$ & 50/46  & 48/48  & 48/48 & 52/44  & 48/48  & 52/44 & 52/44  & 48/48 & (2,0,0,0)  \tabularnewline
 \hline \hline
\end{tabular}
\end{centering}
\end{table*}

Next, we discuss representative examples of magnetic topological semimetals. The antiferrimagnet NpSe with $U$=2 eV, 4 eV and 6 eV is an ESFD with a partially-filled degenerate band at the $\Gamma$ point, where the lowest conduction bands and the highest valence bands meet in a fourfold degeneracy (Fig. \ref{fig1}c).
The antiferrimagnet CeCo$_2$P$_2$ is an enforced semimetal at all the $U$ values used in our calculations.  
For $U=0$ eV and 2 eV, we predict CeCo$_2$P$_2$ to be a Dirac semimetal protected by $C_4$ rotation symmetry. 
Because the Dirac points in CeCo$_2$P$_2$ lie along a high-symmetry line ($\Gamma \mathrm{Z}$) whose little group contains $4mm$, we see that CeCo$_2$P$_2$ is a higher-order topological semimetal that exhibits flat-band-like higher-order Fermi-arc states on mirror-invariant hinges, analogous to the hinge states recently predicted~\cite{wieder2019strong} and experimentally observed~\cite{Cd3As2HingeConfirm} in the non-magnetic Dirac semimetal Cd$_3$As$_2$.
For $U=4$ eV and 6 eV, CeCo$_2$P$_2$ becomes a nodal ring semimetal protected by the mirror symmetry $M_z$.
As detailed in Appendix \ref{app:H2}, the transition between the two enforced semimetal phases is completed by two successively band inversions at $\Gamma$ and Z, which removes the Dirac node and creates the nodal ring, respectively. 
The band structure of the nodal ring semimetal phase at $U$=6 eV is plotted in \cref{fig1}d.
The antiferromagnet MnGeO$_3$ is a $C_3$-rotation-protected Dirac semimetal, in which the number of Dirac nodes changes with the value of $U$. 
For $U$=0 eV, 1 eV, 3 eV and 4 eV, we predict MnGeO$_3$ to have two Dirac nodes along the high-symmetry line $\rm \Gamma F$;
for $U$=2 eV, we observe four Dirac nodes along the same high-symmetry line.
In Fig. \ref{fig1}e, we plot the band structure of MnGeO$_3$ using $U$=4 eV. 
We next predict the non-collinear ferrimagnet Mn$_3$ZnC to be an ES with symmetry-enforced Weyl points coexisting with the Weyl nodal rings  (\cref{fig1}f).
Two of the Weyl points in Mn$_3$ZnC are pinned by the $C_4$-rotation symmetry to the high-symmetry line $\Gamma\rm T$, 
and we observe five nodal rings protected by the mirror symmetry $M_z$ in the $k_z=0$ plane.
In time-reversal-breaking Weyl semimetals, divergent Berry curvature near Weyl points can give rise to a large intrinsic anomalous Hall conductivity \cite{wang2018large,liu2018giant,liu2019magnetic,morali2019fermi,belopolski2019discovery}. 
We thus expect there to be a large anomalous Hall effect in Mn$_3$ZnC.
As detailed in Appendix \ref{app:H4}, we have specifically calculated the the anomalous Hall conductivity of Mn$_3$ZnC to be about $123$ $\rm \Omega^{-1}\cdot cm^{-1}$.

The surface states of the enforced semimetals CeCo$_2$P$_2$ and MnGeO$_3$ are shown in \cref{fig2}b,c, respectively.
Because the bulk states of CeCo$_2$P$_2$ and MnGeO$_3$ have clean Fermi surfaces, the surface states are well separated from the bulk states, and should be observable in experiment.
For the Dirac semimetal MnGeO$_3$, we observe a discontinuous Fermi surface (Fermi-arc) on the surface (\cref{fig2}d). 
In Appendix \ref{app:H}, we provide further details of our surface-state calculations.  

\section{Consistency with previous works}
Our magnetic materials database (https://www.topologicalquantumchemistry.fr/magnetic) includes several topological materials that have previously been reported but whose topology was not known to be protected by symmetry eigenvalues. For example, the non-collinear magnet Mn$_3$Sn in MSG 63.463 ($Cm'cm'$) has been reported as a magnetic Weyl semimetal candidate with six pairs of Weyl points \cite{yang2017topological,kuroda2017evidence}.
In our LDA+U calculation, for $U=$0 eV, 1 eV and 2 eV, we find Mn$_3$Sn to be classified instead as a magnetic topological insulator category with the index $\eta_{4I}=2$. $\eta_{4I}=2$ can correspond to several different topological phases (which we emphasize are not all topological insulators): (1) an axion insulator, (2) a three-dimensional quantum anomalous Hall state with even weak Chern number (not determinable from symmetry eigenvalues) \cite{PhysRevB.101.155130}, or (3) a Weyl semimetal phase with an \emph{even} number of Weyl points in half of the BZ (not determinable from symmetry eigenvalues). 
Thus our calculations on Mn$_3$Sn for $U=0,1,2$ eV  are consistent with the results in refs.\cite{yang2017topological,kuroda2017evidence}. 
We emphasize that if the six Weyl points in half of the Brillouin zone were pairwise annihilated without closing a gap at the inversion-invariant momenta, then the gapped phase would either be an axion insulator or a three-dimensional quantum anomalous Hall state. When U is further increased to 3 eV and 4 eV, a topological phase transition occurs, driving the $\eta_{4I}=2$ phase into a gapless enforced semimetal phase.

\section{Chemical Categories}
In Table~\ref{tab:chemicalclass}, we classify the topological magnetic materials predicted by MTQC into three main chemical categories, and 11 sub-categories, through a consideration of their magnetic ions and chemical bonding. Detailed descriptions of each category are given in the Methods. Of the materials listed in Table~\ref{tab:chemicalclass}, most antiferromagnetic insulators, which are well studied experimentally in the case of the so-called Mott insulators, appear to be trivial. We observe that most of the materials in Table~\ref{tab:chemicalclass} are identified as topological enforced semimetals or ESFDs, which are defined by small densities of states at the Fermi level, and hence lie chemically at the border between insulators and metals.

\begin{table*}
\centering
\begin{tabular}{p{1.5cm}|p{6cm}|p{10.0cm}}
\hline
\hline
Categories & Properties & Materials  \\ 
\hline
I-A & Non-collinear Manganese compounds & Mn$_3$GaC, Mn$_3$ZnC, Mn$_3$CuN, Mn$_3$Sn, Mn$_3$Ge, Mn$_3$Ir, Mn$_3$Pt, Mn$_5$Si$_3$ \\
\hline
I-B & Actinide Intermetallic & UNiGa$_5$, UPtGa$_5$, NpRhGa$_5$, NpNiGa$_5$ \\
 \hline
I-C & Rare earth intermetallic & NdCo$_2$, TbCo$_2$, NpCo$_2$, PrAg
DyCu, NdZn, TbMg, NdMg, Nd$_5$Si$_4$, Nd$_5$Ge$_4$, Ho$_2$RhIn$_8$, Er$_2$CoGa$_8$, Nd$_2$RhIn$_8$,
Tm$_2$CoGa$_8$, Ho$_2$RhIn$_8$, DyCo$_2$Ga$_8$, TbCo$_2$Ga$_8$, Er$_2$Ni$_2$In, CeRu$_2$Al$_{10}$, Nd$_3$Ru$_4$Al$_{12}$, Pr$_3$Ru$_4$Al$_{12}$, ScMn$_6$Ge$_6$, YFe$_4$Ge$_4$, LuFe$_4$Ge$_4$, CeCoGe$_3$ \\
\hline
II-A & Metallic Iron pnictides & LaFeAsO, CaFe$_2$As$_2$, EuFe$_2$As$_2$, BaFe$_2$As$_2$, Fe$_2$As, CaFe$_4$As$_3$, LaCrAsO, Cr$_2$As, CrAs, CrN \\
\hline
II-B & Semiconducting manganese pnictides & BaMn$_2$As$_2$ BaMn$_2$Bi$_2$, CaMnBi$_2$, SrMnBi$_2$, CaMn$_2$Sb$_2$, CuMnAs, CuMnSb, Mn$_2$As \\
\hline
II-C & Rare earth intermetallic compounds with the composition 1:2:2 & PrNi$_2$Si$_2$, YbCo$_2$Si$_2$, DyCo$_2$Si$_2$, PrCo$_2$P$_2$, CeCo$_2$P$_2$, NdCo$_2$P$_2$, DyCu$_2$Si$_2$, CeRh$_2$Si$_2$, UAu$_2$Si$_2$, U$_2$Pd$_2$Sn, U$_2$Pd$_2$In, U$_2$Ni$_2$Sn, U$_2$Ni$_2$In, U$_2$Rh$_2$Sn \\
\hline
II-D & Rare earth ternary compounds of the composition 1:1:1 & CeMgPb, PrMgPb, NdMgPb, TmMgPb \\
\hline
III-A & Semiconducting Actinides/Rare earth Pnictides & HoP, UP, UP$_2$, UAs, NpS, NpSe, NpTe, NpSb, NpBi, U$_3$As$_4$, U$_3$P$_4$ \\
\hline
III-B & Metallic oxides & Ag$_2$NiO$_2$, AgNiO$_2$, Ca$_3$Ru$_2$O$_7$, Double perovskite Sr$_3$CoIrO$_6$ \\
\hline
III-C & Metal to insulator transition compounds & NiS$_2$, Sr$_2$Mn$_3$As$_2$O$_2$ \\
\hline
III-D & Semiconducting and insulating oxides, borates, hydroxides, silicates, phosphate & LuFeO$_3$, PdNiO$_3$, ErVO$_3$, DyVO$_3$, MnGeO$_3$, Tm$_2$Mn$_2$O$_7$, Yb$_2$Sn$_2$O$_7$, 
Tb$_2$Sn$_2$O$_7$, Ho$_2$Ru$_2$O$_7$, Er$_2$Ti$_2$O$_7$, Tb$_2$Ti$_2$O$_7$, Cd$_2$Os$_2$O$_7$, Ho$_2$Ru$_2$O$_7$, Cr$_2$ReO6, NiCr$_2$O$_4$, MnV$_2$O$_4$, Co$_2$SiO$_4$, Fe$_2$SiO$_4$, PrFe$_3$(BO$_3$)$_4$, KCo$_4$(PO$_4$)$_3$, CoPS$_3$, SrMn(VO$_4$)(OH), Ba$_5$Co$_5$ClO$_{13}$, FeI$_2$\\
\hline
\hline
\end{tabular}
\caption{ The magnetic topological materials identified in this work.}\label{tab:chemicalclass}
\end{table*}

\section{Discussion}
A large number of the topological materials predicted in this work (see Appendix \ref{app:E} for a complete tabulation) can readily be synthesized into single crystals for the exploration of their unusual physical properties and the confirmation of their topological electronic structures in different phase categories. These include materials with non-trivial topology over the full range of U values used in our calculations (for example, Mn$_3$Ge, Mn$_3$Sn, Mn$_3$Ir, LuFe$_4$Ge$_2$, and YFe$_4$Ge$_2$), materials sensitive to $U$ (for example, NdCo$_2$ and NdCo$_2$P$_2$), and interaction-driven topological materials (for example,  U$_2$Ni$_2$Sn and CeCuGe$_3$).

We did not find any examples of materials whose entire valence manifolds are fragile topological. However, it is still possible for well isolated bands within the valence manifold to be fragile topological if they can be expressed as a difference of band representations. We find many examples of energetically well isolated fragile branches among the occupied bands. We tabulate all the fragile branches close to the Fermi level in Appendix \ref{app:J}.

We emphasize that there also exist topological insulators and topological semimetals (for example, Weyl semimetals) that cannot be diagnosed through symmetry eigenvalues, which in this work are classified as trivial band representations \cite{bradlyn_topological_2017}.
It is worth mentioning that even the topologically trivial bands may also be interesting if the occupied bands form Wannier functions centred at positions away from the atoms, because a Wannier centre shift in three-dimensional insulators leads to the appearance of topological corner states, like those of quantized ‘quadrupole’ insulators~\cite{wieder2019strong}.
Topological phases characterized by displaced Wannier functions are known as obstructed atomic limits; we leave their high-throughput calculation for future studies.

\section{Conclusion}
We have performed LDA + U calculations on 549 existent magnetic structures and have successfully classified 403 using the machinery of MTQC~\cite{MTQC}. We find that 130 materials (about 32\% of the total) have topological phases as we scan the U parameter. Our results suggest that a large number of previously synthesized magnetic materials are topologically nontrivial. We highlight several ‘high-quality’ magnetic topological materials that should be experimentally examined for topological response effects and surface (and hinge) states.

{\bf Acknowledgements}
We thank U. Schmidt, I. Weidl, W. Shi and Y. Zhang. We acknowledge the computational resources Cobra in the Max Planck Computing and Data Facility (MPCDF), the HPC Platform of ShanghaiTech University and Atlas in the Donostia International Physics Center (DIPC). Y.X. is grateful to D. Liu for help in plotting some diagrammatic sketches. B.A.B., N.R., B.J.W. and Z.S. were primarily supported by a Department of Energy grant (DE-SC0016239), and partially supported by the National Science Foundation (EAGER grant DMR 1643312), a Simons Investigator grant (404513), the Office of Naval Research (ONR; grant N00014-14-1-0330), the NSF-MRSEC (grant DMR-142051), the Packard Foundation, the Schmidt Fund for Innovative Research, the BSF Israel US foundation (grant 2018226), the ONR (grant N00014-20-1-2303) and a Guggenheim Fellowship (to B.A.B.). Additional support was provided by the Gordon and Betty Moore Foundation through grant GBMF8685 towards the Princeton theory programme. L.E. was supported by the Government of the Basque Country (Project IT1301-19) and the Spanish Ministry of Science and Innovation (PID2019-106644GB-I00). M.G.V. acknowledges support from the Diputacion Foral de Gipuzkoa (DFG; grant INCIEN2019-000356) from Gipuzkoako Foru Aldundia and the Spanish Ministerio de Ciencia e Innovación (grant PID2019-109905GB-C21). Y.C. was supported by the Shanghai Municipal Science and Technology Major Project (grant 2018SHZDZX02) and a Engineering and Physical Sciences Research Council (UK) Platform Grant (grant EP/M020517/1). C.F. acknowledges financial support by the DFG under Germany’s Excellence Strategy through the Würzburg-Dresden Cluster of Excellence on Complexity and Topology in Quantum Matter (ct.qmat EXC 2147, project-id 390858490), an ERC Advanced Grant (742068 ‘TOPMAT’). Y.X. and B.A.B. were also supported by the Max Planck Society.

{\bf Author contributions}
B.A.B. conceived this work; Y.X. and M.G.V. performed the first-principles calculations. L.E. wrote the code for calculating the irreducible representations and checking the topologies of materials. Y.X., Z.S., B.J.W. and B.A.B. analysed the calculated results, B.J.W. determined the physical meaning of the topological indices with help from L.E., Z.S. and Y.X. C.F. performed chemical analysis of the magnetic topological materials. N.R. built the topological material database. All authors wrote the main text and Y.X. and Z.S. wrote the Methods and the Supplementary Information.

{\bf Competing interests}

The authors declare no competing interests.

{\bf Correspondence and requests for materials} should be addressed to B.A.B.

{\bf Corresponding authors}

Correspondence to \href{bernevig@princeton.edu}{B. Andrei Bernevig}.

 {
\section{Methods}
\vspace{0.15in}
}
\textbf{Concepts}
Here we give a brief introduction to MTQC \cite{MTQC, bradlyn_topological_2017,elcoro_double_2017,cano_building_2018,vergniory_graph_2017} and the definitions of six topological classes.
A magnetic band structure below the Fermi level is partially described by the irreducible co-representations (irreps) formed by the occupied electronic states at the high-symmetry ${\bf k}$ points, which are defined as the momenta whose little groups - the groups that leave the momenta unchanged - are maximal subgroups of the space group.
If the highest occupied (valence) band and the lowest unoccupied (conduction) band are degenerate at a high-symmetry ${\bf k}$ point, then we refer to the material as an enforced semimetal with Fermi degeneracy (ESFD) \cite{bradlyn_topological_2017}.
Depending on whether the irreps at high-symmetry points satisfy the so-called compatibility relations~\cite{bradlyn_topological_2017,vergniory_graph_2017,po2017symmetry,kruthoff_topological_2017} -- which determine whether the occupied bands must cross with unoccupied bands along high-symmetry lines or planes (whose little groups are non-maximal) -- band structures can then be further classified as insulating (along high-symmetry lines and planes) or enforced semimetals (ES).  ES-classified materials generically feature band crossings along high-symmetry lines or planes.
If a band structure satisfies the compatibility relations, it can be a trivial insulator, whose occupied bands form a BR \cite{bradlyn_topological_2017}, a topological semimetal with crossing nodes at generic momenta [Smith-index semimetal (SISM) or non-symmetry-indicated topological semimetal - a system which satisfies all compatibility relations but exhibits Weyl-type nodes], or a TI.
Some of the topological semimetals and insulators can be diagnosed through their irreps: If the irreps do not match a BR, then the band structure must be a topological insulator or a SISM.
There are two types of topological insulators: Stable TIs \cite{khalaf_symmetry_2017,song_quantitative_2018,Songeaax2007,Luis2019}, which include crystalline and higher-order TIs (TCIs and HOTIs, respectively)~\cite{benalcazar_quantized_2017,schindler_higher-order_2018,schindler_higher-order_2018-1,song_d-2-dimensional_2017,langbehn_reflection-symmetric_2017}, and fragile TIs \cite{po_fragile_2018,cano_topology_2018,bradlyn_disconnected_2019,ahn_failure_2019,song_fragile_2019,Luis2019}.  
Stable TIs remain topological when coupled to trivial or fragile bands, whereas fragile TIs, on the other hand, can be trivialized by being coupled to certain trivial bands, or even other fragile bands~\cite{wieder2018axion,wieder2019strong}.
In the accompanying paper \cite{MTQC}, we explicitly identify all of the symmetry-indicated stable electronic (fermionic) TIs and topological semimetals, specifically detailing the bulk, surface, and hinge states of all symmetry-indicated stable TIs, TCIs, and HOTIs in all 1651 spinful (double) SGs and MSGs.

To summarize, using MTQC, we divide an electronic band structure into one of six topological classes: BRs, ESFD, ES, SISM, stable TI, and fragile TI, among which only BRs are considered to be topologically trivial.  If a band structure satisfies the compatibility relations along high-symmetry lines and planes, and has a nontrivial value of a stable index, then, unlike in the nonmagnetic SGs, it is possible for the bulk to be a topological (semi)metal~\cite{MTQC}.  We label these cases as Smith-index semimetals (SISMs). See Appendix \ref{app:A} for a more detailed description of the six topological classes.

\textbf{Magnetic Materials Database}
We perform high-throughput calculations of the magnetic structures listed on \href{http://webbdcrista1.ehu.es/magndata/}{BCSMD} \cite{gallego_magndata1}. 
BCSMD contains portable structure files, including magnetic structure data and symmetry information, for 707 magnetic structures. 
The magnetic structures of all the materials are determined by neutron scattering experiments. 
We thus consider it reasonable and experimentally motivated to use the crystal and magnetic structures provided on the BCSMD as the initial inputs for {\it ab initio} calculations, instead of letting our theoretical ab-initio codes predict the magnetic ground-state.
We emphasize that predictions of topological magnetic materials based on theoretically calculated magnetic structures, rather than experimentally measured structures, are more likely to predict unphysical (and possibly incorrect) magnetic ground states. 
From the 707 magnetic structures on the BCSMD, we omit 63 structures with lattice-incommensurate magnetism and 95 alloys, as they do not have translation symmetry and hence are not invariant under any MSG. 
We apply {\it ab initio} calculations for the remaining 549 structures. 
These magnetic structures belong to 261 different MSGs, including 29 chiral MSGs and 232 achiral MSGs (chiral MSGs are defined as MSGs without improper rotations or combinations of improper rotations and time reversal; all other MSGs are achiral).  In Appendix \ref{app:B}, we list the number of materials with experimentally obtained magnetic structures in each MSG.

\textbf{Calculation Methods}
We performed {\it ab initio} calculations incorporating spin-orbital coupling (SOC) using VASP \cite{VASP1996}.
Because all of the magnetic materials on BCSMD with translation symmetry contain at least one correlated atom with $3d$, $4d$, $4f$, or $5f$ electrons, we apply a series of LDA+$U$ calculations for each material with different Hubbard-$U$ parameters to obtain a full phase diagram.
For all of the $3d$ valence orbitals and the atom Ru with $4d$ valence orbitals, we take $U$ as 0 eV, 1 eV, 2 eV, 3 eV and 4 eV.
The other atoms with $4d$ valence electrons usually do not exhibit magnetism or have weak correlation effects, and hence are not considered to be correlated in our calculations. Conversely, atoms with $4f$ and $5f$ valence electrons have stronger correlation effects, so we take $U$ for atoms with $4f$ or $5f$ valence electrons to be 0 eV, 2 eV, 4 eV and 6 eV.
If a material has both $d$ and $f$ electrons near the Fermi level, we fix the $U$ parameter of the $d$ electrons as 2 eV, and take $U$ of the $f$ electrons to be 0 eV, 2 eV, 4 eV and 6 eV sequentially. 
We also adopt four other exchange-correlation functionals in the LDA + $U$ scheme to check the consistency between different functionals. Further details of our first-principles calculations are provided in Appendix \ref{app:C},\ref{app:L}, \ref{app:D}and \ref{app:M}.

Of the 549 magnetic structures that we examined, 403 converged self-consistently to a magnetic ground state within an energy threshold of $10^{-5}$eV per cell.   For 324 of the 403 converged materials, magnetic moments matching the experimental values (up to an average error of 50\%) were obtained for at least one of the values of U used to obtain the material phase diagram. We stress that these are good agreements for calculations on these strongly correlated states. However, for the other 79 materials, the calculated magnetic momenta always notably diverged from the experimental values (see Appendix \ref{app:K} for a complete comparison of the experimental and ab initio magnetic moments). The differences can be explained as follows. First, we consider only the spin components, but not the orbital components, of the magnetic moments in our current ab initio calculations. This can result in a large average error for compounds with large spin–orbital coupling. Second, because the average error is defined relative to the experimental moments, the ‘error’ (measured as a percentage) is likely to be larger when the experimental moments are small. In this case, the random, slight changes in the numerically calculated moments have an outsized effect on the reported error percentage. Last but not least, mean-field theory applied in the LDA+$U$ calculations is not a good approximation for some strongly correlated materials, which should be checked further with more advanced methods. Although the prediction of magnetic structure with mean-field theory is sometimes not reliable for strongly correlated materials, it is worth comparing the energy difference between the magnetic structures from neutron scattering and the other possible magnetic structures. In Appendix \ref{app:L}, we selected several topological materials and compared their energies with some possible magnetic configurations and different $U$. We find that their experimental magnetic configurations have the lowest energies, and hence are theoretically favoured. Finally, we have additionally performed self-consistent calculations of the charge density at different values of U, which we used as input for our band structure calculations. In Appendix \ref{app:I}, we provide a complete summary of results.

Considering the possible underestimation of the band gap by generalized gradient approximations (GGA), electronic structures of 23 topological mateirals are further confirmed by the calculations using the modified Becke–Johnson potential \cite{tran2009accurate}. As shown in Appendix \ref{app:D2}, both the features of bands near Fermi level and the topological classes obtained from modified Becke–Johnson potential are consistent with LDA+$U$ calculations.
Because of the limitations of the LDA+$U$ method, we have also performed the more costly LDA+Gutzwiller \cite{PhysRevB.79.075114} calculations in two of the topological materials identified in this work-CeCo$_2$P$_2$ and MnGeO$_3$, both classified as enforced semimetal-to confirm the bulk topology. 
 As shown in Appendix \ref{app:M}, the strong correlations renormalize the quasiparticle spectrum by a factor of quasiparticle weight, but do not change the band topology. The surface-state calculations have been performed using the WannierTools package \ref{WU2017}.

\textbf{Identification of the magnetic irreps}
Using the self-consistent charge density and potentials, we calculate the Bloch wavefunctions at the high-symmetry momenta in the Brillouin zone and then identify the corresponding magnetic irreps using the MagVasp2trace package, which is the magnetic version of Vasp2trace package \cite{gao2020irvsp}. (See Appendix \ref{app:C} for details about MagVasp2trace).
The little group $G_{\bf k}$ of a high symmetry point $\bf k$ is in general isomorphic to an MSG.
For little groups without anti-unitary operations, we calculate the traces of the symmetry representations formed by the wavefunctions, and then decompose the traces into the characters of the small irreps of the little group $G_{\bf k}$.  
For little groups with anti-unitary operations, we calculate only the traces of the unitary operations and decompose the representations into the irreps of the maximal unitary subgroup $G^U_{\bf k}$ of $G_{\bf k}$.
Since anti-unitary operations in general lead to additional degeneracies, specifically enforcing two irreps of $G^U_{\bf k}$ to become degenerate and form a co-representation, we check whether the additional degeneracies hold in the irreps obtained. Because VASP does not impose anti-unitary (magnetic) symmetries, degeneracies labelled by magnetic co-representations may exhibit very small splittings in band structures generated by VASP. In these cases, we reduce the convergence threshold and re-run the self-consistent calculation until the splitting is specifically small ($\le$10\%) compared to the smallest energy gap across all of the high-symmetry momenta.

The algorithm and methods designed in this work are also applicable to future high-throughput searches for magnetic topological materials \cite{frey2020high}.

\textbf{Details of the chemical catagories}
Considering the magnetic ions and chemical bonding of the magnetic materials, we classify the topological magnetic materials predicted in this work into the following 11 chemical categories.

(I-A) 
Non-collinear manganese compounds, which have received considerable recent attention owing to their unusual combination of a large anomalous Hall effect and net-zero magnetic moments. The symmetry of the non-collinear antiferromagnet spin structure allows for a non-vanishing Berry curvature, the origin of the unusual anomalous Hall effect. Examples of non-collinear manganese compounds include the hexagonal Weyl semimetals Mn$_3$Sn, Mn$_3$Ge and the well-studied cubic antiferromagnetic spintronic-material Mn$_3$Ir, as well as the inverse perovskite compounds Mn$_3$Y, which represent ‘stuffed’ versions of the cubic Mn$_3$Y compounds.

(I-B,C) Intermetallic materials, containing rare-earth atoms or actinide atoms, which are typically antiferromagnets. The variation of the Hubbard $U$ changes the band structures slightly in these materials, but not the topological character.

(II-A) The ThCr$_2$Si$_2$ structure and related structures, which have received attention because of the high-temperature iron pnictide superconductors in this group. In these materials, the transition-metal layers and the pnictide layers form square lattices. The square nets of the pnictides act as a driving force for a topological band structure~\cite{PhysRevB.92.205310}. Several of the antiferromagnetic undoped prototypes, such as CaFe$_2$As$_2$, are topological antiferromagnets. This suggests the possibility of topological superconductivity in these materials, like that recently found in FeTe$_{0.55}$Se$_{0.45}$~\cite{wang2018evidence}.

(II-B)  Semiconducting manganese pnictines, which occur when iron is substituted with manganese, leading to materials that are trivial when insulating and gapped, but which become topological antiferromagnets when their gap is closed. By increasing the Hubbard $U$, the antiferromagnetic phases of these compounds can be converted into trivial insulators. The antiferromagnetic insulators and semimetals in this class can also be converted into ferromagnetic metals by doping.

(II-C) Transition metals in combination with rare earth or actinide atoms form compounds of the ThCr$_2$Si$_2$ structure type. Here the antiferromagnetic ordering comes from the Thorium-position in the ThCr$_2$Si$_2$ structure type.

(II-D) Rare earth ternary compounds of the composition 1:1:1.

(III-A,B,C,D) The third class of magnetic materials are ionic compounds, of which most have been experimentally determined to be insulating. Within the density functional approximation, several of the compounds have been identified as topological nontrivial metals, such as oxides, borates, hydroxides, silicates, phosphates and FeI$_2$. By increasing the Hubbard $U$, a topologically trivial gap can be opened in these materials.

Additional data and discussion can be found online in the Supplementary information.
%\bibliographystyle{apsrev4-1}
%\bibliography{ref}

\textbf{Data availability}
All data are available in the Supplementary Information and at https://www.topologicalquantumchemistry.fr/magnetic. The codes required to calculate the character table of magnetic materials are available at https://www.cryst.ehu.es/cryst/checktopologicalmagmat.

\clearpage
\onecolumngrid
\appendix

%\renewcommand{\thesection}{Section \arabic{section}}
%\renewcommand{\thesubsection}{\arabic{section}.\arabic{subsection}}

%\listoffigures
%\listoftables

%\renewcommand{\thesection}{Section \arabic{section}}
%\renewcommand{\thesubsection}{\arabic{section}.\arabic{subsection}}

\section{A brief introduction to Magnetic Topological Quantum Chemistry (MTQC)}\label{app:A}

%\addcontentsline{toc}{section}{Appendices}

The symmetry group property of a band structure is fully described by the multiplicities of the irreducible co-representations (irreps) formed by the occupied bands at all the maximal \textbf{$K$}-points.
In the present paper, we define the 1st band to the $N_e$th band as the ``occupied'' bands, where $N_{e}$ is the number of electrons.
Maximal \textbf{$k$}-points are defined as the high symmetry momenta whose little groups are maximal subgroups of the magnetic space group. The maximal \textbf{$K$}-points of each magnetic space group are supplied in the magnetic vasp2trace package.
We denote the little group at the momentum $K$ as $G_{K}$, the $i$th irrep of $G_K$ as $\rho_{K}^i$, and its multiplicity of $\rho_{K}^i$ formed by the occupied bands as $m(\rho_{K}^i)$.
For example, the Brillouin zone (BZ) of the 2D space group generated from inversion ($I$) and translations has four maximal \textbf{$k$}-points: $\Gamma\ (0,0)$, $\rm X\ (\pi,0)$, $\rm Y\ (0,\pi)$, $\rm M\ (\pi,\pi)$, all of which have the same little group: $C_i=\{E, I\}$. 
Here $E$ is the identity.
$C_i$ has two types of irreps: the even(+) and the odd($-$).
Thus a band structure is characterized by the eight integers $m(\rho_{\rm \Gamma, X, Y, M}^{+,-})$.
For convenience, we introduce the symmetry-data-vector\cite{bradlyn_topological_2017}
\begin{equation}
B = (
m(\rho_{\Gamma}^+), m(\rho_{\Gamma}^-),
m(\rho_{\rm X}^+), m(\rho_{\rm X}^-),
m(\rho_{\rm Y}^+), m(\rho_{\rm Y}^-),
m(\rho_{\rm M}^+), m(\rho_{\rm M}^-) )^T.\label{eq:B-vector}
\end{equation}
The symmetry property of a band structure is fully described by the symmetry-data-vector.

\textbf{Enforced semimetal with Fermi degeneracy.}
In some materials, the highest occupied band and the lowest empty band are degenerate at some maximal \textbf{$K$}-points, and the degeneracy is protected by the MSG. 
We call such states enforced semimetals with Fermi degeneracy (ESFD)\cite{bradlyn_topological_2017,PhysRevLett.118.186401}.
ESFD does not have a well defined symmetry-data-vector $B$.
See FIG. 1c of the main text for examples of ESFD.

For band structures where the occupied bands are gapped from the empty bands along all the high symmetry lines, the multiplicities $m$ necessarily satisfy the compatibility relations \cite{bradlyn_topological_2017,elcoro_double_2017,cano_building_2018,vergniory_graph_2017,po_symmetry-based_2017,watanabe2018structure,kruthoff_topological_2017}.
We consider two maximal \textbf{$k$}-points $K_{1,2}$ and a path $k$ between them.
On the one hand, since $G_{k}$ is a subgroup of $G_{K_1}$, the irreps of $G_k$ formed by the occupied bands in $k$ near to $K_1$ can be obtained by subduction of the irreps of $G_{K_1}$ formed by occupied bands at $K_1$.
On the other hand, the irreps of $G_k$ formed by the occupied bands in $k$ near to $K_2$ can also be obtained by subduction of the irreps at $K_2$.
If the irreps of $G_k$ obtained at the two ends $K_1$ and $K_2$ are not the same, then there must be a symmetry protected level crossing along the path $k$.
In other words, in order to guarantee the path $k$ is gapped, the irreps at $K_1$ and $K_2$ must reduce to the same set of irreps of $G_k$.
This requirement establishes the compatibility relations along $k$.
The full compatibility relations can be obtained by applying this analysis to all the inequivalent paths in the BZ.\cite{elcoro_double_2017}

In the example of 2D space group with inversion $P\bar 1$, all the momenta except $\rm \Gamma$,  $\rm X$, $\rm Y$, $\rm M$ have the same little group: the identity group.
Thus for any two maximal \textbf{$k$}-points, there is only one inequivalent path connecting them, and both even and odd irreps reduce to the identity irrep of the identity group.
The compatibility-relation is nothing but the restriction that the two maximal \textbf{$k$}-points have the same number of occupied bands.
We can write the compatibility relations as
\begin{equation}
m(\rho_{\Gamma}^+) + m(\rho_{\Gamma}^-) = 
m(\rho_{\rm X}^+) + m(\rho_{\rm X}^-) =
m(\rho_{\rm Y}^+) + m(\rho_{\rm Y}^-) =
m(\rho_{\rm M}^+) + m(\rho_{\rm M}^-). \label{eq:comp}
\end{equation}
Since we define the ``occupied'' bands as the 1st band to the $N_e$th band, there are always $N_e$ occupied levels at any momentum and hence Eq.(\ref{eq:comp}) is automatically satisfied.
However, most other magnetic space groups have more compatibility relations than the band number restriction; these compatibility relations can be broken in materials.

\textbf{Enforced Semimetals.}
Band structures breaking the compatibility relations are referred to as enforced semimetals (ESs).
The inversion case is not a good example for ES because the compatibility-relation is satisfied by definition. 
Please see ~\ref{app:H} for example of ES.

We now classify the possible band structures allowed by compatibility relations.
The strategy is that we first enumerate all the atomic insulators and then, for any given band structure from DFT, compare its irreps with those of the atomic insulators.
A band structure must be topologically nontrivial if its irreps are not consistent with any atomic insulator; otherwise  can be either trivial/nontrivial.
Following the terminology of Zak \cite{zak1982band,zak1989berry,michel2001elementary}, we refer to atomic insulators as band representations (BRs) and to the generators of the BRs as elementary BRs (EBRs).

We take the 2D space group with inversion $P\bar 1$ as an example to illustrate the concept of EBRs.
There are four maximal Wyckoff positions in each unit cell, $a\ (0,0)$, $b\ (\frac12,0)$, $c\ (0,\frac12)$, $d\ (\frac12,\frac12)$.
Maximal Wyckoff positions are defined as positions with site-symmetry-groups  which are maximal subgroups of the space group.
In this example, the site-symmetry-groups of $a,b,c,d$ are all isomorphic to $C_i=\{E,I\}$.
Since $C_i$ only has two types of irreps (even and odd), we can add either $s$ orbital (even)/$p$ orbital (odd) at each position. 
We can then obtain eight different EBRs.
To see that they are EBRs, we consider an atomic insulator formed by two orbitals at two general positions, $(x,y)$, $(1-x,1-y)$, which transform into each other under the inversion operation at $d$ position.
We can recombine the two orbitals to form a bonding state and an anti-bonding state at the $d$ position.
Thus this atomic insulator can be generated from two EBRs at the $d$ position.
The symmetry-data-vectors of the eight EBRs can be calculated by acting the symmetry operators on the corresponding Bloch wave functions.
The wave functions are Fourier transformations of the local orbitals
\begin{equation}
|\phi_{\xi, \alpha, \mathbf{k}}\rangle = \frac{1}{\sqrt{N}} \sum_{\mathbf{R}} e^{i\mathbf{k}\cdot\mathbf{(R+t_\alpha)}} |\xi, \mathbf{R+t_\alpha} \rangle
\end{equation}
Here $\xi=\pm$ is the parity of the local orbital, $\alpha=a,b,c,d$ is the Wyckoff position, $\mathbf{t}_\alpha$ is the position vector of the Wyckoff position, $\mathbf{R}$ sums over all lattice vectors, and $N$ is the system size. 
Since $I|\xi, \mathbf{R+t_\alpha} \rangle = \xi |\xi, \mathbf{-R-t_\alpha} \rangle$, we obtain 
\begin{equation}
I|\phi_{\xi, \alpha, \mathbf{k}}\rangle =  \xi \frac{1}{\sqrt{N}} \sum_{\mathbf{R}} e^{i\mathbf{k}\cdot\mathbf{(R+t_\alpha)}} |\xi, \mathbf{-R-t_\alpha} \rangle
=\xi \frac{1}{\sqrt{N}} \sum_{\mathbf{R}^\prime} e^{-i\mathbf{k}\cdot\mathbf{(R^\prime+t_\alpha)}} |\xi, \mathbf{R^\prime+t_\alpha} \rangle,
\end{equation}
where the lattice vector $\mathbf{R}^\prime $ is $ -\mathbf{R}-2\mathbf{t}_\alpha$.
If $\mathbf{k}$ is one of the maximal \textbf{$k$}-points ($\rm \Gamma, X, Y, M$), we can calculate the parity of the Bloch wave function as
\begin{equation}
\langle\phi_{\xi, \alpha, \mathbf{k}}| I|\phi_{\xi, \alpha, \mathbf{k}}\rangle 
= \xi \frac{1}{N} \sum_{\mathbf{R}^\prime} e^{-i2\mathbf{k}\cdot\mathbf{(R^\prime+t_\alpha)}} = \xi e^{-i2\mathbf{k\cdot t_{\alpha}}}.
\end{equation}
We have made use of the fact that $2\mathbf{k}$ is a reciprocal lattice vector and hence $2\mathbf{k}\cdot\mathbf{R}=0\mod 2\pi$.
For $\alpha=a$, the Bloch wave function has the same parity at all the four maximal \textbf{$k$}-points because $e^{-i2\mathbf{k\cdot t_{\alpha}}}=1$.
Thus the EBR induced from the orbital with parity $\pm$ at the $a$ position form the irreps $\rho_{\Gamma}^{\pm}$, $\rho_{\rm X}^{\pm}$, $\rho_{\rm Y}^{\pm}$, $\rho_{\rm M}^{\pm}$.
The symmetry-data-vectors (\ref{eq:B-vector}) of these two EBRs are
\begin{equation}
EBR_{+,a} = (1,0, 1,0, 1,0, 1,0)^T, \qquad
EBR_{-,a} = (0,1, 0,1, 0,1, 0,1)^T. \label{eq:EBR-a}
\end{equation}
For $\alpha=b,c,d$, the Bloch wave function has different parities at the four maximal \textbf{$k$}-points because $e^{-i2\mathbf{k\cdot t_{\alpha}}}$ can be either $1$/$-1$.
For example, for $\alpha=b$, $e^{-i2\mathbf{k\cdot t_{\alpha}}}$ equals to 1 and $-1$ at $\rm \Gamma, Y$ and $\rm X, M$, respectively. 
Thus the EBR induced from the orbital with parity $\pm$ at the $b$ position form the irreps $\rho_{\Gamma}^{\pm}$, $\rho_{\rm X}^{\mp}$, $\rho_{\rm Y}^{\pm}$, $\rho_{\rm M}^{\mp}$.
The corresponding symmetry-data-vectors are
\begin{equation}
EBR_{+,b} = (1,0, 0,1, 1,0, 0,1)^T, \qquad
EBR_{-,b} = (0,1, 1,0, 1,0, 1,0)^T.
\end{equation}
Similarly, one can derive the symmetry-data-vectors of EBRs induced from $c,d$ positions as
\begin{equation}
EBR_{+,c} = (1,0, 1,0, 0,1, 0,1)^T, \qquad
EBR_{-,c} = (0,1, 0,1, 1,0, 1,0)^T,
\end{equation}
\begin{equation}
EBR_{+,d} = (1,0, 0,1, 0,1, 1,0)^T, \qquad
EBR_{-,d} = (0,1, 1,0, 1,0, 0,1)^T. \label{eq:EBR-d}
\end{equation}

\textbf{Stable TI.}
We consider an example where the occupied band form a single odd irrep at $\Gamma$ and three even irreps at $\rm X,Y,M$ respectively.
The corresponding symmetry-data-vector can be written as
\begin{equation}
B_1 = (0,1, 1,0, 1,0, 1,0)^T.
\end{equation}
$B_1$ is not one of the EBRs; It is also not a sum of EBRs,
because all EBRs have even number of odd irreps in Eqs. (\ref{eq:EBR-a}) to (\ref{eq:EBR-d}).
It is also not a sum of EBRs because $B_1$ has only one band but any sum of EBRs has at least two bands.
Thus $B_1$ must be topological.
According to the Fu-Kane-like formula for Chern insulators \cite{fang2012}
\begin{equation}
(-1)^C = \prod_{K=\rm \Gamma,X,Y,M} \prod_{n} \lambda_{n}(K),\label{eq:Chern-number}
\end{equation}
where $C$ is the Chern number, $n$ is the index of occupied bands, and $\lambda_{n}(K)$ is the parity of $n$th band at the momentum $K$, the band structure has an odd Chern number.

One notices that $B_1$ can be written as a linear combination of EBRs with fractional coefficients
\begin{equation}
B_1 = -\frac12 EBR_{-,a} + \frac12 EBR_{-,d} + \frac12 EBR_{-,c} + \frac12 EBR_{-,d},
\end{equation}
but cannot be written as an integer combination of EBRs.
It is a general principle that if a band structure cannot be written as a linear combination of EBRs unless the coefficients are fractional numbers, the band structure must have stable topology.
Such stable topology implied by symmetry eigenvalues is characterized by the stable index (SI) (also referred to as symmetry-based indicator \cite{po_symmetry-based_2017,Luis2019}).
Eq. (\ref{eq:Chern-number}) can be thought as an example of SI.
Readers can refer to supplementary information of Ref.~\cite{Luis2019,song_fragile_2019} for technical details.

\textbf{Smith-index semimetal.}
%For paramagnetic double space groups (or grey groups), If the $B$ vector of a state  satisfies all of the compatibility relations but not equal to any linear combinations of EBRs, we take it as a TI, of which the strong indices are compatible with gapped state. While i
In magnetic space groups, some symmetry-data-vectors are not compatible with gapped state and implies topological Weyl semimetal (WSM), even when all of the compatibility relations are satisfied. In this work, the WSM phase implied by symmetry eigenvalue is named as Smith-index semimetal (SISM).

From the MTQC theory, we have found several MSGs with SI corresponding to WSM phase. All of these MSGs have a minimal subgroup MSG 2.4 ($P\bar1$)/MSG 81.33 ($P\bar4$).
For the MSGs with minimal subgroup $P\bar1$ (with only inversion symmetry), the topologies are described by the stable indices group $\ZZ_4\times\ZZ_2^3$. We found the stable index $\eta_{4I} mod 2$ is the parity of the Chern number difference between $kz = 0$ and $kz = \pi$ planes. Thus $\eta_{4I}=1,3$ correspond to the WSM phase with odd  number of Weyl points in one half Brillouin zone~\cite{MTQC}.
For the MSGs with minimal subgroup $P\bar4$, they have the SI group  $\ZZ_4\times\ZZ_2^2$. We find one of the two $z_2$ indices can be interpreted as~\cite{MTQC}
$\delta_{2S}=\frac{c_{\pi}-c_0}2 mod 2$,
where $c_{0,\pi}$ are the Chern numbers in the $k_{z} = 0,\pi$ planes. Thus, when this $\delta_{2S}$ index is nonzero, $k_z=0,\pi$ planes must have different Chern numbers and hence Weyl nodes must appear in between the two planes.

\textbf{Fragile TI.}
If the $B$ vector of a state cannot be written as a sum of EBRs, but can be written as a difference of EBRs, then the state is at least a fragile TI.~\cite{po_fragile_2018,cano_topology_2018,bradlyn_disconnected_2019,ahn_failure_2019,song_fragile_2019,hwang_fragile_2019,manes_fragile_2019,alexandradinata_crystallographic_2019,Luis2019}
We say ``at least'' because the state can also have a stable topology which cannot be diagnosed through symmetry eigenvalues but through Berry phases.
Now we give an example in the inversion case.
We consider that the occupied bands form two odd irreps at $\Gamma$ and three pairs of even irreps at $\rm X,Y,M$ respectively.
The corresponding symmetry-data-vector is double of $B_1$, \ie
\begin{equation}
B_2 = (0,2, 2,0, 2,0, 2,0)^T.
\end{equation}
Since $B_2 = 2B_1$, we can write the $B_2$ as a linear combination of EBRs with integer coefficients, and one of the coefficients is negative
\begin{equation}
B_2 = -EBR_{-,a} + EBR_{-,d} + EBR_{-,c} + EBR_{-,d}.
\end{equation}
This decomposition implies that, after being coupled to a trivial band forming the $EBR_{-,a}$, $B_2$ becomes trivial because it can be written as a sum of EBRs as $EBR_{-,d} + EBR_{-,c} + EBR_{-,d}$.
Therefore, $B_2$ is at least a fragile TI.
Readers can refer to Ref.~\cite{song_fragile_2019} for more examples and complete classifications of eigenvalue implied fragile TIs. 

\section{Material statistics in the BCSMD}\label{app:B}

Ignoring the magnetic materials with incommensurate structures, there are 644 materials (including 95 alloys) with 261 different MSGs in the Bilbao Crystallographic Server Magnetic database(BCSMD)\cite{gallego_magndata1,perez-mato_symmetry-based_2015}. We provide the number of materials in each MSG in  Table~\ref{tablemat}.  Detailed information about each of the magnetic materials can be obtained on the BCSMD website (http://webbdcrista1.ehu.es/magndata). 
Based on the stable topological classifications of MSGs~\cite{MTQC,watanabe2018structure}, we classify the MSGs into four types. 

\textbf{Type A} The MSGs that have stable topological indices, which are indicated by red color. There are 435 materials in BCSMD with Type A MSGs. 

\textbf{Type B} In this type of MSGs, given the electron number, one can immediately identify whether a material is ESFD. This type of MSGs are indicated by blue color. There are 34 materials with Type B MSGs in BCSMD. 

\textbf{Type C} Among Type B MSGs, some also have stable topological indices, which are indicated by green color. There are 19 materials in BCSMD with Type C MSG.

\textbf{Type D} The other MSGs that do not belong to Type A/Type B are indicated by black color. There are 183 materials with Type D MSG.

We also emphasize that for an ES/ESFD, if the crossing point occurs at a k-point whose little co-group is chiral, the crossing point must necessarily carry a nonzero chiral charge~\cite{PhysRevLett.119.206401,cano2019multifold,chang2018topological}.
The chiral MSGs have been tagged in Table~\ref{tablemat}. 
\LTcapwidth=1.0\textwidth
\renewcommand\arraystretch{1.2}
\begin{longtable*}{p{0.18\columnwidth}c|p{0.18\columnwidth}c|p{0.18\columnwidth}c|p{0.18\columnwidth}c}
\caption{The number of magnetic materials per magnetic space group in BCSMD}\label{tablemat} \\
\hline \hline
MSG & Count & MSG & Count & MSG & Count & MSG & Count \\
\hline \hline
1.3 $P_S1$$^*$ & 4 & 33.144 $Pna2_1$ & 3 & \textcolor{red}{63.462 $Cm'c'm$} & 2 & \textcolor{red}{138.528 $P_c4_2/ncm$} & 1 \\ 
\textcolor{red}{2.4 $P\bar1$} & 4 & 33.147 $Pna'2_1'$ & 2 & \textcolor{red}{63.463 $Cmc'm'$} & 1 & \textcolor{red}{138.529 $P_C4_2/ncm$} & 1 \\ 
2.6 $P\bar1'$ & 3 & 33.148 $Pn'a'2_1$ & 3 & \textcolor{red}{63.464 $Cm'cm'$} & 4 & \textcolor{red}{139.535 $I4'/mmm'$} & 1 \\ 
\textcolor{red}{2.7 $P_S\bar1$} & 34 & 33.149 $P_ana2_1$ & 1 & \textcolor{red}{63.466 $C_cmcm$} & 2 & \textcolor{red}{139.536 $I4'/m'm'm$} & 4 \\ 
4.10 $P_a2_1$$^*$ & 7 & 33.150 $P_bna2_1$ & 1 & \textcolor{red}{63.467 $C_amcm$} & 1 & \textcolor{red}{139.537 $I4/mm'm'$} & 2 \\ 
4.12 $P_C2_1$$^*$ & 1 & 33.154 $P_Cna2_1$ & 3 & \textcolor{green}{63.468 $C_Amcm$} & 1 & \textcolor{red}{140.550 $I_c4/mcm$} & 6 \\ 
4.7 $P2_1$$^*$ & 3 & 35.167 $Cm'm2'$ & 1 & \textcolor{red}{64.476 $Cm'ca'$} & 1 & \textcolor{green}{141.554 $I4_1'/am'd$} & 2 \\ 
4.9 $P2_1'$$^*$ & 3 & 36.174 $Cm'c2_1'$ & 2 & \textcolor{red}{64.479 $C_amca$} & 1 & \textcolor{green}{141.555 $I4_1'/amd'$} & 3 \\ 
5.13 $C2$$^*$ & 1 & 36.176 $Cm'c'2_1$ & 1 & \textcolor{red}{64.480 $C_Amca$} & 13 & \textcolor{red}{141.556 $I4_1'/a'm'd$} & 3 \\ 
5.15 $C2'$$^*$ & 1 & 36.178 $C_amc2_1$ & 4 & 65.483 $Cm'mm$ & 1 & \textcolor{red}{141.557 $I4_1/am'd'$} & 8 \\ 
5.16 $C_c2$$^*$ & 3 & 38.191 $Am'm'2$ & 1 & \textcolor{red}{65.486 $Cmm'm'$} & 2 & \textcolor{red}{142.568 $I4_1'/a'cd'$} & 1 \\ 
6.20 $Pm'$ & 1 & 38.192 $A_amm2$ & 1 & \textcolor{red}{65.489 $C_ammm$} & 2 & 146.10 $R3$$^*$ & 2 \\ 
7.27 $P_ac$ & 1 & 39.201 $A_bbm2$ & 1 & \textcolor{red}{66.500 $C_Accm$} & 5 & 146.12 $R_I3$$^*$ & 2 \\ 
7.29 $P_bc$ & 1 & 41.217 $A_bba2$ & 1 & \textcolor{red}{67.510 $C_Amma$} & 1 & \textcolor{red}{148.17 $R\bar3$} & 5 \\ 
8.35 $C_cm$ & 1 & 42.223 $F_Smm2$ & 1 & 69.523 $Fm'mm$ & 1 & 148.19 $R\bar3'$ & 2 \\ 
8.36 $C_am$ & 4 & 43.227 $Fd'd'2$ & 1 & \textcolor{red}{69.526 $F_Smmm$} & 3 & \textcolor{red}{148.20 $R_I\bar3$} & 1 \\ 
9.39 $Cc'$ & 2 & 45.237 $Ib'a2'$ & 1 & \textcolor{red}{70.530 $Fd'd'd$} & 2 & 152.35 $P3_12'1$$^*$ & 1 \\ 
9.40 $C_cc$ & 3 & 46.243 $Im'a2'$ & 1 & \textcolor{red}{71.536 $Im'm'm$} & 2 & 154.41 $P3_221$$^*$ & 1 \\ 
9.41 $C_ac$ & 3 & \textcolor{red}{49.270 $Pc'cm'$} & 1 & \textcolor{red}{72.543 $Ib'a'm$} & 1 & 154.44 $P_c3_221$$^*$ & 3 \\ 
\textcolor{red}{10.49 $P_C2/m$} & 1 & \textcolor{red}{49.273 $P_cccm$} & 1 & \textcolor{red}{73.551 $Ib'c'a$} & 1 & 155.48 $R_I32$$^*$ & 1 \\ 
\textcolor{red}{11.54 $P2_1'/m'$} & 2 & \textcolor{red}{50.282 $Pb'an'$} & 1 & \textcolor{red}{73.553 $I_cbca$} & 2 & 157.53 $P31m$ & 1 \\ 
\textcolor{red}{11.55 $P_a2_1/m$} & 2 & \textcolor{red}{51.295 $Pmm'a'$} & 1 & \textcolor{red}{74.559 $Imm'a'$} & 1 & \textcolor{red}{157.55 $P31m'$} & 1 \\ 
\textcolor{green}{11.57 $P_C2_1/m$} & 3 & \textcolor{red}{51.298 $P_amma$} & 1 & \textcolor{red}{74.562 $I_bmma$} & 1 & 159.64 $P_c31c$ & 3 \\ 
\textcolor{red}{12.58 $C2/m$} & 1 & \textcolor{red}{52.310 $Pn'n'a$} & 1 & \textcolor{red}{83.50 $P_I4/m$} & 2 & 161.69 $R3c$ & 2 \\ 
12.60 $C2'/m$ & 4 & \textcolor{red}{52.312 $Pn'na'$} & 1 & \textcolor{green}{84.58 $P_I4_2/m$} & 1 & 161.71 $R3c'$ & 2 \\ 
\textcolor{red}{12.62 $C2'/m'$} & 9 & \textcolor{red}{52.315 $P_bnna$} & 1 & \textcolor{red}{85.59 $P4/n$} & 1 & 161.72 $R_I3c$ & 2 \\ 
\textcolor{red}{12.63 $C_c2/m$} & 8 & \textcolor{red}{53.334 $P_Bmna$} & 1 & \textcolor{red}{85.64 $P_c4/n$} & 1 & \textcolor{red}{162.78 $P_c\bar31m$} & 1 \\ 
\textcolor{red}{12.64 $C_a2/m$} & 5 & \textcolor{red}{53.335 $P_Cmna$} & 1 & \textcolor{red}{86.67 $P4_2/n$} & 1 & \textcolor{red}{164.89 $P\bar3m'1$} & 2 \\ 
13.67 $P2'/c$ & 1 & \textcolor{red}{54.350 $P_Bcca$} & 1 & \textcolor{red}{86.73 $P_C4_2/n$} & 3 & 165.94 $P\bar3'c'1$ & 1 \\ 
\textcolor{red}{13.69 $P2'/c'$} & 1 & \textcolor{red}{54.352 $P_Icca$} & 3 & \textcolor{red}{87.75 $I4/m$} & 1 & \textcolor{red}{165.95 $P\bar3c'1$} & 2 \\ 
\textcolor{red}{13.70 $P_a2/c$} & 1 & 55.355 $Pb'am$ & 1 & 87.78 $I4/m'$ & 3 & \textcolor{red}{165.96 $P_c\bar3c1$} & 1 \\ 
\textcolor{green}{13.73 $P_A2/c$} & 3 & 55.356 $Pbam'$ & 1 & \textcolor{red}{88.81 $I4_1/a$} & 1 & \textcolor{red}{166.101 $R\bar3m'$} & 5 \\ 
\textcolor{red}{13.74 $P_C2/c$} & 4 & \textcolor{red}{55.361 $P_cbam$} & 1 & \textcolor{red}{88.86 $I_c4_1/a$} & 1 & \textcolor{red}{166.102 $R_I\bar3m$} & 1 \\ 
\textcolor{red}{14.75 $P2_1/c$} & 9 & \textcolor{red}{56.369 $Pc'c'n$} & 1 & 92.111 $P4_12_12$$^*$ & 1 & \textcolor{red}{166.97 $R\bar3m$} & 2 \\ 
\textcolor{blue}{14.77 $P2_1'/c$} & 3 & \textcolor{red}{56.372 $P_bccn$} & 1 & 92.114 $P4_12_1'2'$$^*$ & 1 & \textcolor{red}{167.103 $R\bar3c$} & 1 \\ 
\textcolor{blue}{14.78 $P2_1/c'$} & 5 & \textcolor{red}{56.373 $P_cccn$} & 2 & 94.132 $P_c4_22_12$$^*$ & 1 & 167.106 $R\bar3'c'$ & 1 \\ 
\textcolor{red}{14.79 $P2_1'/c'$} & 8 & \textcolor{red}{56.374 $'P_Accn'$} & 2 & 96.150 $P_I4_32_12$$^*$ & 1 & \textcolor{red}{167.107 $R\bar3c'$} & 1 \\ 
\textcolor{red}{14.80 $P_a2_1/c$} & 20 & \textcolor{red}{57.389 $P_Abcm$} & 1 & \textcolor{red}{107.231 $I4m'm'$} & 1 & \textcolor{red}{167.108 $R_I\bar3c$} & 5 \\ 
\textcolor{red}{14.81 $P_b2_1/c$} & 1 & \textcolor{red}{57.391 $P_Cbcm$} & 1 & \textcolor{red}{111.255 $P\bar42'm'$} & 1 & \textcolor{red}{173.129 $P6_3$} & 1 \\ 
\textcolor{red}{14.82 $P_c2_1/c$} & 6 & 58.395 $Pn'nm$ & 5 & \textcolor{red}{113.267 $P\bar42_1m$} & 1 & 173.131 $P6_3'$ & 1 \\ 
\textcolor{red}{14.83 $P_A2_1/c$} & 1 & \textcolor{red}{58.398 $Pnn'm'$} & 4 & \textcolor{red}{114.282 $P_I\bar42_1c$} & 1 & \textcolor{red}{174.136 $P_c\bar6$} & 1 \\ 
\textcolor{red}{14.84 $P_C2_1/c$} & 10 & \textcolor{blue}{58.399 $Pn'n'm'$} & 2 & \textcolor{red}{117.305 $P_C\bar4b2$} & 1 & \textcolor{red}{176.145 $P6_3'/m$} & 1 \\ 
\textcolor{red}{15.85 $C2/c$} & 6 & \textcolor{red}{58.404 $P_Innm$} & 1 & \textcolor{red}{119.319 $I\bar4m'2'$} & 1 & 185.197 $P6_3cm$ & 3 \\ 
15.87 $C2'/c$ & 4 & 59.407 $Pm'mn$ & 2 & \textcolor{red}{122.338 $I_c\bar42d$} & 1 & 185.199 $P6_3'c'm$ & 2 \\ 
15.88 $C2/c'$ & 2 & \textcolor{red}{59.409 $Pm'm'n$} & 1 & \textcolor{red}{123.345 $P4/mm'm'$} & 1 & 185.200 $P6_3'cm'$ & 1 \\ 
\textcolor{red}{15.89 $C2'/c'$} & 11 & \textcolor{red}{59.410 $Pmm'n'$} & 1 & \textcolor{red}{124.360 $P_c4/mcc$} & 4 & \textcolor{red}{185.201 $P6_3c'm'$} & 3 \\ 
\textcolor{red}{15.90 $C_c2/c$} & 28 & \textcolor{red}{59.416 $P_Immn$} & 1 & \textcolor{green}{125.367 $P4'/nbm'$} & 1 & \textcolor{red}{186.207 $P6_3m'c'$} & 1 \\ 
\textcolor{red}{15.91 $C_a2/c$} & 3 & 60.419 $Pb'cn$ & 2 & \textcolor{red}{125.373 $P_C4/nbm$} & 1 & \textcolor{red}{188.220 $P_c\bar6c2$} & 1 \\ 
18.19 $P2_12_1'2'$$^*$ & 1 & \textcolor{red}{60.422 $Pb'c'n$} & 1 & \textcolor{red}{126.384 $P_c4/nnc$} & 1 & 189.223 $P\bar6'2'm$ & 1 \\ 
18.22 $P_B2_12_12$$^*$ & 1 & \textcolor{red}{60.431 $P_Cbcn$} & 1 & \textcolor{red}{126.386 $P_I4/nnc$} & 1 & 189.224 $P\bar6'2m'$ & 1 \\ 
19.25 $P2_12_12_1$$^*$ & 2 & \textcolor{red}{61.433 $Pbca$} & 2 & \textcolor{red}{127.394 $P4'/m'bm'$} & 2 & \textcolor{red}{192.252 $P_c6/mcc$} & 2 \\ 
19.27 $P2_1'2_1'2_1$$^*$ & 1 & \textcolor{blue}{61.437 $Pb'c'a'$} & 3 & \textcolor{blue}{127.395 $P4/m'b'm'$} & 1 & \textcolor{red}{193.259 $P6_3'/m'cm'$} & 1 \\ 
19.28 $P_c2_12_12_1$$^*$ & 1 & \textcolor{red}{61.439 $P_Cbca$} & 1 & \textcolor{red}{127.397 $P_C4/mbm$} & 1 & \textcolor{red}{193.260 $P6_3/mc'm'$} & 3 \\ 
19.29 $P_C2_12_12_1$$^*$ & 1 & \textcolor{red}{62.441 $Pnma$} & 11 & \textcolor{red}{128.408 $P_c4/mnc$} & 1 & \textcolor{red}{194.268 $P6_3'/m'm'c$} & 1 \\ 
20.34 $C22'2_1'$$^*$ & 1 & 62.443 $Pn'ma$ & 2 & \textcolor{red}{128.410 $P_I4/mnc$} & 5 & \textcolor{green}{203.26 $Fd\bar3$} & 1 \\ 
20.37 $C_A222_1$$^*$ & 1 & 62.444 $Pnm'a$ & 4 & \textcolor{red}{129.416 $P4'/n'm'm$} & 3 & \textcolor{red}{205.33 $Pa\bar3$} & 2 \\ 
26.66 $Pmc2_1$ & 2 & 62.445 $Pnma'$ & 5 & 129.419 $P4/n'm'm'$ & 1 & \textcolor{green}{216.77 $F_S\bar43m$} & 1 \\ 
26.68 $Pm'c21'$ & 2 & \textcolor{red}{62.446 $Pn'm'a$} & 9 & \textcolor{red}{130.432 $P_c4/ncc$} & 2 & \textcolor{red}{222.103 $P_In\bar3n$} & 1 \\ 
26.72 $P_bmc2_1$ & 3 & \textcolor{red}{62.447 $Pnm'a'$} & 3 & \textcolor{red}{131.440 $P4_2'/m'm'c$} & 1 & \textcolor{green}{224.113 $Pn\bar3m'$} & 4 \\ 
27.82 $P_ccc2$ & 1 & \textcolor{red}{62.448 $Pn'ma'$} & 5 & \textcolor{red}{132.456 $P_c4_2/mcm$} & 1 & \textcolor{green}{227.131 $Fd\bar3m'$} & 1 \\ 
29.101 $Pc'a2_1'$ & 4 & 62.449 $Pn'm'a'$ & 4 & \textcolor{red}{134.481 $P_C4_2/nnm$} & 3 & \textcolor{red}{228.139 $F_Sd\bar3c$} & 3 \\ 
29.104 $P_aca2_1$ & 5 & \textcolor{red}{62.450 $P_anma$} & 5 & \textcolor{red}{135.492 $P_c4_2/mbc$} & 2 & \textcolor{red}{229.143 $Im\bar3m'$} & 1 \\ 
29.105 $P_bca2_1$ & 1 & \textcolor{red}{62.452 $P_cnma$} & 1 & \textcolor{red}{136.499 $P4_2'/mnm'$} & 2 & \textcolor{red}{230.148 $Ia\bar3d'$} & 1 \\ 
29.110 $P_Ica2_1$ & 1 & \textcolor{red}{62.453 $P_Anma$} & 1 & \textcolor{blue}{136.503 $P4_2/m'n'm'$} & 1 & &   \\ 
31.129 $P_bmn2_1$ & 3 & 63.459 $Cm'cm$ & 1 & \textcolor{red}{136.506 $P_I4_2/mnm$} & 1 & &   \\ 
32.137 $Pb'a2'$ & 1 & 63.461 $Cmcm'$ & 1 & \textcolor{red}{138.525 $P4_2/nc'm'$} & 1 & &   \\ 
\hline \hline
\footnotesize{$^*$ Chiral MSG}\\
\end{longtable*}

In the magnetic material database, all of the materials have distinct chemical formulae/different MSGs except for the 15 materials tabulated in Table \ref{tab:diff}. The 15 compounds are reported having the same chemical formulae and MSGs but different  magnetic moments in two independent neutron experiments. The differences between them have been described in Table \ref{tab:diff}. These differences consist in the experimental temperature/lattice parameters. In this work, we have performed the $ab initio$ calculations for all of them.
\LTcapwidth=1.0\textwidth
\begin{longtable}{|c|c|c|c|c|p{0.35\columnwidth}<{\centering}|p{0.35\columnwidth}}
\caption{The 15 compounds that have the same chemical formulae and MSG but with different magnetic moments are tabulated together. }\label{tab:diff}
\label{tab:diff} \\
\hline\hline
No. & Chemical Formula  &  MSG & BSCID & ($M_x,M_y,M_z$)($\mu_{B}$) & Differences \\
\hline
\hline
\multirow{2}{*}{1}   & \multirow{2}{*}{CoSe2O5} & \multirow{2}{*}{60.419}  & 0.119 & Co(3.1,0.0,0.8)  & \multirow{2}{*}{Small canting along $z$ axis} \\
\cline{4-5}
 & & & 0.161 & Co(3,0,0) &  \\
\hline
\multirow{2}{*}{ 2 }  & \multirow{2}{*}{Er2BaNiO5} & \multirow{2}{*}{15.90  }& 1.15 & Er(7.89,0,0.25), Ni(-1.4,0,-0.64) &  Small difference between the magnetic moments \\
 \cline{4-5}
 & & & 1.53 & Er(7.23,0,0.32),Ni(-1.38,0,-0.18)  &  \\
\hline
\multirow{2}{*}{ 3 }  & \multirow{2}{*}{Cr2TeO6  } & \multirow{2}{*}{58.395 }& 0.76 & Cr(1,0,0) &  Experimental temperature is different; $T=93K$ for BCSID-58.395, $T=4.2K$ for BCSID-0.143 \\
\cline{4-5}
 & & & 0.143 & Cr(2.45,0,0)  &  \\
\hline
\multirow{2}{*}{ 4 }  & \multirow{2}{*}{Cr2WO6   } & \multirow{2}{*}{58.395 }& 0.75 & Cr(1,0,0) &  Experimental temperature is different; $T=45K$ for BCSID-58.395, $T=4.2K$ for BCSID-0.143 \\
\cline{4-5}
 & & & 0.144 & Cr(2.14,0,0)  &  \\
\hline
\multirow{2}{*}{ 5 }  & \multirow{2}{*}{Ho2Ru2O7 } & \multirow{2}{*}{141.557}& 0.49 & Ru(0.56,0.56,0.9) &  Experimental temperature is different; $T=20K$ for BCSID-141.557, $T=0.1K$ for BCSID-0.51  \\
\cline{4-5}
 & & & 0.51 & Ho(-4.26,-4.26,-1.84),Ru(0.22,0.22,1.77)  &  \\
\hline
\multirow{2}{*}{ 6 }  & \multirow{2}{*}{Ni2SiO4  } & \multirow{2}{*}{14.82  }& 1.203 & Ni(1,0,1) &  Small difference on the lattice parameter and experimental temperature \\
\cline{4-5}
 & & &  1.204 & Ni(1.82,0,-0.9)  &  \\
\hline
\multirow{2}{*}{ 7 }  & \multirow{2}{*}{ScMn6Ge6 } & \multirow{2}{*}{192.252}& 1.110 & Mn(0,0,1.96) &  Experimental temperature is different; $T=309K$ for BCSID-1.110, $T=149$ for BCSID-1.225 \\
\cline{4-5}
 & & &  1.225 & (0,0,2.08)  &  \\
\hline
\multirow{2}{*}{ 8 }  & \multirow{2}{*}{Sr2IrO4  } & \multirow{2}{*}{54.352 }& 1.3 & Ir(0.24,0,0) & Small canting along $y$ axis \\
\cline{4-5}
 & & &   1.77 & Ir(0.202,0.048,0)  &  \\
\hline
\multirow{2}{*}{ 9 }  & \multirow{2}{*}{U2Rh2Sn  } & \multirow{2}{*}{135.492}& 1.103 & U(0,0,0.53) &   Small poloarization on Rh \\
\cline{4-5}
 & & &  1.207 & U(0,0,0.5),Rh(0.04,0.04,0)  &  \\
\hline
\multirow{2}{*}{ 10}  & \multirow{2}{*}{Mn2O3    } & \multirow{2}{*}{61.433 }& 0.40 & Mn1(2.6,0,-1.6),Mn2(3.4,0,0.7) &  Experimental temperature is different; $T=2K$ for BCSID-0.40, $T=40K$ for BCSID-0.41 \\
\cline{4-5}
 & & &  0.41 & Mn1(2.4,0,-1.4),Mn2(3.0,0,0.8)  &  \\
\hline
\multirow{2}{*}{ 11}  & \multirow{2}{*}{Co4Nb2O9 } & \multirow{2}{*}{15.88  }& 0.196 & Co1(3.7,1.85,1.42),Co2(2.78,1.39,0.97) &  Small difference on the lattice parameter and experimental temperature  \\
\cline{4-5}
 & & &  0.197 & Co1(2.677,1.312,0),Co(2.842,1.953,0)  &  \\
\hline
\multirow{2}{*}{ 12}  & \multirow{2}{*}{HoMnO3   } & \multirow{2}{*}{185.197}& 0.32 & Mn(1.72,3.44,0) &  Experimental temperature is different; $T=32K$ for BCSID-0.32, $T=1.7K$ for BCSID-0.33 \\
\cline{4-5}
 & & &  0.33 & Mn(1.76,3.52,0),Ho(0,0,2.87)  &  \\
\hline
\multirow{2}{*}{ 13}  & \multirow{2}{*}{FeI2     } & \multirow{2}{*}{12.62  }& 3.14 & Fe(0,0,1) &  Different lattice parameters \\
\cline{4-5}
 & & & 1.0.13 & Fe(0,0,1)  &  \\
\hline
\multirow{2}{*}{ 14}  & \multirow{2}{*}{Co2SiO4  } & \multirow{2}{*}{62.441 }& 0.218 & Co1(0.94,3.14,0.47),Co2(0,3.64,0) & Small difference on the experimental temperature \\
\cline{4-5}
 & & &   0.219 & Co1(1.2,3.64,0.57),Co2(0,3.35,0)  &  \\
\hline
\multirow{2}{*}{ 15}  & \multirow{2}{*}{CuMnSb   } & \multirow{2}{*}{16.72  }& 1.233 & Mn(2.53,1.39,2.53) &  Small difference on the lattice parameter and experimental temperature \\
\cline{4-5}
 & & & 1.265 & Mn(2.25,2.25,2.25)  & \\
\hline
\hline
\end{longtable}

\LTcapwidth=1.0\textwidth
%\clearpage

\section{Computational methods}\label{app:C}
\subsection{Convention setting of the magnetic unit cell}
We read the crystalline parameters and magnetic moments from the magnetic structure files, whose datatype are 'mcif', provided by \href{http://webbdcrista1.ehu.es/magndata/}{BCSMD}.
In the BCSMD website, lattice parameters of the magnetic unit cell are in the convention called working setting ($\vec a$, $\vec b$, $\vec c$) and it can be
 transformed to the standard convention ($\vec a_s$, $\vec b_s$, $\vec c_s$) by the transformation matrix $T_s=\{T|\vec \tau\}$ as,
 
\begin{equation}
(\vec a_s,\vec b_s,\vec c_s) = T \cdot (\vec a, \vec b, \vec c) + \vec \tau
 \end{equation}
 where the transformation matrix $T_s=\{T|\vec \tau\}$ of each material has been supplied in the BCSMD website.
 
While, in the {\it ab initio} calculations, we adopt the primitive magnetic unit cell. The primitive lattice vectors ($\vec p_1$, $\vec p_2$, $\vec p_3$) can be obtained by transforming the lattice vectors in standard convention ($\vec a_s$, $\vec b_s$, $\vec c_s$) with the transformation matrix $M_X$, 
 \begin{equation}
(\vec p_1, \vec p_2, \vec p_3) = (\vec a_s,\vec b_s,\vec c_s) \cdot  M_X 
 \end{equation}
where $M_X$ has been supplied in the VASP2trace package (www.cryst.ehu.es/cryst/checktopologicalmat)  and $X$ is the lattice type of the magnetic unit cell.

\subsection{Parameters setting in {\it ab initio} calculations}

We  perform all of the first-principle calculations using the Vienna {\it ab initio} simulation package(VASP); the generalized gradient approximation (GGA) with the Perdew-Burke-Ernzerhof (PBE) type exchange-correlation potential was adopted. For each material, we set the cutoff energy for plane wave basis as 1.2 times larger than 
the suggested value in the pseudo-potential files. 
In the {\it ab initio} calculations, the initial magnetic moments are set to the experimental values provided by BCSMD
website. The convergence accuracy of self-consistent calculations is $10^{-5}$eV and spin-orbital coupling (SOC) has been included. For magnetic cells containing less than 50 atoms, the Brillouin zone (BZ) sampling 
is performed by using k grids with a 9$\times$9$\times$9 mesh
in self-consistent calculations. We reduce the grids to 5$\times$5$\times$5 if there are more than 50 atoms in the magnetic primitive cell to save calculational costs.  
We implement the {\it ab initio} calculations on the MPG supercomputer Cobra and Draco with 960 CPU cores in total and the supercomputer in ShanghaiTech University with 560 CPU cores. 
For benchmarking, we calculate the simple compound CeCo$_2$P$_2$ (with 10 atoms
per magnetic primitive cell) on the Cobra supercomputer with 80 Skylake cores at 2.4GHz. The time used is 15 min 34s for the self consistent calculations and 15 min 45s for the band structure calculations with 240 k points.
For the complex compound Sr$_3$CoIrO$_6$ with 66 atoms per magnetic primitive cell, it costs 7h 50min for the self consistent calculations and 6h 42 min 53s for the band structure calculation with 200 k points. 

Since all of the magnetic materials contain at least one correlated element, we also perform the L(S)DA+U calculations for all of the magnetic materials using the VASP. 
For the L(S)DA+U calculations, we adopt the simplified (rotationally invariant) approach and set the Hubbard U as 1, 2, 3, 4 eV for the $d$ electrons and 2, 4, 6 eV for the $f$ electrons. For the materials which have both $d$ and $f$
electron, we set U of $d$ electron as 2 eV and the U of  $f$ electron as 2, 4, 6 eV.

Similar with the TQC for paramagnetic materials, we also provide the maximal $\bf k$ vectors  for each magnetic space group in the BCS website.
Based on the self-consistent charge density files, we calculate the wave functions at the magnetic maximal $\bf k$ vectors and obtain the characters using the MagVASP2trace package.

\subsection{Magnetic VASP2trace package}
In TQC, the in house VASP2trace package \cite{gao2020irvsp} is used to calculate the character tables of paramagnetic materials. It read the unitary symmetry operators from the output files of VASP and can identify the space group. While the anti-unitary symmetries are absent and VASP2trace cannot identify the magnetic space groups (MSGs). 

In the MTQC, we revise the VASP2trace package to calculate the character tables of magnetic materials and supply the symmetry file for each MSG. The magnetic VASP2trace (MagVASP2trace) \cite{magwebsite} reads the magnetic symmetries from the symmetry files that we supply, instead of reading them from the output files of VASP. The symmetry file contains both unitary operations and anti-unitary operations. Both SO(3) and SU(2) matrix in the symmetry files are written in the basis of primitive lattice vectors.

MagVASP2trace adopts both SO(3) and SU(2) matrix in the convention used in the BCS website (https://www.cryst.ehu.es/) and generate the \textbf{trace.txt} file that contains all of the magnetic symmetry operators and the character tables of the occupied bands at the magnetic maximal \textbf{k} vectors.

\subsection{Construction of Wannier tight-binding Hamiltonian and surface states calculation}
We construct the tight-binding Hamiltonians of NpBi, CeCo$_2$P$_2$, MnGeO$_3$ and Mn$_3$ZnC using the  Wannier90 package~\cite{Souza2001Maximally-PRB}. We generate the maximally localized Wannier functions (MLWFs) for 5$p$ orbitals on Bi, 5$f$ and 6$d$ orbitals on Np for the magnetic TI NpBi. For the magnetic NLSM CeCo$_2$P$_2$, the MLWFs for 3$p$ orbitals on P, 3$d$ orbitals on Co, 4$f$ and 5$d$ orbitals on Ce are constructed. For the magnetic DSM MnGeO$_3$, we generate the MLWFs for 4$s$ orbitals on Ge, 2$p$ orbitals on O, and 3$d$ orbitals on Mn. For the ferrimagnetic ES Mn$_3$ZnC, we generate the MLWFs for $4s$, $4p$ and $3d$ orbitals on Zn, $2p$ orbitals on C and $3d$ orbitals on Mn. 

The surface states are calculated with the Green's function method using the WannierTools package~\cite{sancho1985highly,WU2017}, and the results are shown in FIG. 2 of main text and FIG.~\ref{appfig1}-\ref{appfig2}.

\section{Comparison of the ground state energy  between different magnetic configurations of several compounds}\label{app:L}
\begin{figure}[htbp] 
\centering\includegraphics[width=7.0in]{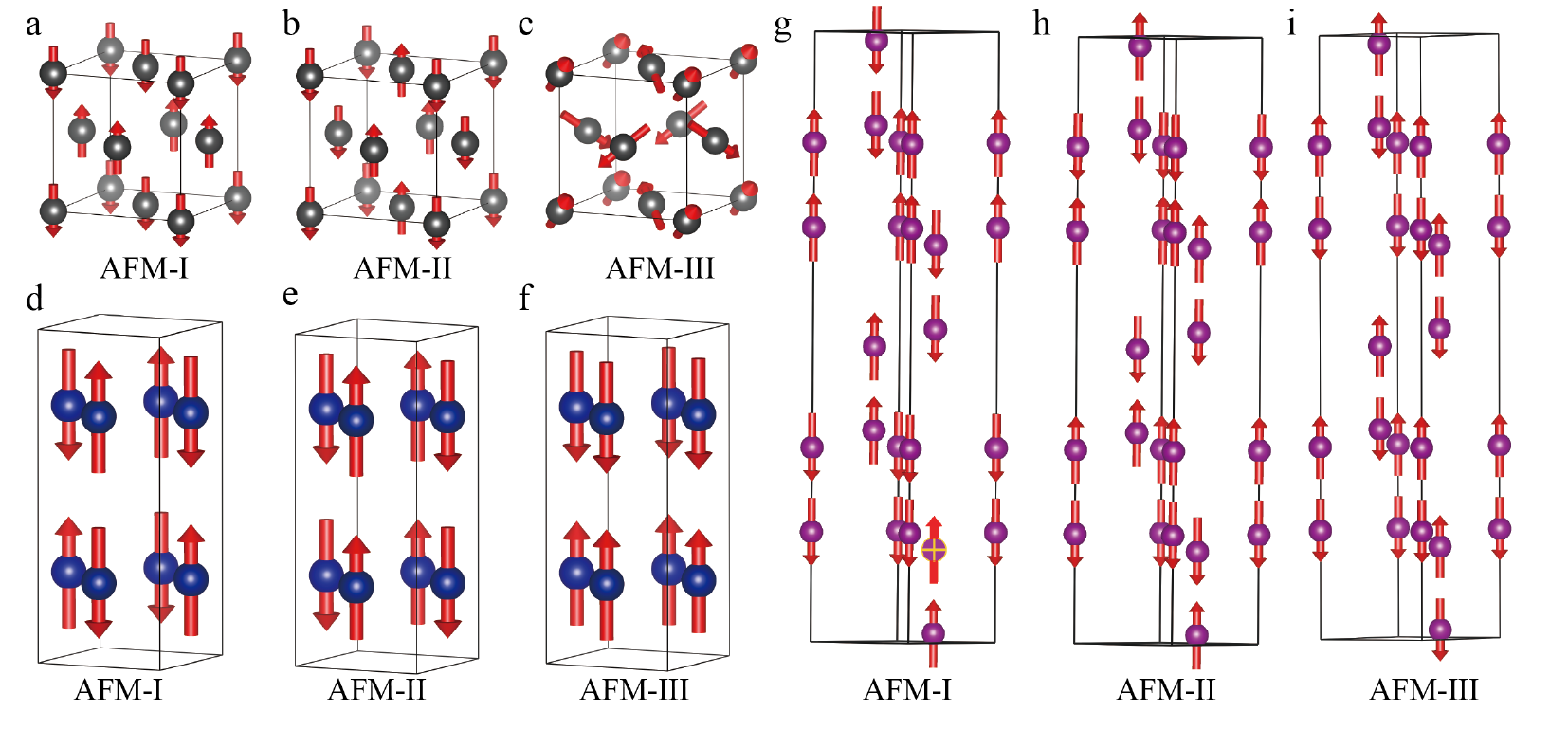} 
\caption{Three possible magnetic structures for (a-c)NpBi, (d-f)CeCo$_2$P$_2$ and (g-i) MnGeO$_3$, where the AFM-III phase in (c)(f)(i) are the ones from neutron scattering experiments. }\label{compare} 
\end{figure}

We select the three magnetic topological materials NpBi with BCSID-3.7, CeCo$_2$P$_2$ with BCSID-1.253 and MnGeO$_3$ with BCSID-0.125 to compare the energy difference between different magnetic structures, respectively.
As shown in Figure~\ref{compare}, there are three possible magnetic structures for each material, where AFM-I and AFM-II are the assumed configurations and the AFM-III phase is the one obtained from neutron experiments. 
The relative ground state energies at each U for the three materials are tabulated in Table \ref{mc-table}. For NpBi and CeCo$_2$P$_2$, AFM-III phase always has the lowest ground state energy at different U. For MnGeO$_3$, there is only one exception, i.e. the AFM-I phase of MnGeO$_3$ with U=0, that has lower energy than the AFM-III phase. With increasing U, the experimental AFM-III phase lowers its energy to become the lowest. 

The comparisons in Table \ref{mc-table} indicate that the magnetic configurations obtained from neutron experiments are favorable with the lowest ground state energy.

\LTcapwidth=1.0\textwidth
\begin{longtable}{c|c|c|c|c}
\caption{The relative ground state energy of NpBi, CeCo$_2$P$_2$  and MnGeO$_3$ in three possible magnetic structures with different U added. The magnetic structures are shown in Figure~\ref{compare}, where the magnetic configurations AFM-III are obtained from neutron scattering experiments.}\label{mc-table}
\label{tab:compare} \\
\hline
\hline
Materials (BCSID)  &  U(eV)	& E(AFM-I) (eV)	 & E(AFM-II) (eV)	& E(AFM-III) (eV) \\
\hline\hline
\multirow{4}{*}{NpBi (BCSID: 3.7)}  &  0 &  0.079	& 0.091 & 0 \\
\cline{2-5}
  &  2	& 2.108	& 2.125 & 1.995 \\
\cline{2-5}
  &  4	& 3.049	& 3.05 & 2.999 \\
\cline{2-5}
  &  6	& 3.538	& 3.543 & 3.525 \\
\hline
\hline
\multirow{4}{*}{CeCo$_2$P$_2$ (BCSID: 1.253)}  &  0	 & 0.388	& 0.388 & 0 \\
\cline{2-5}
  &  2	& 9.159	 & 9.243  & 8.468 \\
  &  4	& 10.666 & 10.729 & 9.973 \\
  &  6	& 11.955 & 12.098 & 11.263 \\
\hline
\hline
\multirow{5}{*}{MnGeO$_3$ (BCSID: 0.125)}   &  0 & -0.4 & 0.164 & 0 \\
\cline{2-5}
  &  1	& 6.845  & 6.532	 & 6.437 \\
\cline{2-5}
  &  2	& 12.713 & 12.151 & 12.068  \\
\cline{2-5}
  &  3	& 17.703 & 17.132 & 17.068 \\
\cline{2-5}
  &  4	& 22.095 & 21.589 & 21.529 \\
\hline
\hline
\end{longtable}

\section{Comparisons between different exchange-correlation potentials}\label{app:D}
\subsection{Band structure calculations with GGA functional}
\begin{figure}[htbp] 
\centering\includegraphics[width=5.0in]{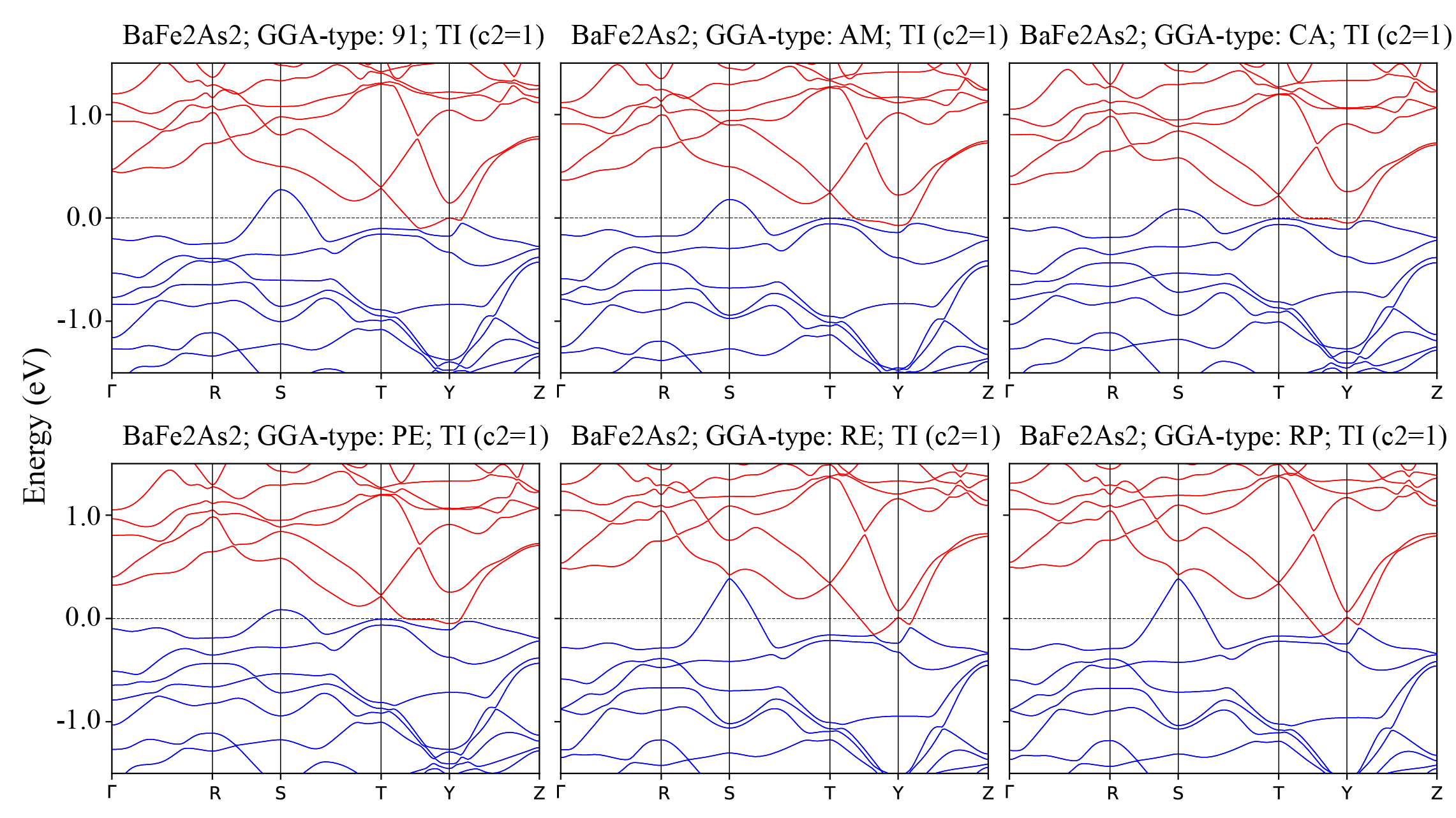} 
\caption{The band structures and topology of BaFe$_2$As$_2$ obtained by different exchange-correlation functional methods. The topology is maintained for different methods, indicating a TI with topological index $c_2=1$. The Hubbard U of $3d$ electron is set to 1 eV.}\label{sm-BaFeAs} 
\end{figure}

\begin{figure}[htbp] 
\centering\includegraphics[width=5.0in]{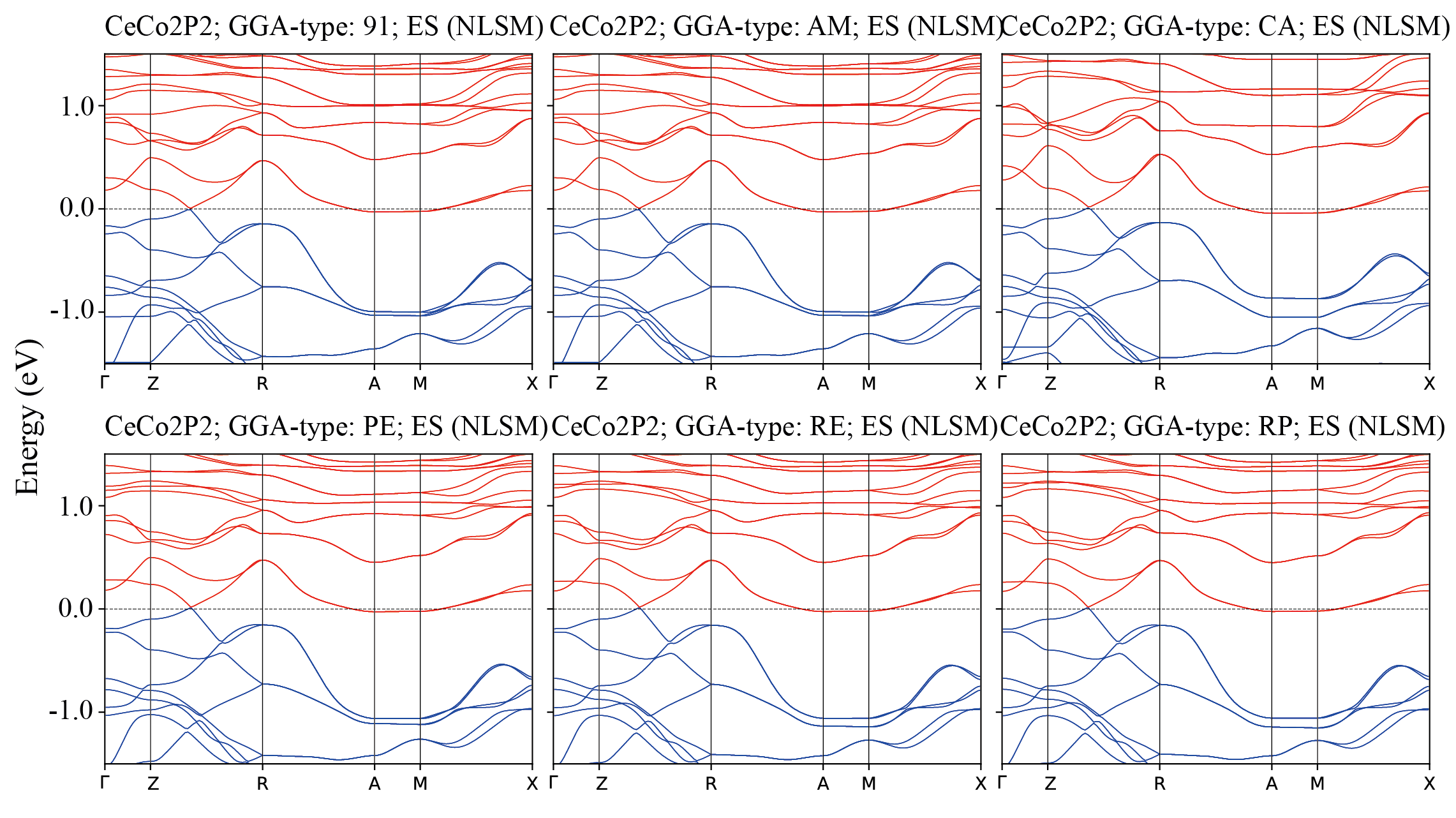} 
\caption{The band structures and topology of CeCo$_2$P$_2$ obtained by different exchange-correlation functional methods. The topology is maintained ES (also NLSM) for the different methods. The Hubbard U of $3d$ and $4f$ electron are set to 2 and 6 eV, respectively.}\label{sm-CeCoP} 
\end{figure}

\begin{figure}[htbp] 
\centering\includegraphics[width=5.0in]{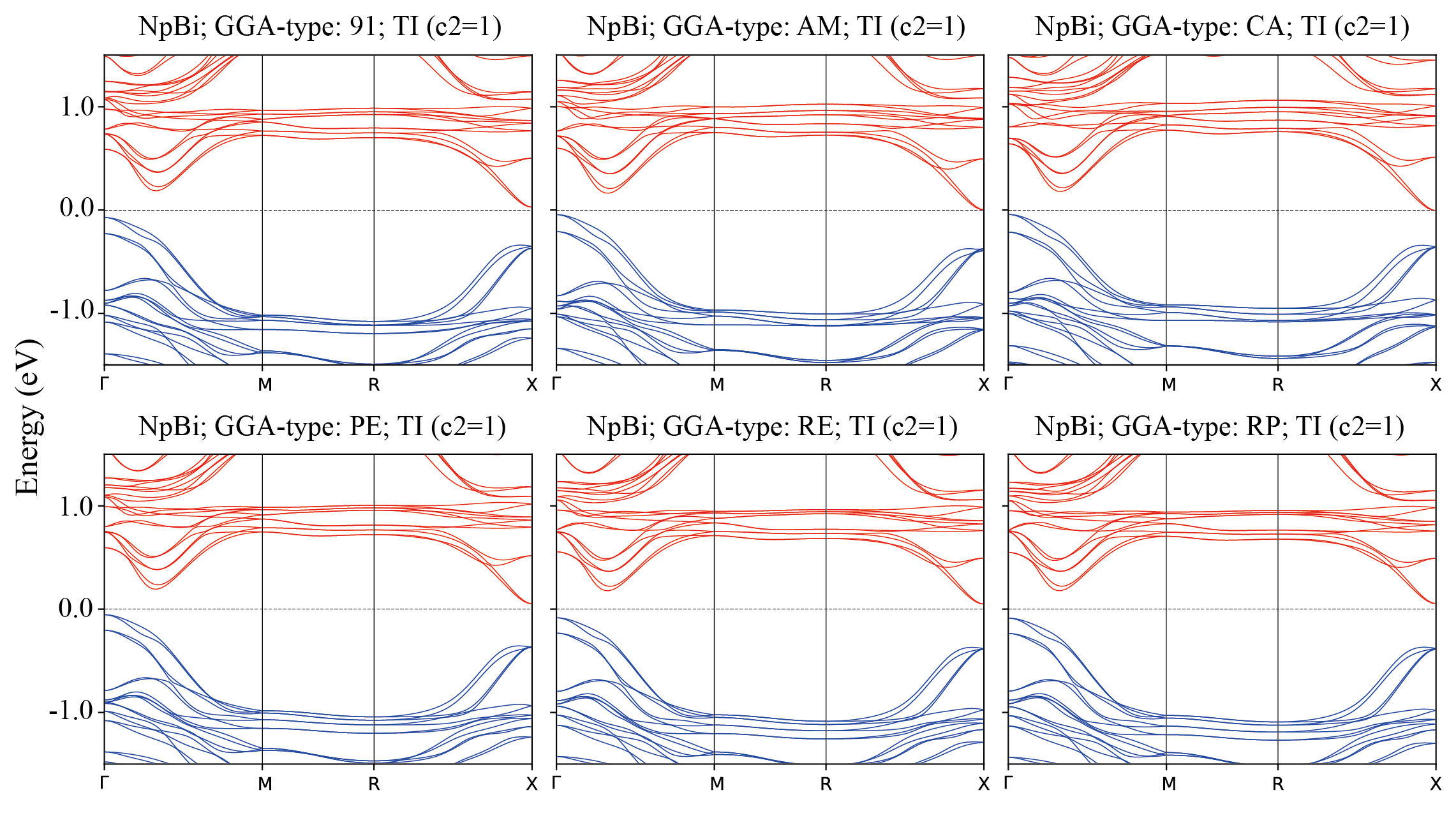} 
\caption{The band structures and topology of NpBi obtained by different exchange-correlation functional methods. The topology is maintained TI with topological index $c_2=1$ for the different methods. The Hubbard U of $5f$ electron is set to 2 eV.}\label{sm-NpBi} 
\end{figure}

\begin{figure}[htbp] 
\centering\includegraphics[width=5.0in]{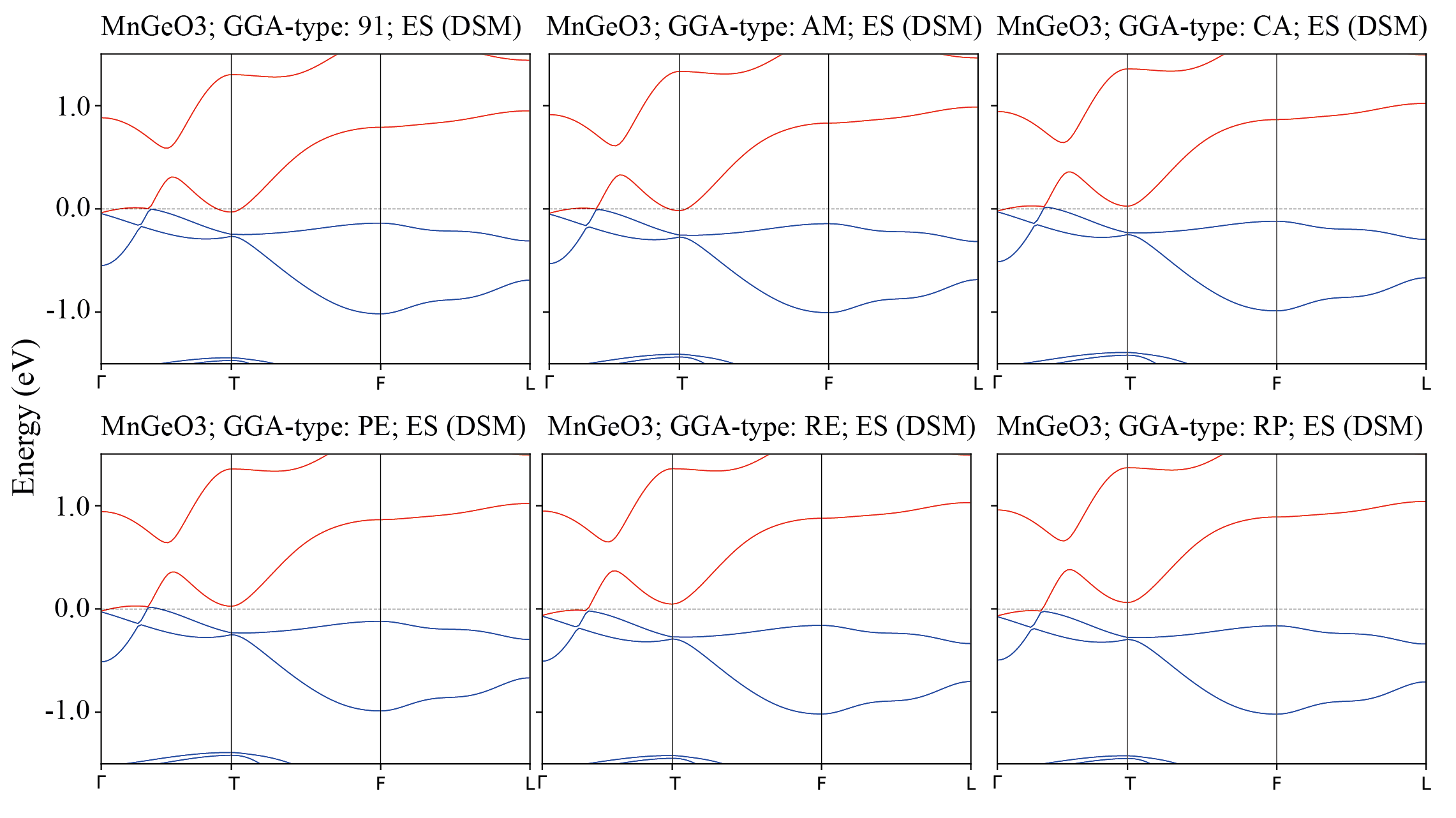} 
\caption{The band structures and topology of MnGeO$_3$ obtained by different exchange-correlation functional methods. The topology is maintained ES (also DSM) for the different methods. The Hubbard U of $3d$ electron is set to 4 eV.}\label{sm-MnGeO} 
\end{figure}

We adopt five different exchange-correlation functional methods, including Perdew-Wang 91 (91), AM05 (AM), revised PBE (RE), revised PBE with Pade Approximation (RP) and Ceperley-Alder functional (CA), to check the topologies and the band structures that obtained by the PBE (PE) method.  
The topology and band structure comparisons of the topological materials 
BaFe$_2$As$_2$, CeCo$_2$P$_2$, NpBi, and MnGeO$_3$ are 
shown in the FIG.~\ref{sm-BaFeAs}-\ref{sm-MnGeO}. 
The comparisons indicate that different exchange-correlation functional methods have minor effect on the band structures but do not change the topologies of these materials.

\subsection{Band structure calculations with meta-GGA functional}\label{app:D2}

To further check the band structures and topologies obtained from LDA+U calculations, we have also performed
ab initio calculations with the modified Becke-Johnson (mBJ) \cite{tran2009accurate} potential for 23 topological compounds. They are Mn$_3$Ir (BCSID-0.108), Mn$_3$Sn (BCSID-0.200), Mn$_3$Pt (BCSID-1.143), MnGeO$_3$ (BCSID-0.125), Mn$_2$As (BCSID-1.132), CaFe$_2$As$_2$ (BCSID-1.52), Cd$_2$Os$_2$O$_7$ (BCSID-0.2), NiCr$_2$O$_4$ (BCSID-0.4), PbNiO$_3$ (BCSID-0.21), LuFeO$_3$ (BCSID-0.117), LuFe$_4$Ge$_2$ (BCSID-0.140), NiS$_2$ (BCSID-0.150), Mn$_3$Ge (BCSID-0.203), Co$_2$SiO$_4$ (BCSID-0.218), CrN (BCSID-1.28), ScMn$_6$Ge$_6$ (BCSID-1.110), CaCo$_2$P$_2$ (BCSID-1.252), CeCo$_2$P$_2$ (BCSID-1.253), GdIn$_3$ (BCSID-1.81), Mn$_3$ZnC (BCSID-2.19), NpBi (BCSID-3.7), NpSe (BCSID-3.10) and NpSb (BCSID-3.12).

\begin{figure}[htbp] 
\centering\includegraphics[width= 4.2in]{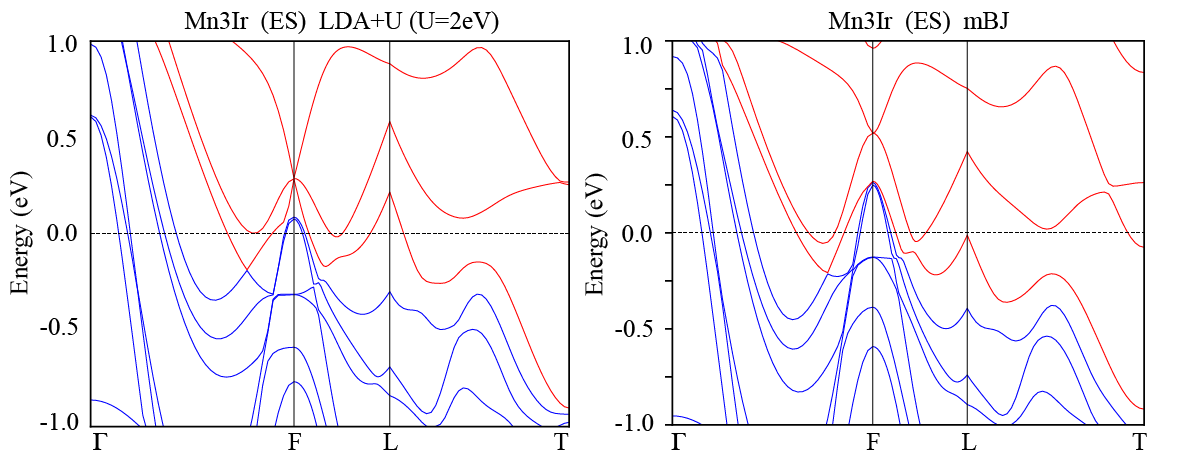} 
\caption{Band structures of the ES Mn$_3$Ir obtained from LDA+U ($U=2eV$)  and mBJ methods. }
\label{mbj-0.108} 
\end{figure}

\begin{figure}[htbp] 
\centering\includegraphics[width= 4.2in]{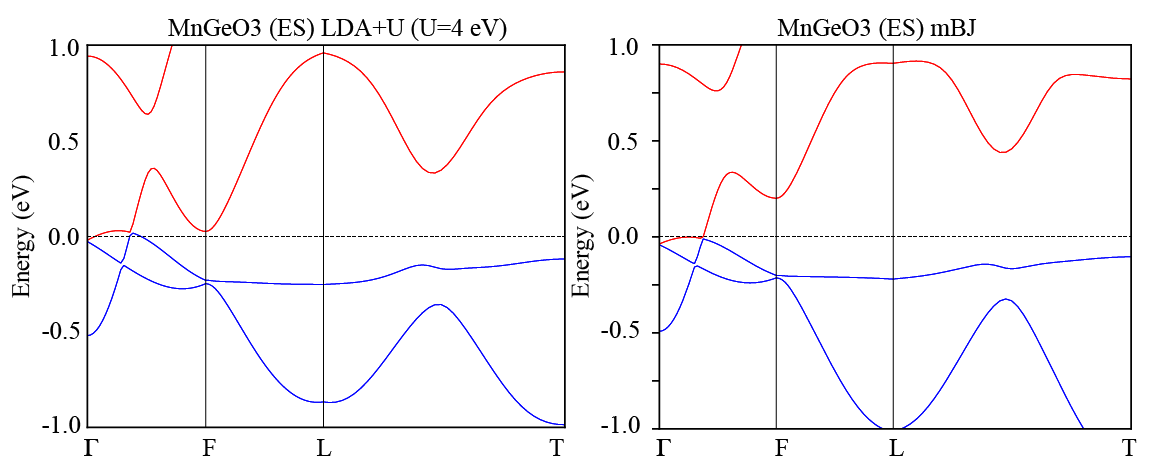} 
\caption{Band structures of the ES MnGeO$_3$ obtained from LDA+U ($U=4eV$)  and mBJ methods. }
\label{mbj-0.125} 

\end{figure}

\begin{figure}[htbp] 
\centering\includegraphics[width=  4.0in]{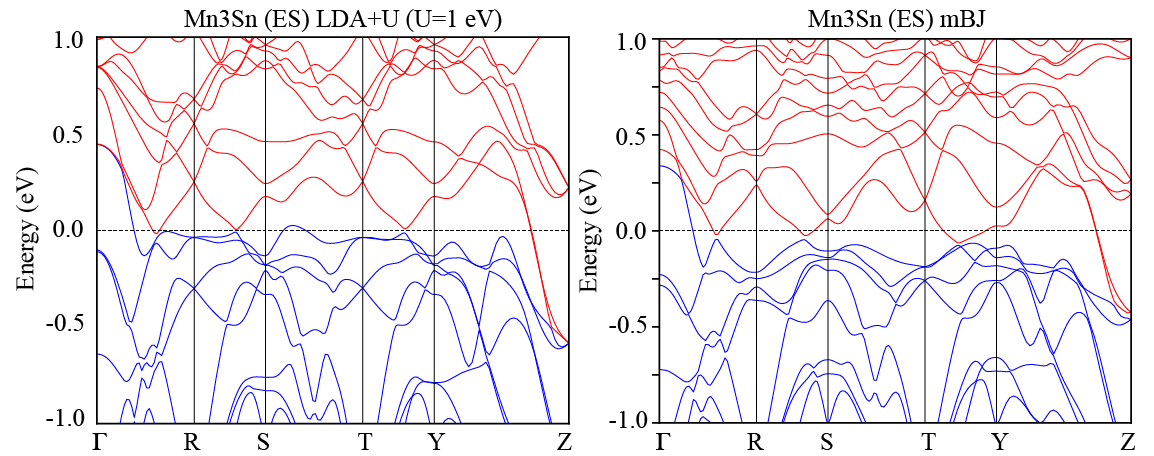} 
\caption{Band structures of the ES Mn$_3$Sn obtained from LDA+U ($U=1eV$)  and mBJ methods. }
\label{mbj-0.200} 
\end{figure}

\begin{figure}[htbp] 

\centering\includegraphics[width=  4.2in]{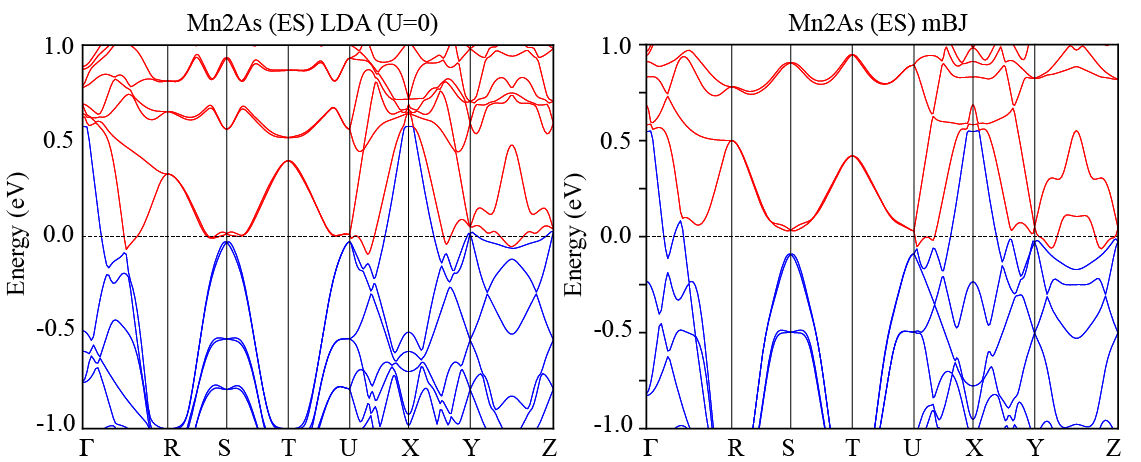} 
\caption{Band structures of the ES Mn$_2$As obtained from LDA+U ($U=0eV$)  and mBJ methods. }
\label{mbj-1.132} 
\end{figure}

\begin{figure}[htbp] 

\centering\includegraphics[width=  4.2in]{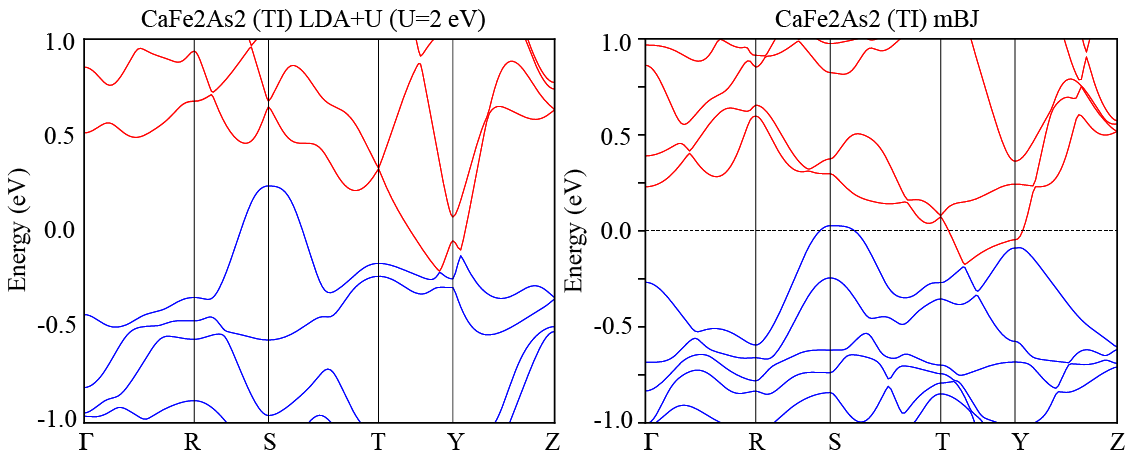} 
\caption{Band structures of the TI CaFe$_2$As$_2$ obtained from LDA+U ($U=2eV$)  and mBJ methods. }
\label{mbj-1.52} 
\end{figure}

\begin{figure}[htbp] 
\centering\includegraphics[width=  4.2in]{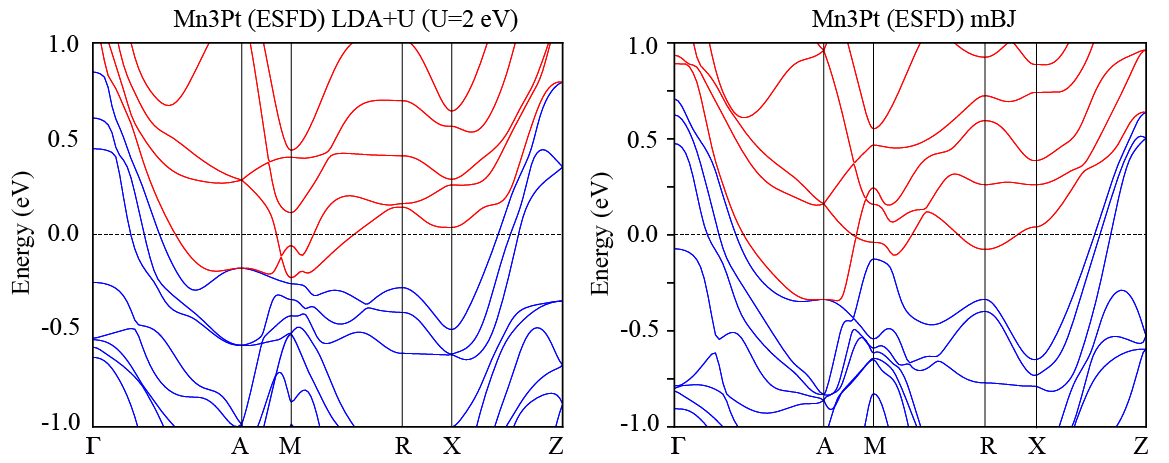} 
\caption{Band structures of the ESFD Mn$_3$Pt obtained from LDA+U ($U=2eV$)  and mBJ methods. }
\label{mbj-1.143} 
\end{figure}

\begin{figure}[htbp] 
\centering\includegraphics[width=  4.2in]{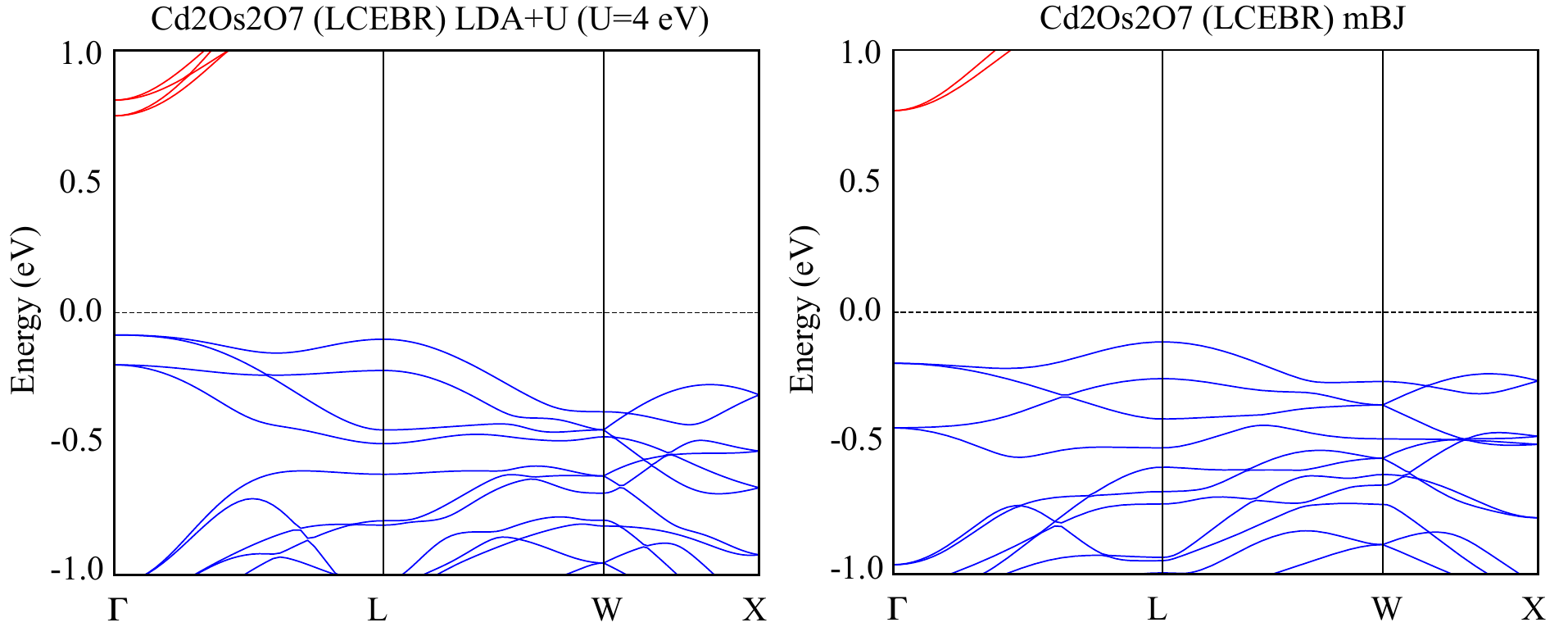} 
\caption{Band structures of Cd$_2$Os$_2$O$_7$ obtained from LDA+U ($U=2eV$) with LCEBR phase and mBJ method with LCEBR phase. }
\label{mbj-0.2} 
\end{figure}

\begin{figure}[htbp] 
\centering\includegraphics[width=  4.2in]{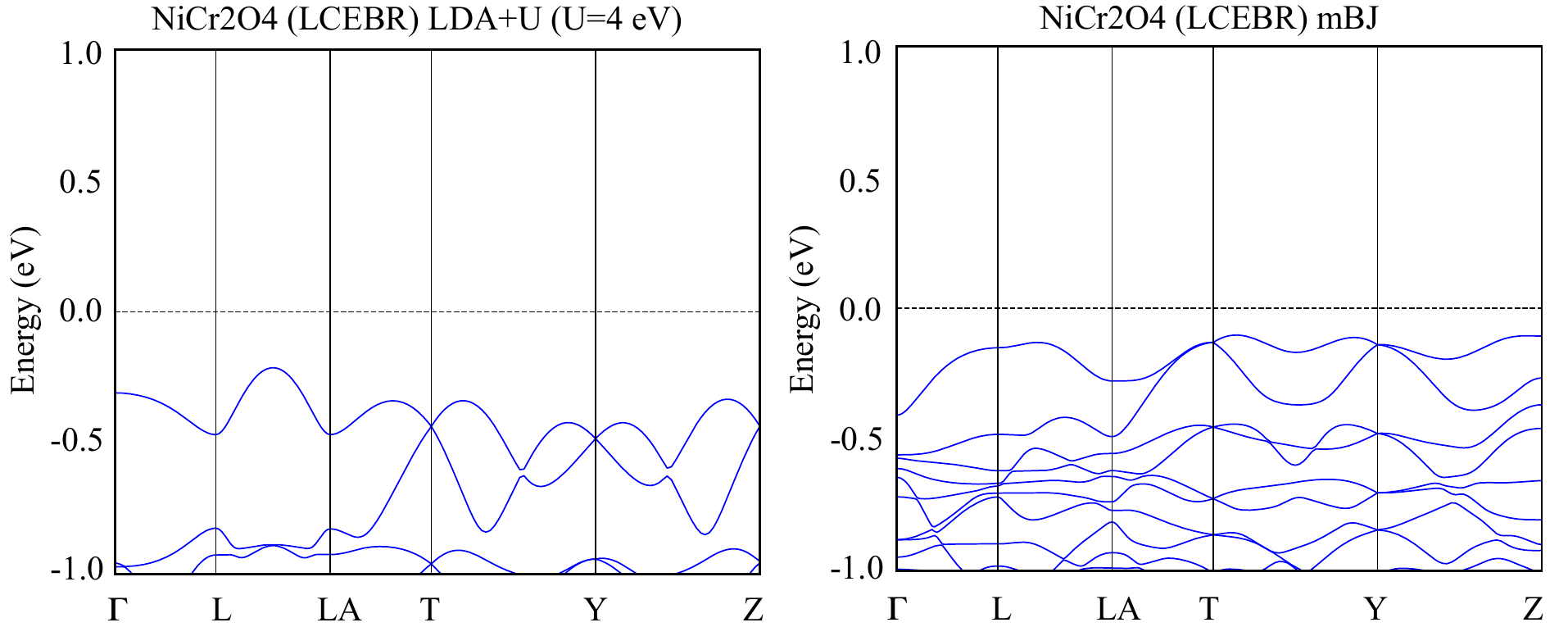} 
\caption{Band structures of the NiCr$_2$O$_4$ obtained from LDA+U ($U=4eV$) with LCEBR phase and mBJ method with LCEBR phase. }
\label{mbj-0.4} 
\end{figure}

\begin{figure}[htbp] 
\centering\includegraphics[width=  4.2in]{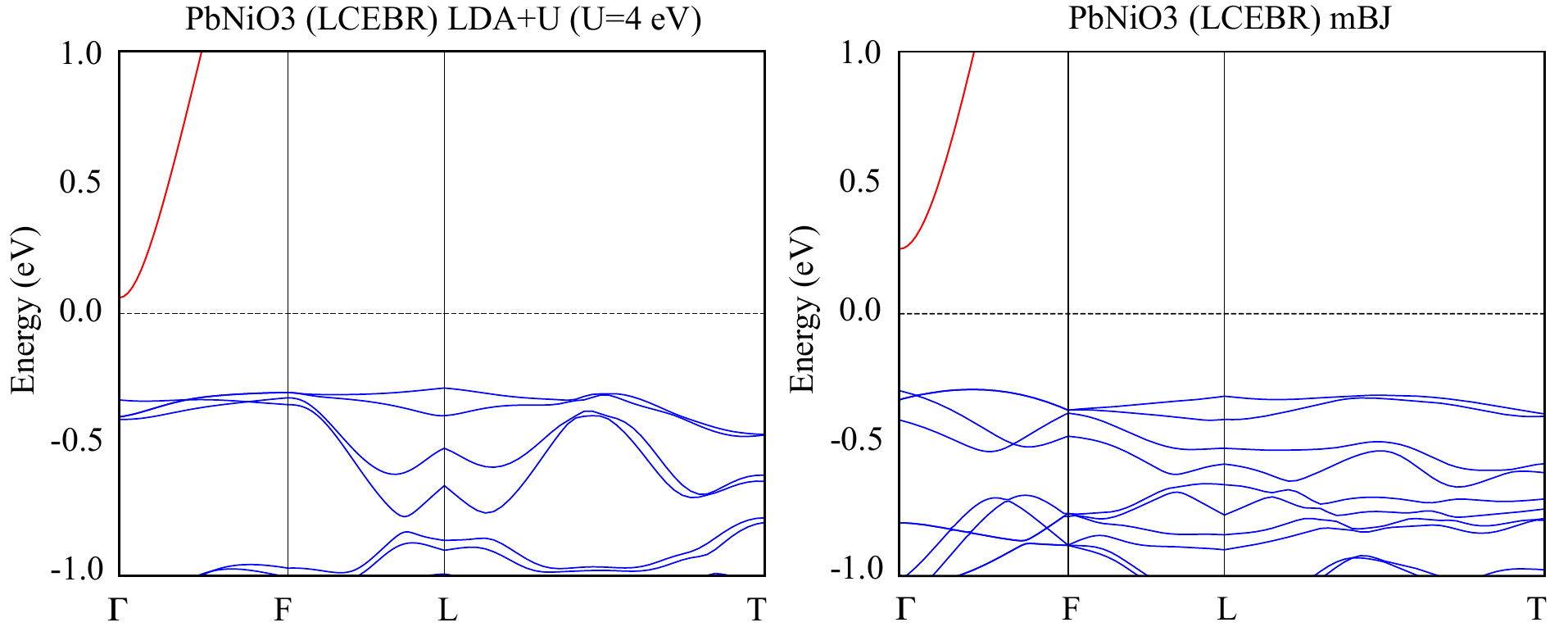} 
\caption{Band structures of PbNiO$_3$ obtained from LDA+U ($U=4eV$) with LCEBR phase and mBJ methods with LCEBR phase. }
\label{mbj-0.21} 
\end{figure}

\begin{figure}[htbp] 
\centering\includegraphics[width=  4.2in]{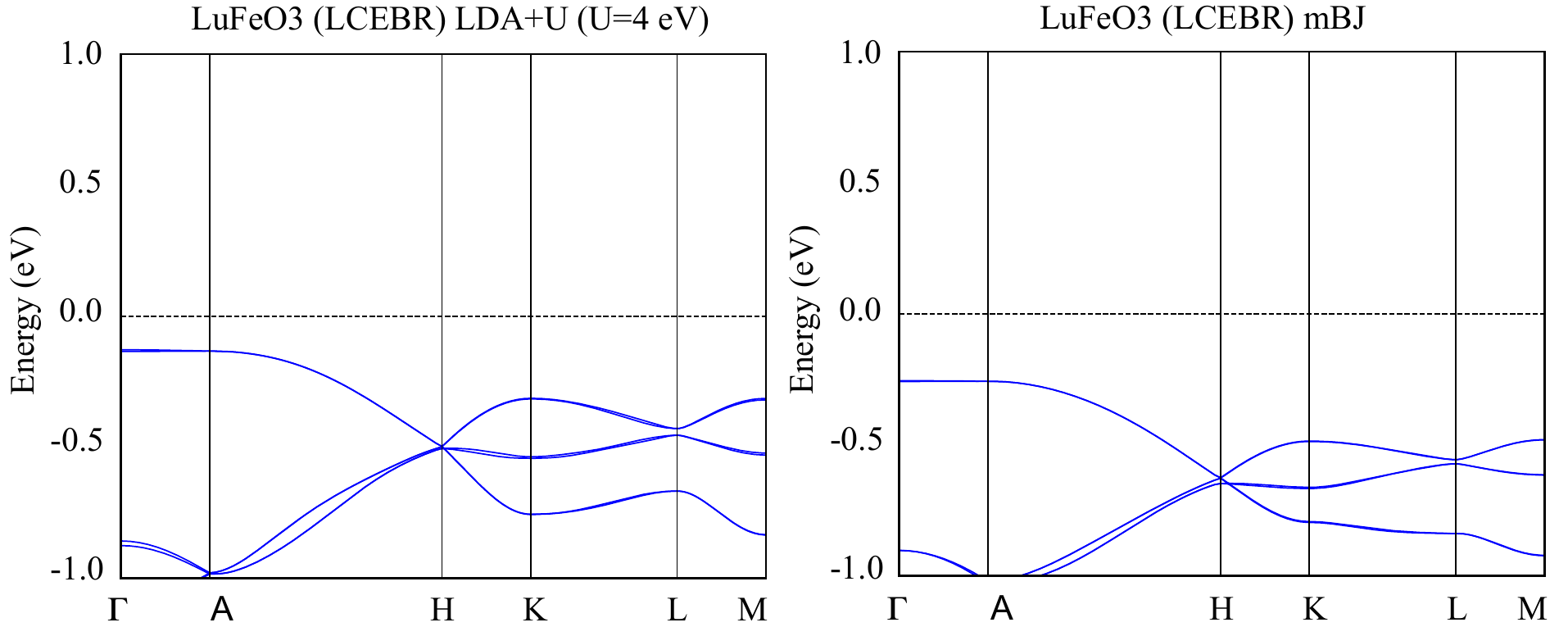} 
\caption{Band structures of the LuFeO$_3$ obtained from LDA+U ($U=4eV$) with LCEBR phase and mBJ methods with LCEBR phase. }
\label{mbj-0.117} 
\end{figure}

\begin{figure}[htbp] 
\centering\includegraphics[width=  4.2in]{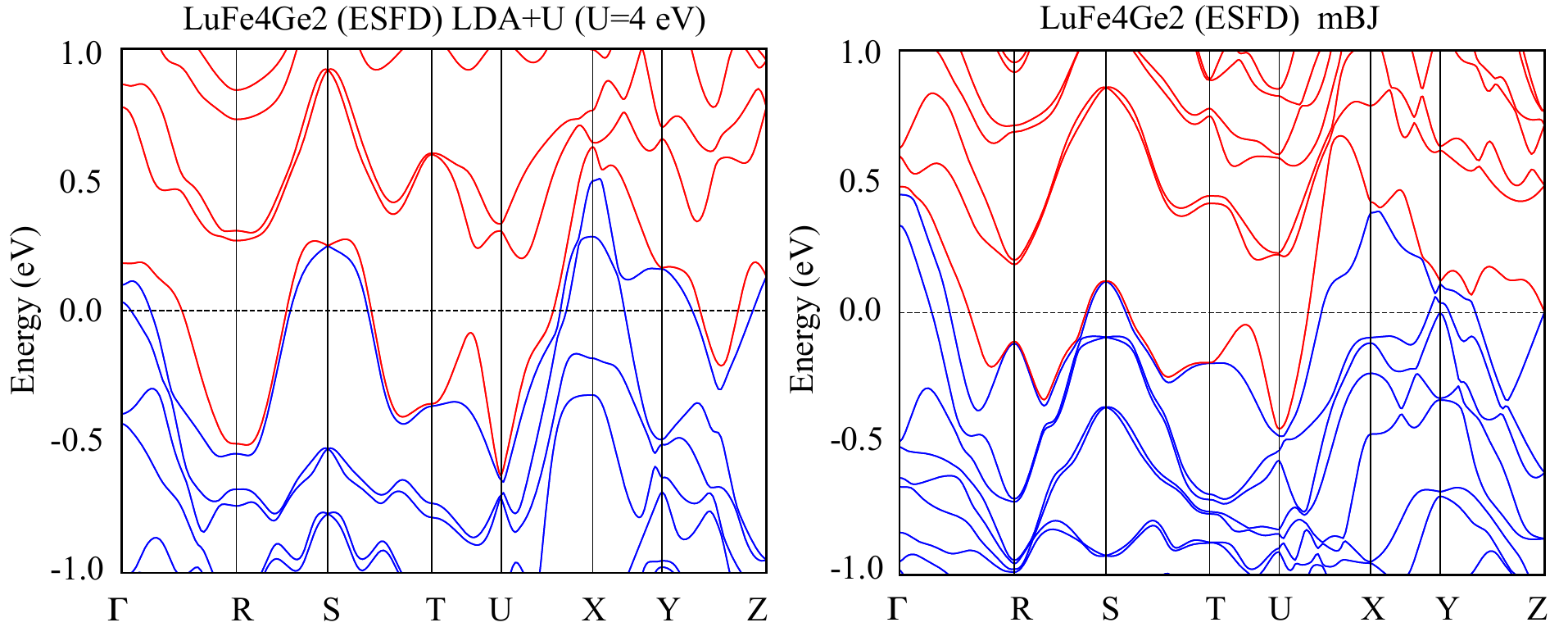} 
\caption{Band structures of the ESFD LuFe$_4$Ge$_2$ obtained from LDA+U ($U=4eV$)  and mBJ methods. }
\label{mbj-0.140} 
\end{figure}

\begin{figure}[htbp] 
\centering\includegraphics[width=  4.2in]{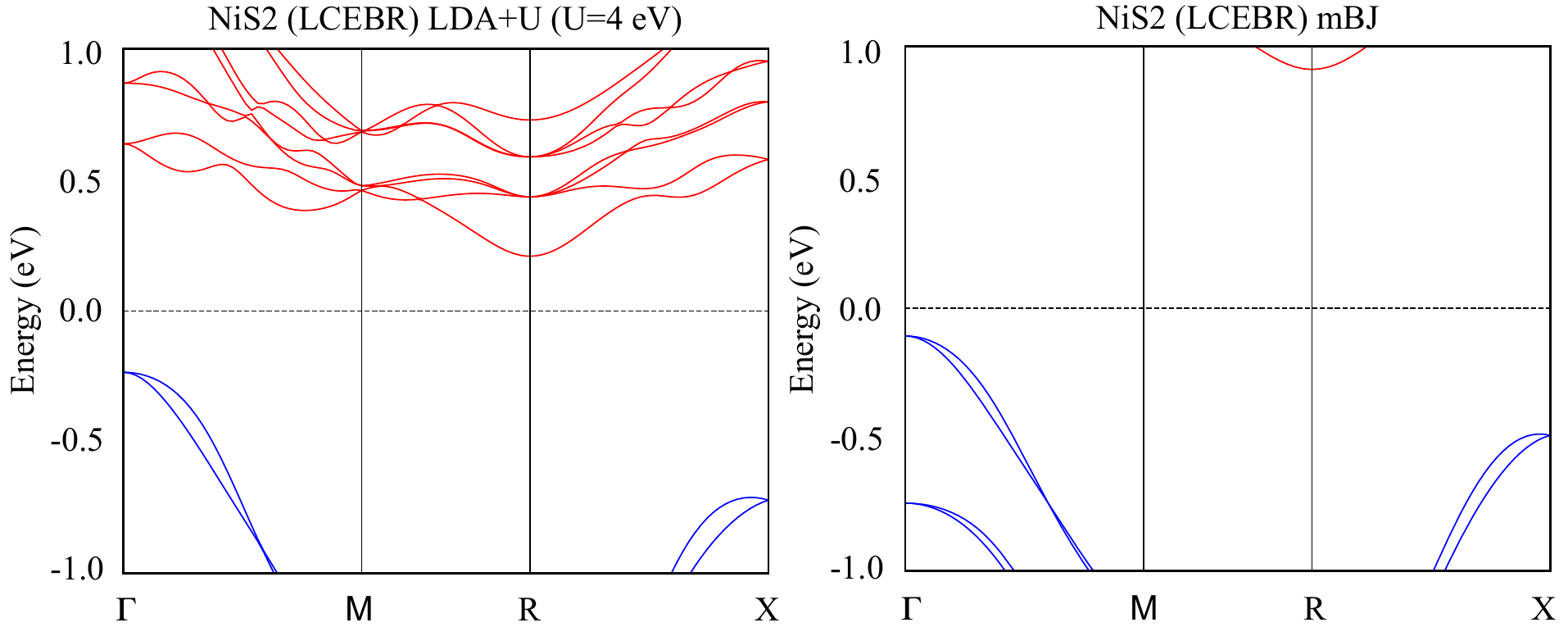} 
\caption{Band structures of NiS$_2$ obtained from LDA+U ($U=2eV$) with LCEBR phase and mBJ methods with LCEBR phase. }
\label{mbj-0.150} 
\end{figure}

\begin{figure}[htbp] 
\centering\includegraphics[width=  4.2in]{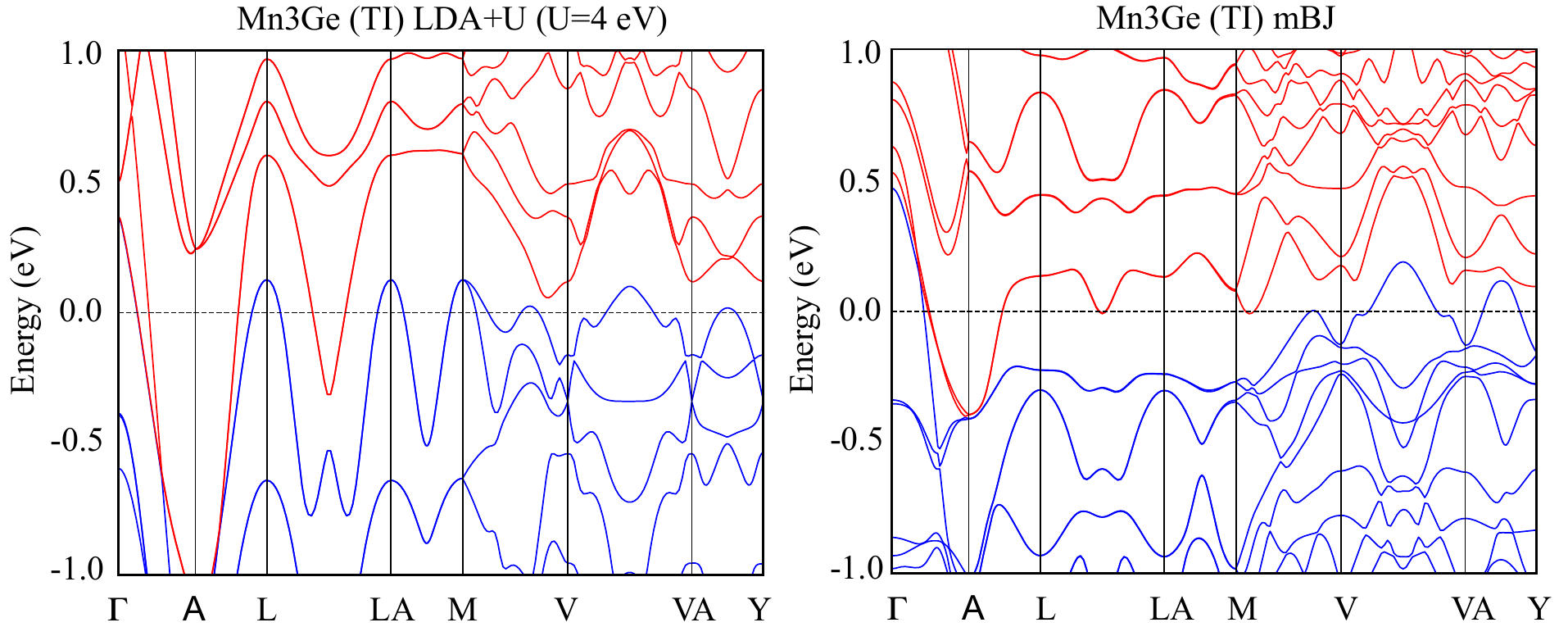} 
\caption{Band structures of the TI Mn$_3$Ge obtained from LDA+U ($U=4eV$)  and mBJ methods. }
\label{mbj-0.203} 
\end{figure}

\begin{figure}[htbp] 
\centering\includegraphics[width=  4.2in]{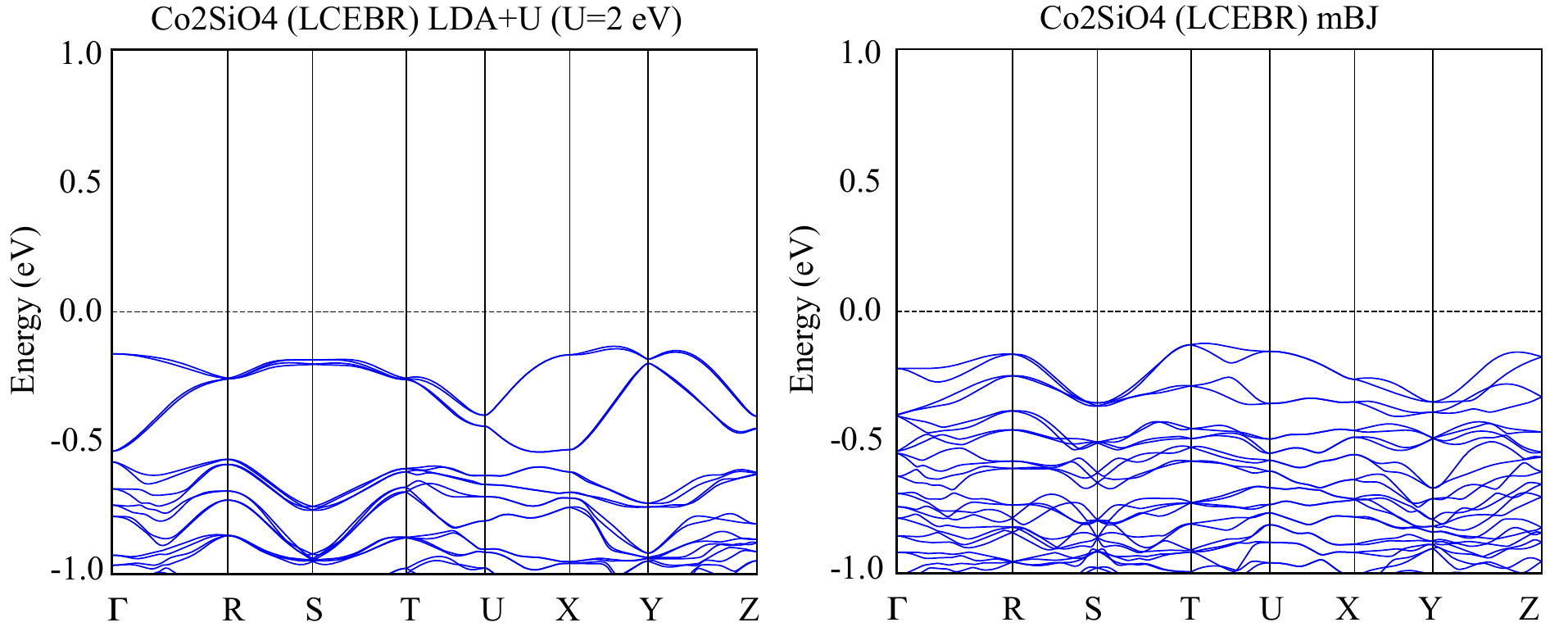} 
\caption{Band structures of Co$_2$SiO$_4$ obtained from LDA+U ($U=2eV$) with LCEBR phase and mBJ methods with LCEBR phase.  }
\label{mbj-0.218} 
\end{figure}

\begin{figure}[htbp] 
\centering\includegraphics[width=  4.2in]{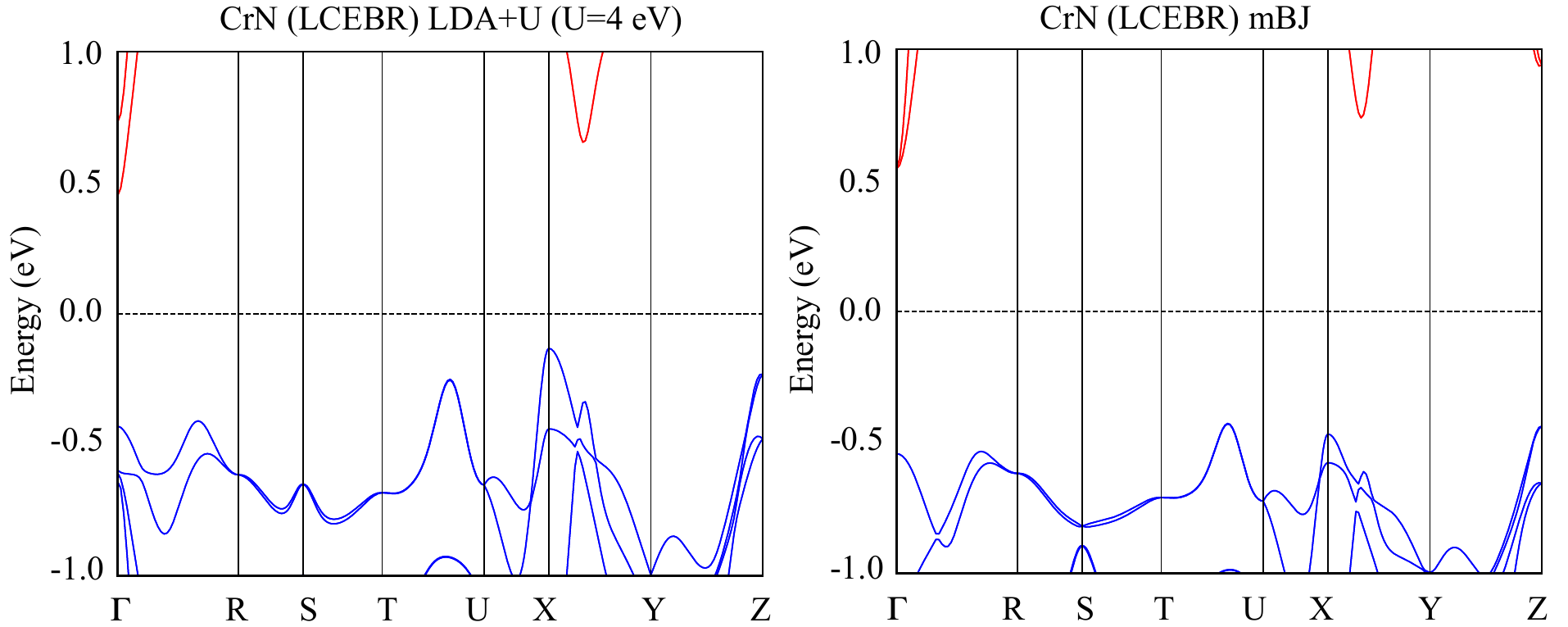} 
\caption{Band structures of CrN obtained from LDA+U ($U=2eV$) with LCEBR phase and mBJ methods with LCEBR phase. }
\label{mbj-1.28} 
\end{figure}

\begin{figure}[htbp] 
\centering\includegraphics[width=  4.2in]{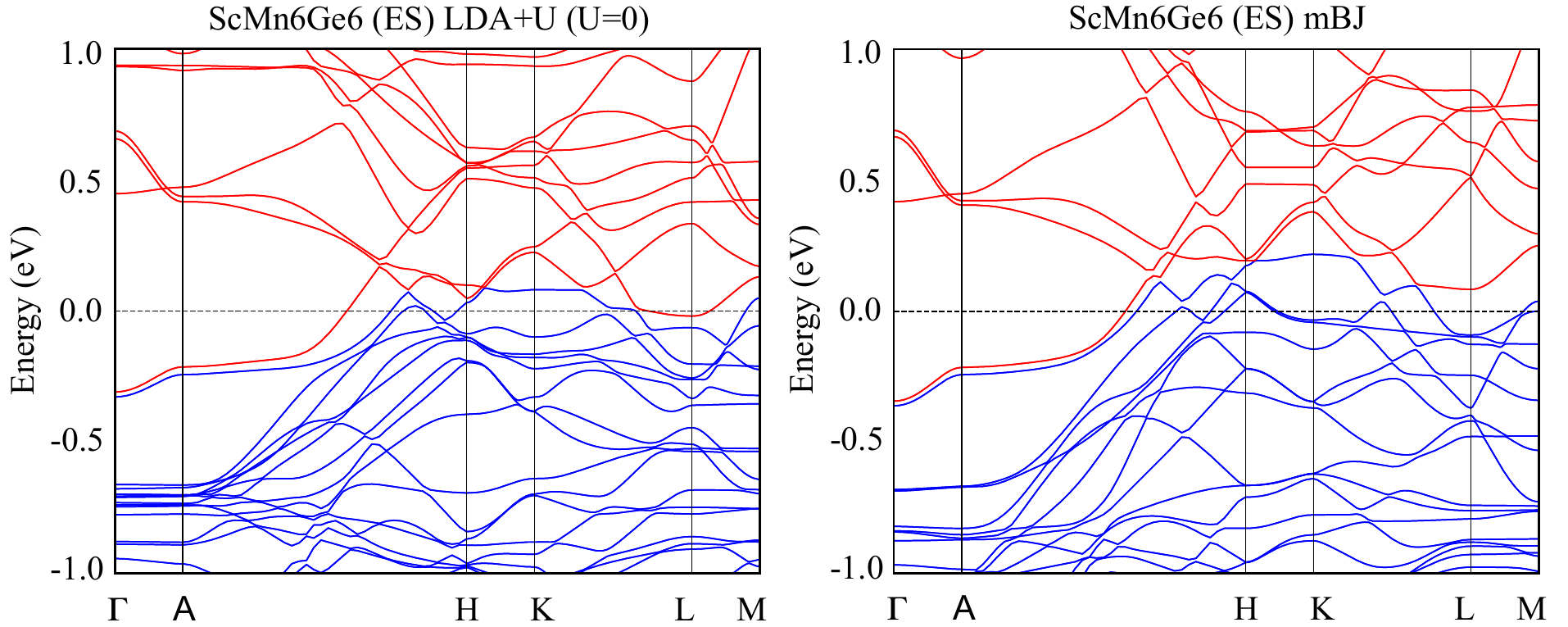} 
\caption{Band structures of the ES ScMn$_6$Ge$_6$ obtained from LDA+U ($U=0$)  and mBJ methods. }
\label{mbj-1.110} 
\end{figure}

Apart from NpSe (BCSID-3.10), we find slightly different band structures with MBJ and LDA+U methods. However, comparing these 2 band structures and its topology, we can always find a value of U that reproduces the MBJ calculations. As shown in FIG. \ref{mbj-0.108}$-$\ref{mbj-3.12}, we have found the correct value of U for each compound.  Using the correct value of U, we can reproduce the the band structures and topology at the Fermi level consistent with the results obtained from mBJ.


\begin{figure}[htbp] 
\centering\includegraphics[width=  4.2in]{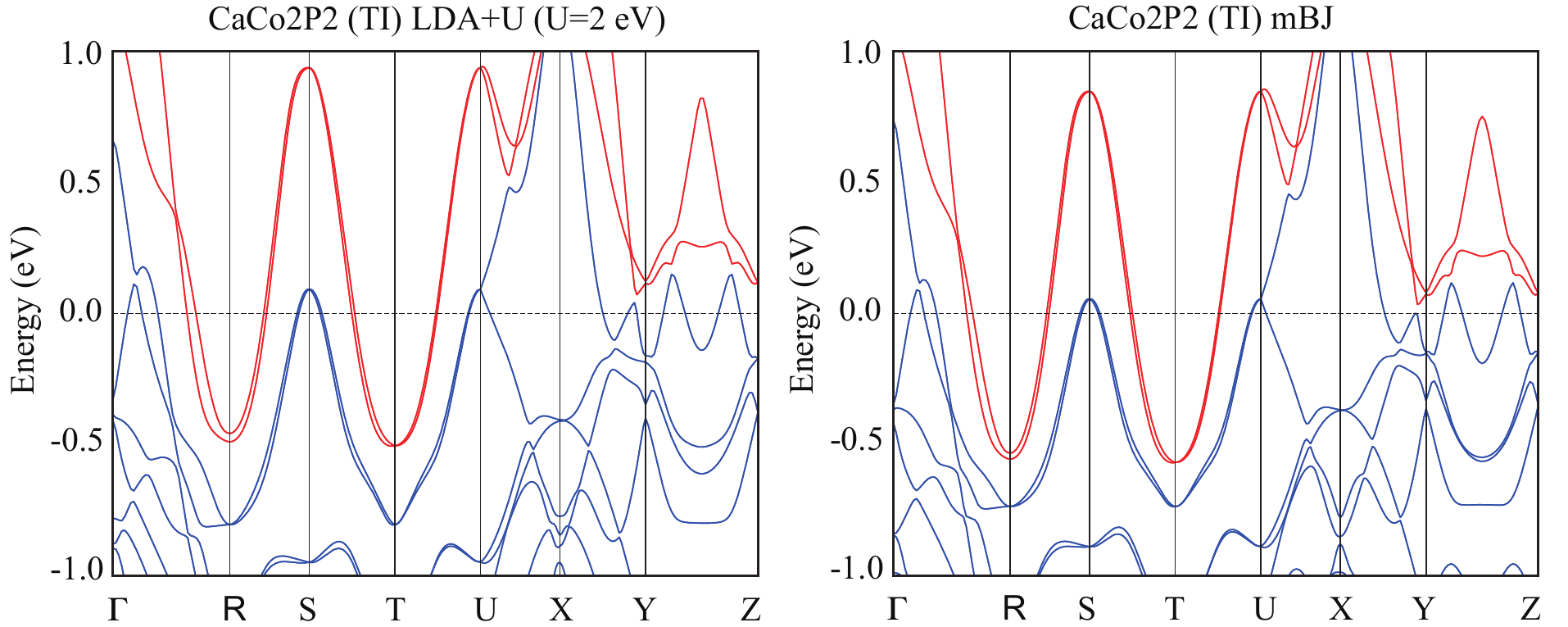} 
\caption{Band structures of the TI CaCo$_2$P$_2$ obtained from LDA+U ($U=2eV$)  and mBJ methods. }
\label{mbj-1.253} 
\end{figure} 

\begin{figure}[htbp] 
\centering\includegraphics[width=  4.2in]{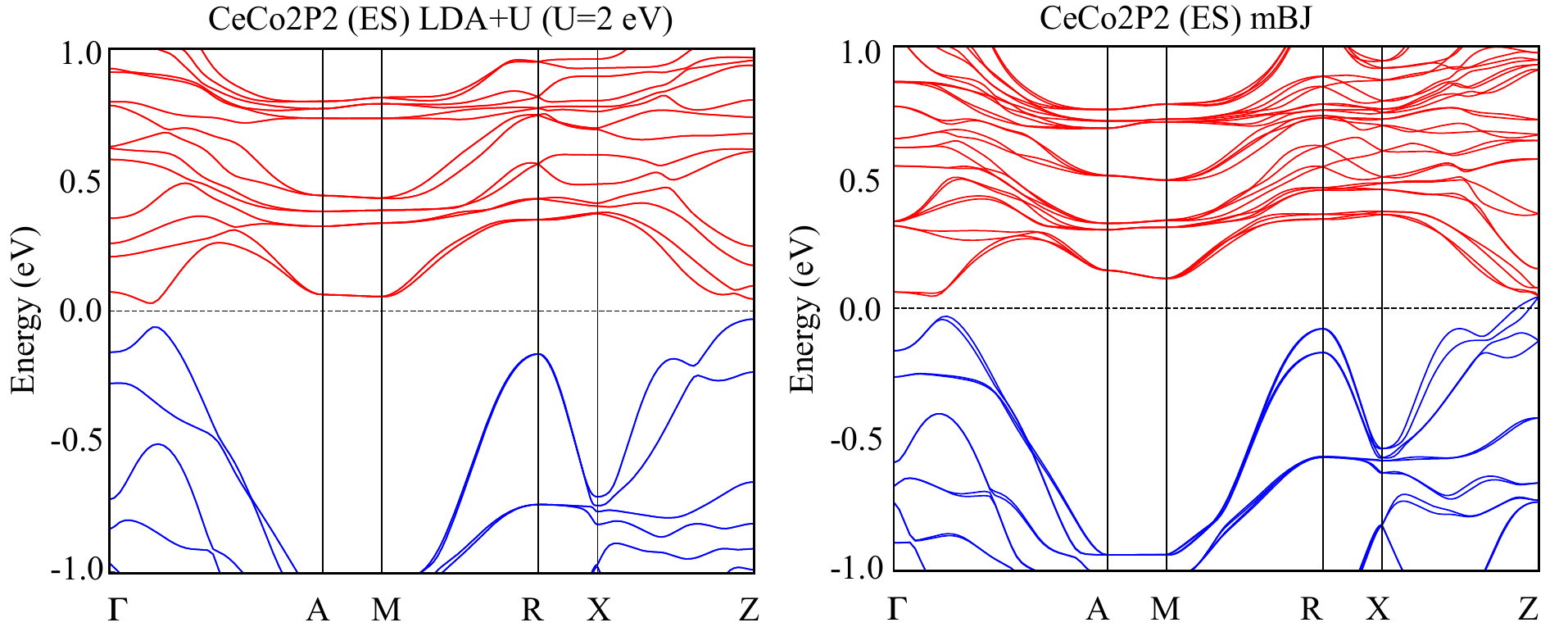} 
\caption{Band structures of the ESFD CeCo$_2$P$_2$ obtained from LDA+U ($U=2eV$)  and mBJ methods. }
\label{mbj-1.253} 
\end{figure} 

\begin{figure}[htbp] 
\centering\includegraphics[width=  4.2in]{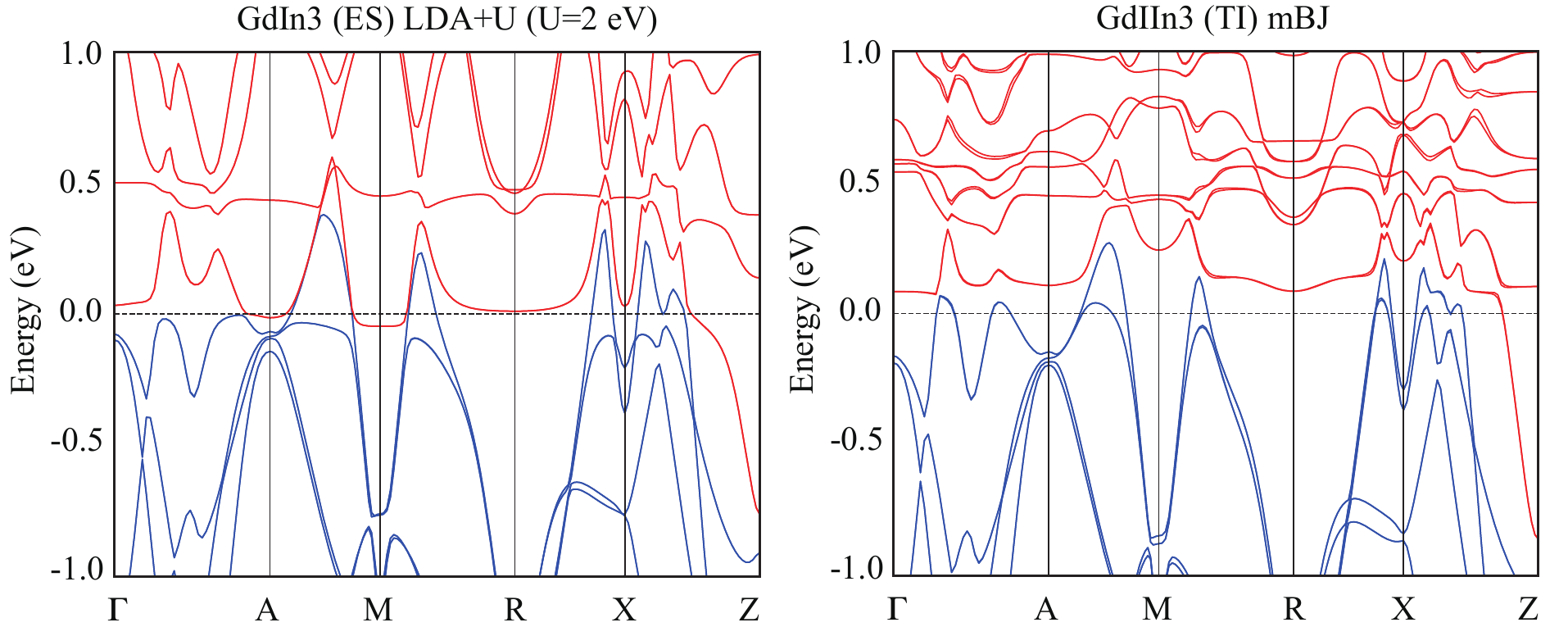} 
\caption{Band structures of the ES GdIn$_3$ obtained from LDA+U ($U=2eV$)  and mBJ methods. }
\label{mbj-1.253} 
\end{figure} 

\begin{figure}[htbp] 
\centering\includegraphics[width=  4.2in]{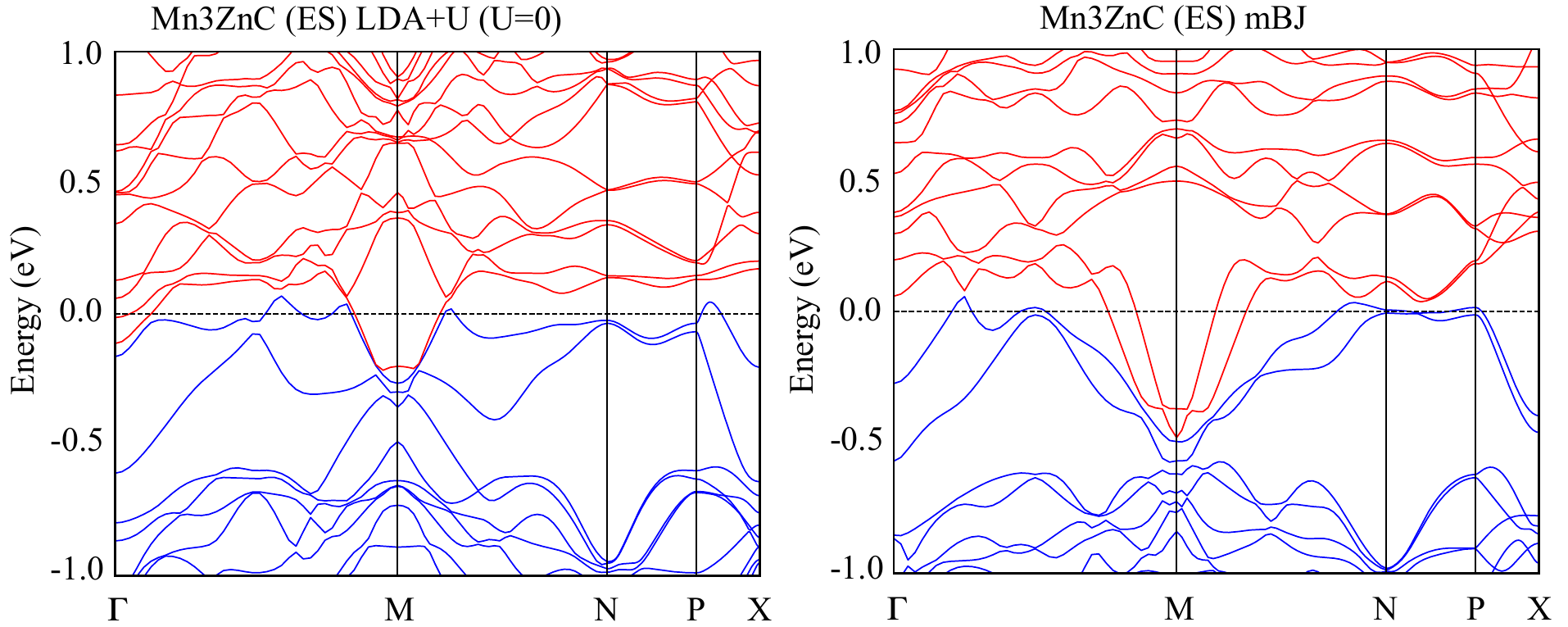} 
\caption{Band structures of the ES Mn$_3$ZnC obtained from LDA+U ($U=2eV$)  and mBJ methods. }
\label{mbj-2.19} 
\end{figure} 

\begin{figure}[htbp] 
\centering\includegraphics[width=  4.2in]{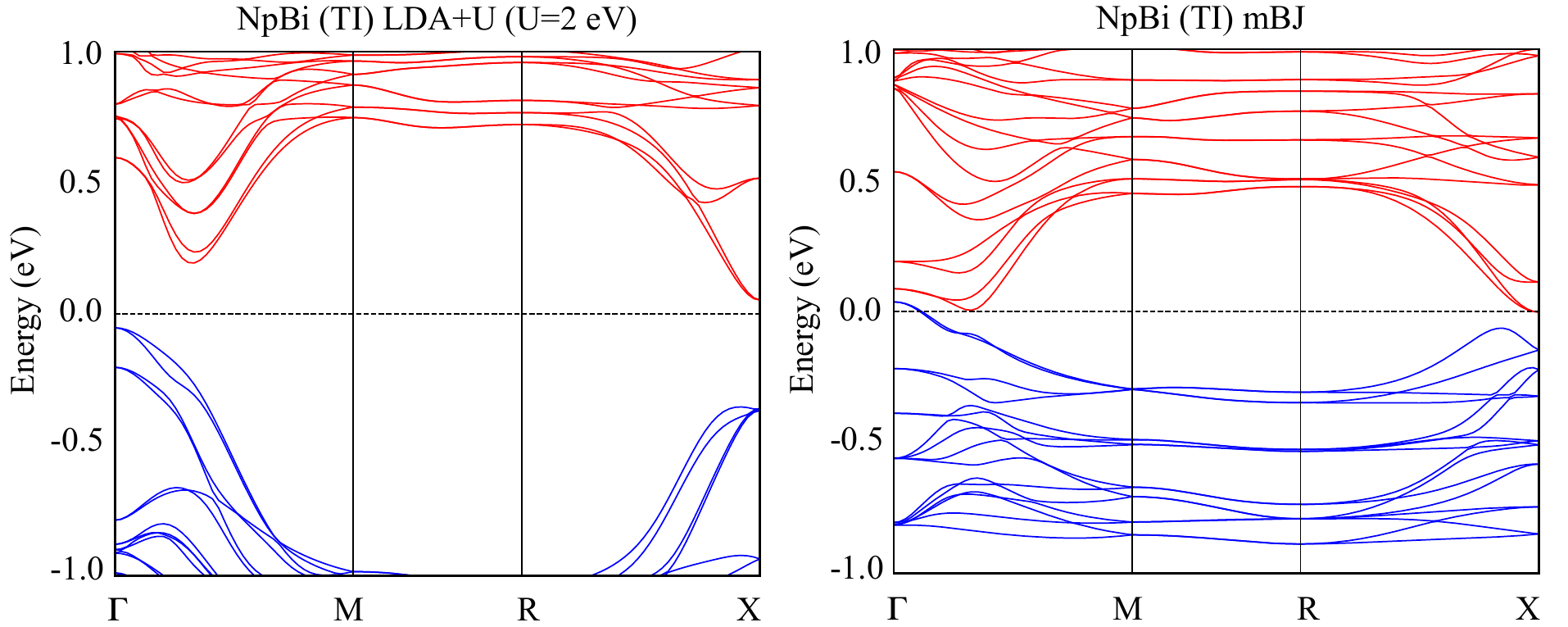} 
\caption{Band structures of the TI NpBi obtained from LDA+U ($U=2eV$)  and mBJ methods. }
\label{mbj-3.7} 
\end{figure} 

\begin{figure}[htbp] 
\centering\includegraphics[width=  4.2in]{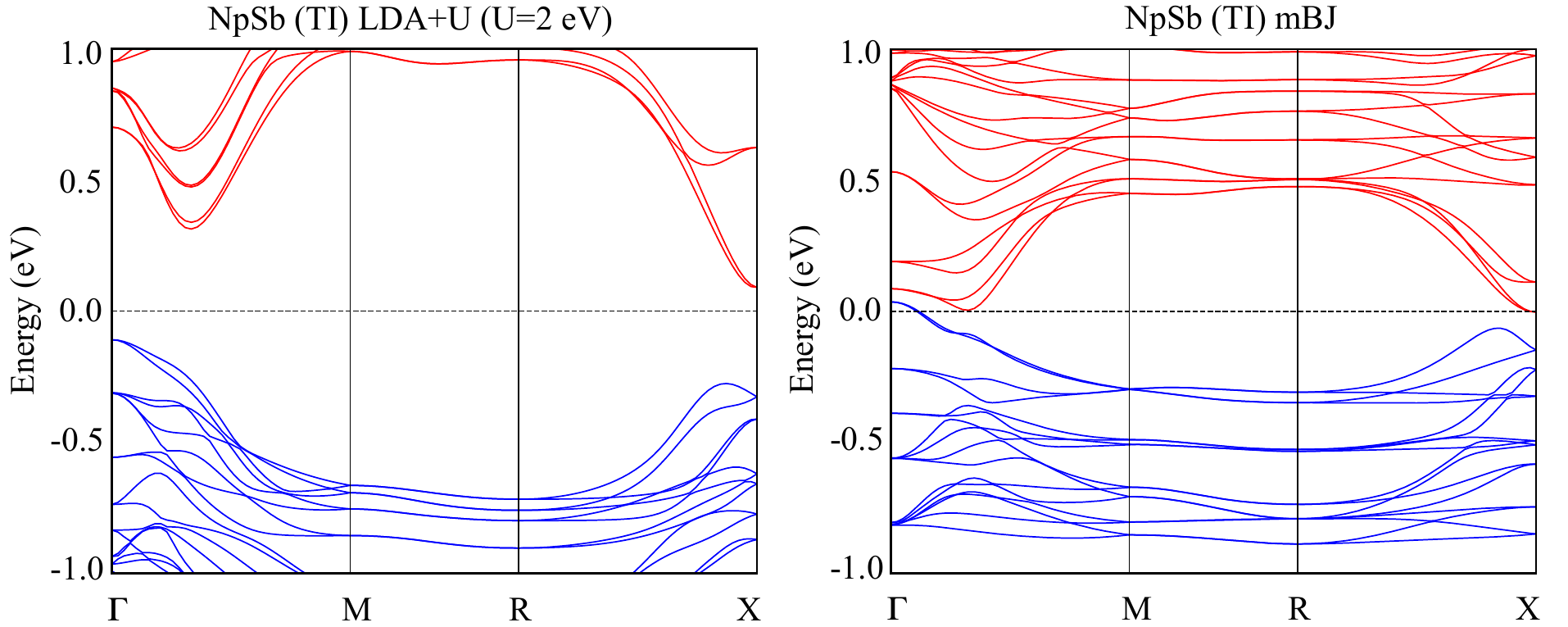} 
\caption{Band structures of the TI NpSb obtained from LDA+U ($U=2eV$)  and mBJ methods. }
\label{mbj-3.12} 
\end{figure} 
\clearpage
\begin{figure}[htbp] 
\centering\includegraphics[width=  4.2in]{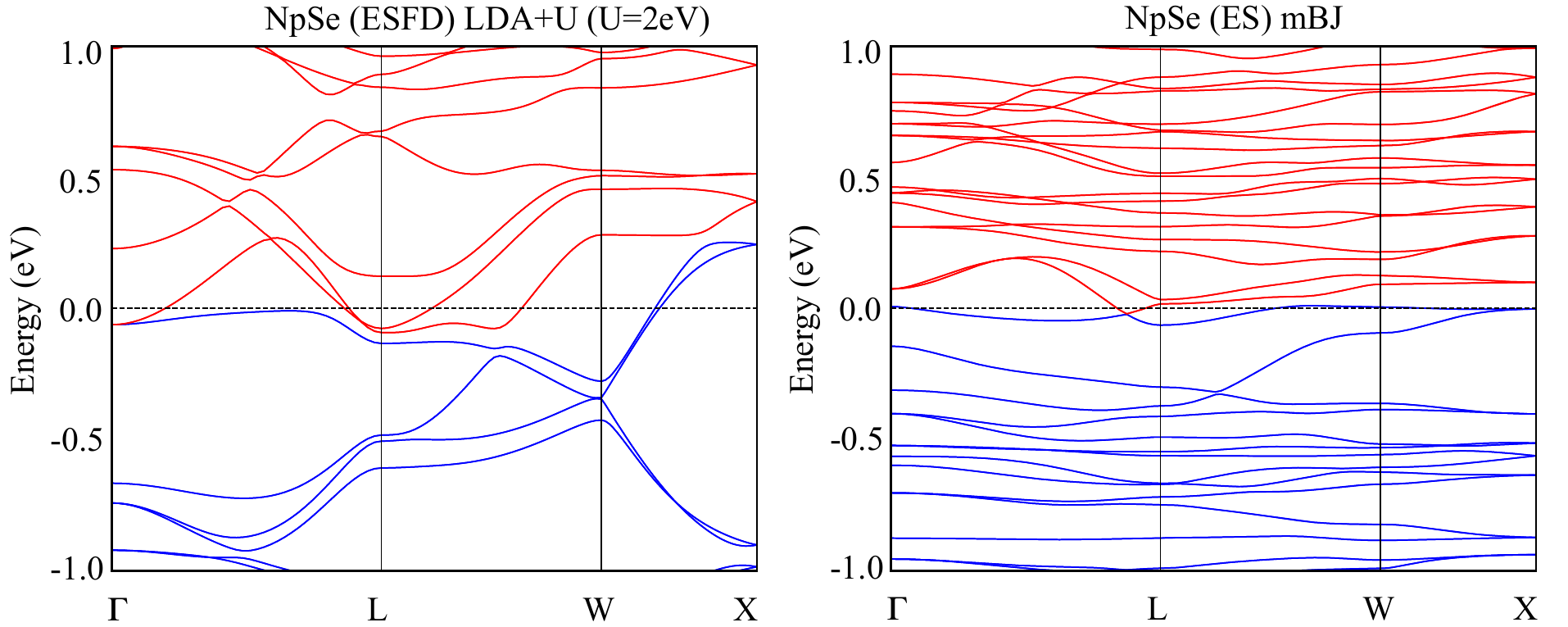} 
\caption{Band structures of NpSe obtained from LDA+U ($U=2eV$) with ESFD phase  and mBJ methods with ES phase. In LDA+U calculations, the irreps of the bands at $\Gamma$ point on the Fermi level is 2-fold degenerate and half-filled. So, it's in ESFD phase. While, in mBJ calculations, the valence band that crossing the Fermi level is 1-dimentional. It's in the ES phase with a symmetry protected crossing point on the $\Gamma L$ path.}
\label{mbj-3.10} 
\end{figure} 
%\clearpage
\section{Comparisons between LDA+U and LDA+Gutzwiller methods}\label{app:M}
To further check the robustness of the results obtained from LDA+U calculations, we have performed the LDA+Gutzwiller calculations \cite{PhysRevB.79.075114} for the two stable magnetic topological semimetals, MnGeO$_3$ and CeCo$_2$P$_2$. 

LDA+Gutzwiller is a many-body technics combined with DFT calculations,
which has been successfully applied to predict correlated topological materials \cite{PhysRevLett.110.096401,PhysRevLett.112.016403, PhysRevX.7.011027}. Similar to other post-LDA methods, i.e. LDA+U and LDA+DMFT, the total Hamiltonian adopt in LDA+Gutzwiller can be written as,

\begin{equation}
H_{total}=H_{LDA} + H_{int} + H_{dc}
\label{Ham}
\end{equation}
with $H_{LDA}$ being the non-interacting Hamiltonian obtained by LDA+SOC, the atomic spin orbital coupling and $H_{int}$ being the  interacting Hamiltonian. The last term in Eq.~(\ref{Ham}) is the double counting Hamiltonian, which needs to be included to remove the local interaction energy treated by LDA already in the mean field manner. In the present study, the Kanamori-type interaction and the fully localized limit scheme for the double counting energy~\cite{held2007} are adopted.

\begin{figure}[htbp] 
\centering\includegraphics[width=4.0in]{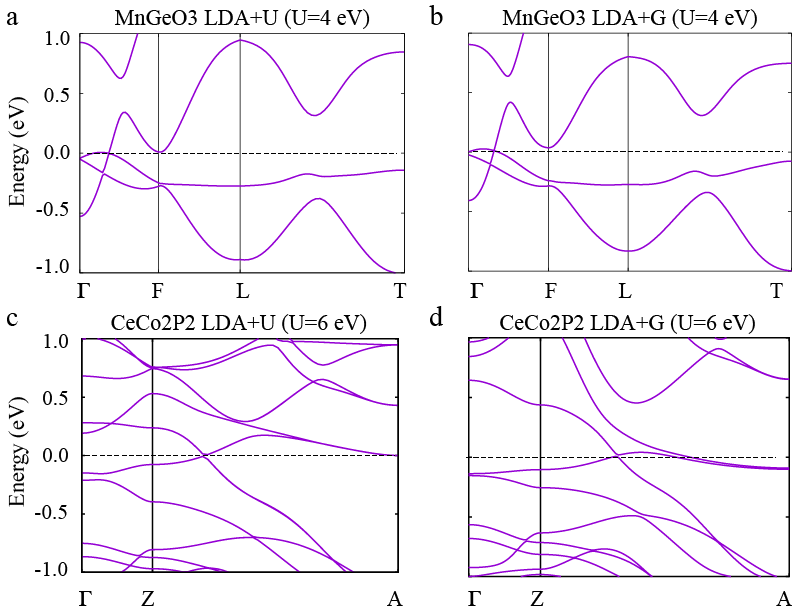} 
\caption{(a) Electronic band structures of MnGeO$_3$ obtained from LDA+U and (b) LDA+Gutzwiller (b) with the on-site Coulomb interaction $U=4 eV$ and the Hund's coupling $J=0.8 eV$. From LDA+Gutzwiller calculations, the quasi-particle weight of the $d$ electron on Mn is about 0.86 and the magnetic moments on Mn is about 4.8 $\mu_B$(c)-(d) Electronic band structures of CeCo$_2$P$_2$ obtained from LDA+U and LDA+Gutzwiller, respectively. The on-site interaction of $f$ orbitals is taken as $U=6 eV$. From LDA+Gutzwiller calculations, the quasi-particle weight of the $f$ electron on Ce is about 0.25. The magnetic moments on Ce and Co are 0.0 and 0.9 $\mu_B$, respectively.}
\label{ldag} 
\end{figure} 

In the LDA+Gutzwiller method, the Gutzwiller type wave function $\ket{\psi_{G}}=\hat{P}\ket{\psi_{0}}$ has been proposed for the trial wave function to minimize the ground state energy, where $\ket{\psi_{0}}$ is the non-interacting wave function and $\hat{P}$ is the local projector applied to adjust the probability of the local atomic configuration (in the many-particle Fock space). In addition, the Gutzwiller approximation is applied to evaluate the ground state energy and an effective Hamiltonian $H_{eff}\approx \hat{P}H_{LDA}\hat{P}$ describing the quasi-particle dispersion can be obtained. For detailed description for the method please refer to references~\cite{PhysRevB.79.075114,PhysRevB.85.035133,PhysRevB.83.245139}.

In the LDA+Gutzwiller calculation, we adopt the same parameter U as LDA+U calculation. For MnGeO$_3$, the quasi-particle weight of the $d$ electron on Mn is about 0.8 and the local magnetic moment on Mn is about 4.6 $\mu_B$/Mn which is consistent with LDA+U calculation (4.3 $\mu_B$/Mn). Compared with the band structure obtained from LDA+U calculations in Figure \ref{ldag}(a), the quasi-particle band structure in LDA+Gutzwiller are renormalized by a factor of 0.86, as shown in Figure \ref{ldag}(b).

For CeCo$_2$P$_2$, the quasi-particle weight of the $f$ electron on Ce is about 0.25 and the occupation of $f$ electron is about 1.0/Ce. Compared with the band structure obtained from LDA+U calculations in Figure \ref{ldag}(c), the quasi-particle band structure in LDA+Gutzwiller are strongly renormalized by a factor of 0.25, as shown in Figure \ref{ldag}(d). Although the large-size renormalization on $f$ orbitals changes the quasi-particle bands a lot, the symmetry enforced band crossing along $ZA$ path is stable.

Both comparisons for MnGeO$_3$ and CeCo$_2$P$_2$ indicate that strong correlations only renormalize the band width by a factor of quasi-particle weight but don't change the topologies for the stable topological materials MnGeO$_3$ and CeCo$_2$P$_2$. 

%\clearpage
\section{Topological phase diagrams of the topological materials that predicted by MTQC}\label{app:E}
We tabulate the topological categories at different $U$'s for all the magnetic materials.
In the Table~\ref{tabledtopo} and \ref{tableftopo}, each material list is represented by different colors based on their phase transition trends with increasing $U$ 
The tables contain the material identification number in BCSMD (BCSID), chemical formula (Formula), magnetic space group (MSG), the correlated atoms (CA) that exhibit added $U$ in the {\it ab initio} calculations, topological phases with different $U$ and the link to our plotted band structures (BS).

In Tables~\ref{tabledtopo} and \ref{tableftopo}, the interaction parameter $U$ is varied in the range of 0, 1, 2, 3, 4 eV for $d$ electron and 0, 2, 4, 6 eV for $f$ electron, respectively. 
Upon adding and increasing U, there are 49 of the 130 topological materials that have stable topology (remain the same topology for all $U$), and 49 materials have nontrivial topology for weak correlation, while becoming topological trivial with strong correlations. There are 20 materials whose topologies are sensitive to the interaction and have topological phase transitions between TI and ES in a small interaction range. There are only 5 materials that belong to trivial class in the weak correlated case and become topological nontrivial when the correlation is strong enough. The trend of 'Topological $\rightarrow$ Trivial' upon large U is clear in our data.

The \textcolor{green}{green} color stands for nontrivial topology stable in the whole range of $U$ considered in the calculations.
The \textcolor{blue}{blue} color stands for topology nontrivial for weak correlation effect, but trivial with strong correlation.
The \textcolor{yellow}{yellow} color stands for topology sensitive to correlations and for topological phase transitions between TI and ES in a small interaction range.
The \textcolor{red}{red} color stands for topology trivial in weak correlations, but nontrivial with increasing $U$. 
The \textcolor{black}{grey} color stands for the cases where the self consistent calculations are not converged 
and the topological phase diagrams are not completed. We have separated the material list into two parts: one for $d$ electron with Coulomb in the range of $0\sim4$ eV, another for $f$ electron with Coulomb in the range of $0\sim6$ eV. The materials are ordered by the MSG. We also tag the ESs/ESFDs with chiral MSGs, in which the crossing points carry nonzero chiral charges. 

We emphasis that if a symmetry-data-vector cannot be written as an integer combination of EBRs (LCEBR) and the compatibility relations are satisfied, it is diagnosed as a TI in Table~\ref{tabledtopo} and \ref{tableftopo}. This TI can be stable TI/SISM with stable topological index. In ~\ref{app:F}, we interpret all the TIs by their topological indices.
We also find that some of the ES phases can be changed to TI/SISM phase by symmetry breaking. For example, the ES phase ($U=3,4eV$) of Mn$_3$Sn (with BCSID-0.200) can be calssified as SISM phase (with indice $\eta_{4I}=3$) if its MSG 63.464 ($Cm\prime cm\prime$) is subducted to the minimal subgroup MSG 2.4 ($P\bar1$).

\LTcapwidth=1.0\textwidth
\renewcommand\arraystretch{1.2}
\begin{longtable*}{|c|c|c|c|c|c|c|c|c|p{0.1\columnwidth}<{\centering}|}
\caption{Topological phase diagram of the magnetic materials that have transition elements.. The interaction parameter $U$ of $d$ electrons on the correlated atoms have been set to 0, 1, 2, 3 and 4 eV. For the material EuFe$_2$As$_2$ (BCS-ID: 2.1), since the local magnetic moments on Eu are about 7.0 $\mu_B$, which is fully spin-polarized, we take the U of $f$ electron on Eu as 4 eV and the U of $d$ electron on Fe as 0, 1, 2, 3, 4 eV.  }\label{tabledtopo} \\ 
\hline
BCS-ID & Formula & MSG & CA & U=0 & U=1 & U=2 & U=3 & U=4 & BS \\ 
\hline
\hline
\rowcolor{Blue} 0.165 & SrMn(VO4)(OH) & 4.7($P2_1$)$^*$ & Mn,V & ES & LCEBR & LCEBR & LCEBR & LCEBR & Table [\ref{table0-165}] \\ 
\hline
\rowcolor{Blue} 1.264 & CoPS3 & 11.57($P_C2_1/m$) & Co & TI & LCEBR & LCEBR & LCEBR & LCEBR & Table [\ref{table1-264}] \\ 
\hline
\rowcolor{Green} 0.203 & Mn3Ge & 12.62($C2'/m'$) & Mn & TI & TI & TI & TI & TI & Table [\ref{table0-203}] \\ 
\hline
\rowcolor{Blue} 1.0.13 & FeI2 & 12.62($C2'/m'$) & Fe & TI & LCEBR & LCEBR & LCEBR & LCEBR & Table [\ref{table1-0-13}] \\ 
\hline
\rowcolor{Blue} 1.201 & Cr2ReO6 & 14.80($P_a2_1/c$) & Cr & TI & LCEBR & LCEBR & TBD & TBD & Table [\ref{table1-201}] \\ 
\hline
\rowcolor{Blue} 1.49 & Ag2NiO2 & 15.90($C_c2/c$) & Ni & TI & LCEBR & LCEBR & LCEBR & LCEBR & Table [\ref{table1-49}] \\ 
\hline
\rowcolor{Blue} 1.50 & AgNiO2 & 18.22($P_B2_12_12$)$^*$ & Ni & ES & ES & LCEBR & LCEBR & LCEBR & Table [\ref{table1-50}] \\ 
\hline
\rowcolor{Blue} 1.263 & Ca3Ru2O7 & 33.154($P_Cna2_1$) & Ru & ES & ES & LCEBR & LCEBR & LCEBR & Table [\ref{table1-263}] \\ 
\hline
\rowcolor{Blue} 0.85 & KCo4(PO4)3 & 58.398($Pnn'm'$) & Co & ES & LCEBR & LCEBR & LCEBR & LCEBR & Table [\ref{table0-85}] \\ 
\hline
\rowcolor{Green} 0.140 & LuFe4Ge2 & 58.399($Pn'n'm'$) & Fe & ESFD & ESFD & ESFD & ESFD & ESFD & Table [\ref{table0-140}] \\ 
\hline
\rowcolor{Green} 0.27 & YFe4Ge2 & 58.399($Pn'n'm'$) & Fe & ESFD & ESFD & ESFD & ESFD & ESFD & Table [\ref{table0-27}] \\ 
\hline
\rowcolor{Red} 1.252 & CaCo2P2 & 59.416($P_Immn$) & Co & LCEBR & TI & TI & TI & TI & Table [\ref{table1-252}] \\ 
\hline
\rowcolor{Blue} 1.88 & Mn5Si3 & 60.431($P_Cbcn$) & Mn & TI & ES & LCEBR & LCEBR & LCEBR & Table [\ref{table1-88}] \\ 
\hline
\rowcolor{Blue} 2.1 & EuFe2As2 & 61.439($P_Cbca$) & Eu,Fe & LCEBR & ES & TI & TI & LCEBR & Table [\ref{table2-1}] \\ 
\hline
\rowcolor{Blue} 0.218 & Co2SiO4 & 62.441($Pnma$) & Co & ES & LCEBR & LCEBR & LCEBR & LCEBR & Table [\ref{table0-218}] \\ 
\hline
\rowcolor{Blue} 0.219 & Co2SiO4 & 62.441($Pnma$) & Co & ES & ES & LCEBR & LCEBR & LCEBR & Table [\ref{table0-219}] \\ 
\hline
\rowcolor{Blue} 0.221 & Fe2SiO4 & 62.441($Pnma$) & Fe & ES & LCEBR & LCEBR & LCEBR & LCEBR & Table [\ref{table0-221}] \\ 
\hline
\rowcolor{Blue} 1.130 & Cr2As & 62.450($P_anma$) & Cr & LCEBR & TI & LCEBR & LCEBR & LCEBR & Table [\ref{table1-130}] \\ 
\hline
\rowcolor{Yellow} 1.131 & Fe2As & 62.450($P_anma$) & Fe & TI & TI & ES & TI & TI & Table [\ref{table1-131}] \\ 
\hline
\rowcolor{Green} 1.132 & Mn2As & 62.450($P_anma$) & Mn & ES & ES & ES & ES & ES & Table [\ref{table1-132}] \\ 
\hline
\rowcolor{Blue} 1.28 & CrN & 62.450($P_anma$) & Cr & TI & TI & LCEBR & LCEBR & LCEBR & Table [\ref{table1-28}] \\ 
\hline
\rowcolor{Green} 0.199 & Mn3Sn & 63.463($Cmc'm'$) & Mn & ES & ES & ES & ES & ES & Table [\ref{table0-199}] \\ 
\hline
\rowcolor{Green} 0.200 & Mn3Sn & 63.464($Cm'cm'$) & Mn & TI & TI & TI & ES & ES & Table [\ref{table0-200}] \\ 
\hline
\rowcolor{Blue} 1.16 & BaFe2As2 & 64.480($C_Amca$) & Fe & TI & TI & TI & TI & LCEBR & Table [\ref{table1-16}] \\ 
\hline
\rowcolor{Blue} 1.52 & CaFe2As2 & 64.480($C_Amca$) & Fe & TI & TI & TI & LCEBR & LCEBR & Table [\ref{table1-52}] \\ 
\hline
\rowcolor{Yellow} 2.15 & Mn3Ni20P6 & 65.486($Cmm'm'$) & Mn,Ni & TBD & ES & ES & TI & ES & Table [\ref{table2-15}] \\ 
\hline
\rowcolor{Blue} 0.4 & NiCr2O4 & 70.530($Fd'd'd$) & Ni,Cr & ES & LCEBR & LCEBR & LCEBR & LCEBR & Table [\ref{table0-4}] \\ 
\hline
\rowcolor{Blue} 1.125 & LaFeAsO & 73.553($I_cbca$) & Fe & TI & LCEBR & LCEBR & LCEBR & LCEBR & Table [\ref{table1-125}] \\ 
\hline
\rowcolor{Red} 1.176 & YbCo2Si2 & 73.553($I_cbca$) & Co & LCEBR & TI & TI & TI & TI & Table [\ref{table1-176}] \\ 
\hline
\rowcolor{Green} 2.5 & Mn3CuN & 85.59($P4/n$) & Mn & ES & ES & ES & ES & ES & Table [\ref{table2-5}] \\ 
\hline
\rowcolor{Blue} 0.64 & MnV2O4 & 88.81($I4_1/a$) & Mn,V & ES & LCEBR & LCEBR & LCEBR & LCEBR & Table [\ref{table0-64}] \\ 
\hline
\rowcolor{Blue} 1.85 & alpha-Mn & 114.282($P_I\bar 42_1c$) & Mn & ES & LCEBR & LCEBR & LCEBR & TBD & Table [\ref{table1-85}] \\ 
\hline
\rowcolor{Green} 1.143 & Mn3Pt & 132.456($P_c4_2/mcm$) & Mn & ESFD & ESFD & ESFD & ESFD & ESFD & Table [\ref{table1-143}] \\ 
\hline
\rowcolor{Green} 1.146 & LaCrAsO & 138.528($P_c4_2/ncm$) & Cr & TI & TI & TI & TI & TI & Table [\ref{table1-146}] \\ 
\hline
\rowcolor{Green} 0.212 & Sr2Mn3As2O2 & 139.536($I4'/m'm'm$) & Mn & ESFD & ESFD & ESFD & ESFD & ESFD & Table [\ref{table0-212}] \\ 
\hline
\rowcolor{Green} 2.19 & Mn3ZnC & 139.537($I4/mm'm'$) & Mn & ES & ES & ES & ES & ES & Table [\ref{table2-19}] \\ 
\hline
\rowcolor{Green} 0.125 & MnGeO3 & 148.19($R\bar 3'$) & Mn & ES & ES & ES & ES & ES & Table [\ref{table0-125}] \\ 
\hline
\rowcolor{Blue} 0.21 & PbNiO3 & 161.69($R3c$) & Ni & ESFD & ESFD & ESFD & LCEBR & LCEBR & Table [\ref{table0-21}] \\ 
\hline
\rowcolor{Blue} 1.0.5 & Sr3CoIrO6 & 165.95($P\bar 3c'1$) & Co & ES & LCEBR & LCEBR & LCEBR & LCEBR & Table [\ref{table1-0-5}] \\ 
\hline
\rowcolor{Green} 0.108 & Mn3Ir & 166.101($R\bar 3m'$) & Mn & ES & ES & ES & ES & ES & Table [\ref{table0-108}] \\ 
\hline
\rowcolor{Green} 0.109 & Mn3Pt & 166.101($R\bar 3m'$) & Mn & ES & ES & ES & ES & ES & Table [\ref{table0-109}] \\ 
\hline
\rowcolor{Green} 0.177 & Mn3GaN & 166.97($R\bar 3m$) & Mn & ES & ES & ES & ES & ESFD & Table [\ref{table0-177}] \\ 
\hline
\rowcolor{Blue} 1.153 & Mn3GaC & 167.108($R_I\bar 3c$) & Mn & ES & LCEBR & LCEBR & LCEBR & LCEBR & Table [\ref{table1-153}] \\ 
\hline
\rowcolor{Blue} 0.117 & LuFeO3 & 185.201($P6_3c'm'$) & Fe & TI & LCEBR & LCEBR & LCEBR & LCEBR & Table [\ref{table0-117}] \\ 
\hline
\rowcolor{Green} 1.110 & ScMn6Ge6 & 192.252($P_c6/mcc$) & Mn & ES & ES & ES & ES & ES & Table [\ref{table1-110}] \\ 
\hline
\rowcolor{Green} 1.225 & ScMn6Ge6 & 192.252($P_c6/mcc$) & Mn & ES & ES & ES & ES & ES & Table [\ref{table1-225}] \\ 
\hline
\rowcolor{Red} 0.118 & Ba5Co5ClO13 & 194.268($P6_3'/m'm'c$) & Co & LCEBR & LCEBR & LCEBR & LCEBR & ES & Table [\ref{table0-118}] \\ 
\hline
\rowcolor{Blue} 0.150 & NiS2 & 205.33($Pa\bar 3$) & Ni & TI & LCEBR & LCEBR & LCEBR & LCEBR & Table [\ref{table0-150}] \\ 
\hline
\rowcolor{Blue} 0.2 & Cd2Os2O7 & 227.131($Fd\bar 3m'$) & Os & ES & LCEBR & LCEBR & LCEBR & LCEBR & Table [\ref{table0-2}] \\ 
\hline
\end{longtable*}
\footnotesize{$^*$ Chiral MSG. The crossing points in the ES/ESFD phase with Chiral MSG must carry nonzero chiral charges.~\cite{cano2019multifold,chang2018topological}} \\

\begin{longtable*}{|c|c|c|c|c|c|c|c|c|}
\caption{Topological phase diagram of the magnetic materials that have Rare-earth elements. The interaction parameter $U$ of $f$ electrons on Rare-earth elements have been set to 0, 2, 4 and 6 eV. If the material also have transition elements, we  take $U$ of $d$ electron as 2 eV.}\label{tableftopo} \\ 
\hline
\centering BCS-ID & Formula & MSG & CA & U=0 & U=2 & U=4 & U=6 & BS \\ 
\hline
\rowcolor{Green} 1.206 & Dy2Fe2Si2C & 2.7($P_S\bar 1$) & Dy,Fe & TI & TI & TI & TI & Table. [\ref{table1-206}] \\ 
\hline
\rowcolor{Blue} 0.104 & ErVO3 & 11.54($P2_1'/m'$) & Er,V & LCEBR & TI & LCEBR & LCEBR & Table. [\ref{table0-104}] \\ 
\hline
\rowcolor{Blue} 0.106 & DyVO3 & 11.54($P2_1'/m'$) & Dy,V & TI & LCEBR & LCEBR & LCEBR & Table. [\ref{table0-106}] \\ 
\hline
\rowcolor{Black} 1.22 & DyCu2Si2 & 12.63($C_c2/m$) & Dy,Cu & TBD & TBD & LCEBR & ES & Table. [\ref{table1-22}] \\ 
\hline
\rowcolor{Green} 1.140 & PrMgPb & 13.73($P_A2/c$) & Pr & TI & TI & TI & TI & Table. [\ref{table1-140}] \\ 
\hline
\rowcolor{Blue} 1.141 & NdMgPb & 13.73($P_A2/c$) & Nd & TI & TI & LCEBR & LCEBR & Table. [\ref{table1-141}] \\ 
\hline
\rowcolor{Blue} 0.105 & ErVO3 & 14.75($P2_1/c$) & Er,V & ES & LCEBR & LCEBR & LCEBR & Table. [\ref{table0-105}] \\ 
\hline
\rowcolor{Green} 0.174 & Pr3Ru4Al12 & 15.89($C2'/c'$) & Pr,Ru & TI & TI & TI & TI & Table. [\ref{table0-174}] \\ 
\hline
\rowcolor{Yellow} 0.226 & NdCo2 & 15.89($C2'/c'$) & Nd,Co & ES & TI & TI & TI & Table. [\ref{table0-226}] \\ 
\hline
\rowcolor{Green} 2.10 & HoP & 15.89($C2'/c'$) & Ho & TI & TI & TI & TI & Table. [\ref{table2-10}] \\ 
\hline
\rowcolor{Blue} 1.211 & Dy2O2S & 15.90($C_c2/c$) & Dy & TI & LCEBR & LCEBR & LCEBR & Table. [\ref{table1-211}] \\ 
\hline
\rowcolor{Blue} 1.216 & Nd2BaNiO5 & 15.90($C_c2/c$) & Nd,Ni & TI & LCEBR & LCEBR & LCEBR & Table. [\ref{table1-216}] \\ 
\hline
\rowcolor{Blue} 1.217 & Tb2BaNiO5 & 15.90($C_c2/c$) & Tb,Ni & TI & LCEBR & LCEBR & LCEBR & Table. [\ref{table1-217}] \\ 
\hline
\rowcolor{Blue} 1.43 & PrNiO3 & 36.178($C_amc2_1$) & Pr,Ni & LCEBR & LCEBR & ES & LCEBR & Table. [\ref{table1-43}] \\ 
\hline
\rowcolor{Black} 0.26 & TmAgGe & 38.191($Am'm'2$) & Tm & ES & TBD & TBD & TBD & Table. [\ref{table0-26}] \\ 
\hline
\rowcolor{Green} 2.12 & TbMg & 49.270($Pc'cm'$) & Tb & ES & ES & ES & ES & Table. [\ref{table2-12}] \\ 
\hline
\rowcolor{Yellow} 1.139 & Ho2RhIn8 & 49.273($P_cccm$) & Ho & ES & TI & TI & ES & Table. [\ref{table1-139}] \\ 
\hline
\rowcolor{Green} 2.11 & TbMg & 51.295($Pmm'a'$) & Tb & ES & ES & ES & ES & Table. [\ref{table2-11}] \\ 
\hline
\rowcolor{Green} 1.222 & Er2CoGa8 & 51.298($P_amma$) & Er,Co & ES & ES & ES & ES & Table. [\ref{table1-222}] \\ 
\hline
\rowcolor{Black} 1.150 & PrAg & 53.334($P_Bmna$) & Pr & LCEBR & TBD & TI & TBD & Table. [\ref{table1-150}] \\ 
\hline
\rowcolor{Blue} 1.8 & CeRu2Al10 & 57.391($P_Cbcm$) & Ce,Ru & TI & TI & TI & LCEBR & Table. [\ref{table1-8}] \\ 
\hline
\rowcolor{Blue} 0.187 & CeMnAsO & 59.407($Pm'mn$) & Ce,Mn & ES & LCEBR & LCEBR & LCEBR & Table. [\ref{table0-187}] \\ 
\hline
\rowcolor{Green} 0.185 & Nd5Ge4 & 62.447($Pnm'a'$) & Nd & ES & ES & TBD & ES & Table. [\ref{table0-185}] \\ 
\hline
\rowcolor{Yellow} 1.179 & NdCoAsO & 62.450($P_anma$) & Nd,Co & ES & TI & ES & ES & Table. [\ref{table1-179}] \\ 
\hline
\rowcolor{Green} 0.149 & Nd3Ru4Al12 & 63.462($Cm'c'm$) & Nd,Ru & ES & ES & ES & ES & Table. [\ref{table0-149}] \\ 
\hline
\rowcolor{Green} 0.173 & Pr3Ru4Al12 & 63.462($Cm'c'm$) & Pr,Ru & ES & ES & ES & ES & Table. [\ref{table0-173}] \\ 
\hline
\rowcolor{Black} 3.3 & Ho2RhIn8 & 63.464($Cm'cm'$) & Ho & ES & ES & TBD & TBD & Table. [\ref{table3-3}] \\ 
\hline
\rowcolor{Red} 1.200 & U2Ni2Sn & 63.466($C_cmcm$) & U,Ni & LCEBR & TI & TI & TI & Table. [\ref{table1-200}] \\ 
\hline
\rowcolor{Yellow} 1.262 & NpRhGa5 & 63.466($C_cmcm$) & Np & LCEBR & TI & ES & LCEBR & Table. [\ref{table1-262}] \\ 
\hline
\rowcolor{Blue} 1.195 & Er2Ni2In & 63.467($C_amcm$) & Er,Ni & LCEBR & LCEBR & TI & LCEBR & Table. [\ref{table1-195}] \\ 
\hline
\rowcolor{Yellow} 1.188 & CeRh2Si2 & 64.480($C_Amca$) & Ce & TI & LCEBR & ES & TI & Table. [\ref{table1-188}] \\ 
\hline
\rowcolor{Yellow} 1.223 & Tm2CoGa8 & 65.489($C_ammm$) & Tm,Co & TI & ES & ES & ES & Table. [\ref{table1-223}] \\ 
\hline
\rowcolor{Black} 1.142 & CeMgPb & 67.510($C_Amma$) & Ce & LCEBR & TI & TBD & TBD & Table. [\ref{table1-142}] \\ 
\hline
\rowcolor{Green} 1.0.12 & UAu2Si2 & 71.536($Im'm'm$) & U & ES & ES & ES & ES & Table. [\ref{table1-0-12}] \\ 
\hline
\rowcolor{Green} 2.28 & NpNiGa5 & 74.559($Imm'a'$) & Np,Ni & ES & ES & ES & ES & Table. [\ref{table2-28}] \\ 
\hline
\rowcolor{Green} 0.184 & Nd5Si4 & 92.114($P4_12_1'2'$)$^*$ & Nd & ES & ES & ES & ES & Table. [\ref{table0-184}] \\ 
\hline
\rowcolor{Red} 1.0.11 & CeCoGe3 & 107.231($I4m'm'$) & Ce,Co & LCEBR & ES & ES & TBD & Table. [\ref{table1-0-11}] \\ 
\hline
\rowcolor{Green} 2.26 & PrCo2P2 & 123.345($P4/mm'm'$) & Pr,Co & ES & ES & ES & ES & Table. [\ref{table2-26}] \\ 
\hline
\rowcolor{Green} 1.162 & NdMg & 124.360($P_c4/mcc$) & Nd & ES & ES & ES & ES & Table. [\ref{table1-162}] \\ 
\hline
\rowcolor{Yellow} 1.251 & NdCo2P2 & 124.360($P_c4/mcc$) & Nd,Co & ES & ES & ES & TI & Table. [\ref{table1-251}] \\ 
\hline
\rowcolor{Yellow} 1.255 & UPtGa5 & 124.360($P_c4/mcc$) & U & TI & ES & ES & ES & Table. [\ref{table1-255}] \\ 
\hline
\rowcolor{Green} 1.261 & NpRhGa5 & 124.360($P_c4/mcc$) & Np & ES & ES & ES & ES & Table. [\ref{table1-261}] \\ 
\hline
\rowcolor{Green} 2.14 & NdMg & 125.373($P_C4/nbm$) & Nd & ES & ES & ES & ES & Table. [\ref{table2-14}] \\ 
\hline
\rowcolor{Green} 1.253 & CeCo2P2 & 126.386($P_I4/nnc$) & Ce,Co & ES & ES & ES & ES & Table. [\ref{table1-253}] \\ 
\hline
\rowcolor{Green} 0.80 & U2Pd2In & 127.394($P4'/m'bm'$) & U & ESFD & ESFD & ESFD & ESFD & Table. [\ref{table0-80}] \\ 
\hline
\rowcolor{Green} 0.81 & U2Pd2Sn & 127.394($P4'/m'bm'$) & U & TI & TI & TI & TI & Table. [\ref{table0-81}] \\ 
\hline
\rowcolor{Green} 1.81 & GdIn3 & 127.397($P_C4/mbm$) & Gd & ES & ES & ES & ES & Table. [\ref{table1-81}] \\ 
\hline
\rowcolor{Green} 1.102 & U2Ni2In & 128.408($P_c4/mnc$) & U,Ni & ES & ES & ES & ES & Table. [\ref{table1-102}] \\ 
\hline
\rowcolor{Green} 1.160 & UP & 128.410($P_I4/mnc$) & U & ES & ES & ES & ES & Table. [\ref{table1-160}] \\ 
\hline
\rowcolor{Green} 1.187 & TbRh2Si2 & 128.410($P_I4/mnc$) & Tb & ES & ES & ES & ES & Table. [\ref{table1-187}] \\ 
\hline
\rowcolor{Yellow} 1.208 & UAs & 128.410($P_I4/mnc$) & U & ES & ES & ES & TI & Table. [\ref{table1-208}] \\ 
\hline
\rowcolor{Green} 1.21 & DyCo2Si2 & 128.410($P_I4/mnc$) & Dy,Co & ES & ES & ES & ES & Table. [\ref{table1-21}] \\ 
\hline
\rowcolor{Red} 0.186 & CeMnAsO & 129.416($P4'/n'm'm$) & Ce,Mn & LCEBR & TI & TI & TI & Table. [\ref{table0-186}] \\ 
\hline
\rowcolor{Green} 1.215 & UP2 & 130.432($P_c4/ncc$) & U & ES & ES & ES & ES & Table. [\ref{table1-215}] \\ 
\hline
\rowcolor{Green} 2.13 & UP & 134.481($P_C4_2/nnm$) & U & TI & TI & TI & TI & Table. [\ref{table2-13}] \\ 
\hline
\rowcolor{Yellow} 2.20 & UAs & 134.481($P_C4_2/nnm$) & U & TI & ES & TI & TI & Table. [\ref{table2-20}] \\ 
\hline
\rowcolor{Blue} 2.6 & Nd2CuO4 & 134.481($P_C4_2/nnm$) & Nd & TI & LCEBR & LCEBR & LCEBR & Table. [\ref{table2-6}] \\ 
\hline
\rowcolor{Green} 1.103 & U2Rh2Sn & 135.492($P_c4_2/mbc$) & U & ES & ES & ES & ES & Table. [\ref{table1-103}] \\ 
\hline
\rowcolor{Green} 1.207 & U2Rh2Sn & 135.492($P_c4_2/mbc$) & U & ES & ES & ES & ES & Table. [\ref{table1-207}] \\ 
\hline
\rowcolor{Blue} 1.254 & UNiGa5 & 140.550($I_c4/mcm$) & U,Ni & ES & TI & TI & LCEBR & Table. [\ref{table1-254}] \\ 
\hline
\rowcolor{Yellow} 1.82 & Nd2RhIn8 & 140.550($I_c4/mcm$) & Nd & LCEBR & ES & ES & ES & Table. [\ref{table1-82}] \\ 
\hline
\rowcolor{Green} 1.87 & TbCo2Ga8 & 140.550($I_c4/mcm$) & Tb,Co & ES & ES & ES & ES & Table. [\ref{table1-87}] \\ 
\hline
\rowcolor{Yellow} 0.126 & NpCo2 & 141.556($I4_1'/a'm'd$) & Np,Co & ES & TBD & TI & TI & Table. [\ref{table0-126}] \\ 
\hline
\rowcolor{Blue} 0.151 & Tm2Mn2O7 & 141.557($I4_1/am'd'$) & Tm,Mn & ES & LCEBR & LCEBR & LCEBR & Table. [\ref{table0-151}] \\ 
\hline
\rowcolor{Green} 0.227 & NdCo2 & 141.557($I4_1/am'd'$) & Nd,Co & ES & ES & ES & ES & Table. [\ref{table0-227}] \\ 
\hline
\rowcolor{Blue} 0.48 & Tb2Sn2O7 & 141.557($I4_1/am'd'$) & Tb & ES & LCEBR & LCEBR & LCEBR & Table. [\ref{table0-48}] \\ 
\hline
\rowcolor{Blue} 0.49 & Ho2Ru2O7 & 141.557($I4_1/am'd'$) & Ho,Ru & ES & LCEBR & LCEBR & ES & Table. [\ref{table0-49}] \\ 
\hline
\rowcolor{Blue} 0.51 & Ho2Ru2O7 & 141.557($I4_1/am'd'$) & Ho,Ru & TI & LCEBR & LCEBR & LCEBR & Table. [\ref{table0-51}] \\ 
\hline
\rowcolor{Blue} 1.161 & PrFe3(BO3)4 & 155.48($R_I32$)$^*$ & Pr,Fe & LCEBR & TBD & ES & LCEBR & Table. [\ref{table1-161}] \\ 
\hline
\rowcolor{Blue} 0.169 & U3As4 & 161.71($R3c'$) & U & ES & ES & LCEBR & LCEBR & Table. [\ref{table0-169}] \\ 
\hline
\rowcolor{Blue} 0.170 & U3P4 & 161.71($R3c'$) & U & ES & ES & LCEBR & LCEBR & Table. [\ref{table0-170}] \\ 
\hline
\rowcolor{Yellow} 0.228 & TbCo2 & 166.101($R\bar 3m'$) & Tb,Co & TI & ES & TI & TI & Table. [\ref{table0-228}] \\ 
\hline
\rowcolor{Blue} 0.77 & Tb2Ti2O7 & 166.101($R\bar 3m'$) & Tb,Ti & ES & LCEBR & LCEBR & LCEBR & Table. [\ref{table0-77}] \\ 
\hline
\rowcolor{Yellow} 3.8 & NdZn & 222.103($P_In\bar 3n$) & Nd & ESFD & ES & ES & ESFD & Table. [\ref{table3-8}] \\ 
\hline
\rowcolor{Green} 3.12 & NpSb & 224.113($Pn\bar 3m'$) & Np & TI & TI & TI & TI & Table. [\ref{table3-12}] \\ 
\hline
\rowcolor{Blue} 3.2 & UO2 & 224.113($Pn\bar 3m'$) & U & ES & LCEBR & LCEBR & LCEBR & Table. [\ref{table3-2}] \\ 
\hline
\rowcolor{Green} 3.7 & NpBi & 224.113($Pn\bar 3m'$) & Np & TI & TI & TI & TI & Table. [\ref{table3-7}] \\ 
\hline
\rowcolor{Yellow} 3.10 & NpSe & 228.139($F_Sd\bar 3c$) & Np & TI & ESFD & ESFD & ESFD & Table. [\ref{table3-10}] \\ 
\hline
\rowcolor{Yellow} 3.11 & NpTe & 228.139($F_Sd\bar 3c$) & Np & TI & ESFD & ESFD & ESFD & Table. [\ref{table3-11}] \\ 
\hline
\rowcolor{Yellow} 3.9 & NpS & 228.139($F_Sd\bar 3c$) & Np & ES & ESFD & ESFD & ESFD & Table. [\ref{table3-9}] \\ 
\hline
\rowcolor{Green} 3.6 & DyCu & 229.143($Im\bar 3m'$) & Dy & TBD & ES & ES & ES & Table. [\ref{table3-6}] \\ 
\hline
\hline
\end{longtable*}
\footnotesize{$^*$ Chiral MSG. The crossing points in the ES/ESFD phase with Chiral MSG must carry nonzero chiral charges.~\cite{cano2019multifold,chang2018topological}} \\

Some of the topological comounds in Table \ref{tableftopo} contain both $3d$ element and $4f/5f$ element, where we set the U value of $d$ electron as 2eV. For comparisons, we have also considered the empirical U values for the $3d$ electron and identified the topological phase diagram using MTQC theory. In the Table \ref{compare_u3d}, we chose 15 materials, each of which contains the $3d$ element V, Co, Ni/Mn. The empirical U values of V, Co, Ni and Mn are set as 3.25, 3.7, 6.2 and 3.9eV, respectively\cite{PhysRevB.73.195107}. As tabulated in Table \ref{compare_u3d}, the results indicate that interaction of $3d$ electron almost doesn't change the topological phase diagram for the compounds containing $4f/5f$ element. 
\LTcapwidth=1.0\textwidth
\begin{longtable*}{c|c|c|c|c|c|c|c}
\caption{Comparisions of the topological phase diagrams for some of the topological compounds containing both $3d$ element and $4f/5f$ element. For each material, the U value of $3d$ element ($U_{3d}$) is set as $2eV$ and an empirical value. The U value of $4f/5f$ element ($U_f$) is set as 2, 4 and 6eV. The cases that have different topologies when $U_{3d}$ is taken 2eV and an empirical value are marked by red color.}\label{compare_u3d} \\
 \hline
BCSID & Formula & MSG & CA & $U_{3d} (eV) $ & $U_f=2eV$ & $U_f=4eV$ & $U_f=6eV$ \\
\hline
\hline
  \multirow{2}{*}{2.26}  & \multirow{2}{*}{PrCo2P2 } &\multirow{2}{*}{ 123.345($P4/mm'm'$)    } & \multirow{2}{*}{Pr,Co } & 2 & ES & ES & ES  \\
  \cline{5-8}
  &&&&3.7&ES & ES & ES  \\
  \hline\hline
   \multirow{2}{*}{1.251} & \multirow{2}{*}{NdCo2P2 } &\multirow{2}{*}{ 124.360($P_c4/mcc$)    } & \multirow{2}{*}{Nd,Co } & 2 & ES & ES & \textcolor{red}{TI}  \\
  \cline{5-8}
   &&&&3.7&ES & ES & \textcolor{red}{ES}  \\
  \hline\hline
  \multirow{2}{*}{1.253} & \multirow{2}{*}{CeCo2P2 } &\multirow{2}{*}{ 126.386($P_I4/nnc$)    } & \multirow{2}{*}{Ce,Co}  & 2 & ES & ES & ES  \\
  \cline{5-8}
  &&&&3.7&ES & ES & ES  \\
  \hline\hline
   \multirow{2}{*}{1.102} & \multirow{2}{*}{U2Ni2In } &\multirow{2}{*}{ 128.408($P_c4/mnc$)    } & \multirow{2}{*}{U,Ni }  & 2 & ES & ES & ES  \\
  \cline{5-8}
   &&&&6.2&ES & ES & ES  \\
  \hline\hline
  \multirow{2}{*}{1.21 } & \multirow{2}{*}{DyCo2Si2} &\multirow{2}{*}{ 128.410($P_I4/mnc$)    } & \multirow{2}{*}{Dy,Co}  & 2 & ES & ES & ES  \\
  \cline{5-8}
  &&&&3.7&ES & ES & ES  \\
  \hline\hline
   \multirow{2}{*}{0.105} & \multirow{2}{*}{ErVO3   } &\multirow{2}{*}{ 14.75($P2_1/c$)        } & \multirow{2}{*}{Er,V }  & 2 & LCEBR & LCEBR & LCEBR \\
  \cline{5-8}
   &&&&3.25&LCEBR & LCEBR & LCEBR \\
  \hline\hline
  \multirow{2}{*}{1.254} & \multirow{2}{*}{UNiGa5  } &\multirow{2}{*}{ 140.550($I_c4/mcm$)    } & \multirow{2}{*}{U,Ni }  & 2 & TI & TI & \textcolor{red}{LCEBR} \\
  \cline{5-8}
  &&&&6.2&TI & TI & \textcolor{red}{TI} \\
  \hline\hline
   \multirow{2}{*}{0.151 } & \multirow{2}{*}{Tm2Mn2O7} &\multirow{2}{*}{ 141.557($I4_1/am'd'$)  } & \multirow{2}{*}{Tm,Mn} & 2 & LCEBR & LCEBR & LCEBR \\
  \cline{5-8}
   &&&&3.9&LCEBR & LCEBR & LCEBR \\
  \hline\hline
  \multirow{2}{*}{0.227} & \multirow{2}{*}{NdCo2   } &\multirow{2}{*}{ 141.557($I4_1/am'd'$)  } & \multirow{2}{*}{Nd,Co}  & 2 & ES & ES & ES \\
  \cline{5-8}
  &&&&3.7&ES & ES & ES  \\
  \hline\hline
   \multirow{2}{*}{1.222} & \multirow{2}{*}{Er2CoGa8} &\multirow{2}{*}{ 51.298($P_amma$)       } & \multirow{2}{*}{Er,Co}  & 2 & ES & ES & ES  \\
  \cline{5-8}
   &&&&3.7&ES & ES & ES  \\
  \hline\hline
  \multirow{2}{*}{0.187} & \multirow{2}{*}{CeMnAsO } &\multirow{2}{*}{ 59.407($Pm'mn$)        } & \multirow{2}{*}{Ce,Mn}  & 2 & LCEBR & LCEBR & LCEBR \\
  \cline{5-8}
  &&&&3.9&LCEBR & LCEBR & LCEBR \\
  \hline\hline
   \multirow{2}{*}{1.200} & \multirow{2}{*}{U2Ni2Sn } &\multirow{2}{*}{ 63.466($C_cmcm$)       } & \multirow{2}{*}{U,Ni }  & 2 & \textcolor{red}{TI} & TI & TI \\
  \cline{5-8}
   &&&&6.2&\textcolor{red}{LCEBR} & TI & TI \\
  \hline\hline
  \multirow{2}{*}{1.195} & \multirow{2}{*}{Er2Ni2In} &\multirow{2}{*}{ 63.467($C_amcm$)       } & \multirow{2}{*}{Er,Ni}  & 2 & LCEBR & \textcolor{red}{TI} & LCEBR \\
  \cline{5-8}
  &&&&6.2&LCEBR & \textcolor{red}{LCEBR} & LCEBR \\
  \hline\hline
   \multirow{2}{*}{1.223} & \multirow{2}{*}{Tm2CoGa8} &\multirow{2}{*}{ 65.489($C_ammm$)       } & \multirow{2}{*}{Tm,Co}  & 2 & ES & ES & ES \\
  \cline{5-8}
   &&&&3.7&ES & ES & ES \\
  \hline\hline
  \multirow{2}{*}{2.28 } & \multirow{2}{*}{NpNiGa5 } &\multirow{2}{*}{ 74.559($Imm'a'$)       }& \multirow{2}{*}{Np,Ni }  & 2 & ES & ES & ES \\
  \cline{5-8}
  &&&&6.2&ES & ES & ES \\
  \hline\hline
\end{longtable*}

\section{Physical interpretations for the TI classified by MTQC}\label{app:F}
In MTQC theory~\cite{MTQC}, in order to have a physical interpretation for the SI, we  reduce the SI of MSG to one of its subgroup, and then interpret the SI of the subgroup using the layer construction approach \cite{song_quantitative_2018}. 
In this work, all the involved SI are subduced onto MSG 2.4 ($P \bar1$), MSG 47.24 ($Pmmm$), MSG 81.33 ($P\bar4$), MSG 83.43 ($P4/m$),/MSG 143.1 ($P3$). 
In the following, we provide the definitions of SI of these groups.

\subsection{Definitions for the stable indices of MSG 2.4}\label{app:F1}
\textbf{MSG 2.4 ($P\bar1$)} has the SI group $\ZZ_4\times \ZZ_2^3$.
The four SI $(\eta_{4I},z_{2I,1},z_{2I,2},z_{2I,3})$ are defined as
\begin{equation}
\eta_{4I} = \sum_K n_K^- = \sum_{K} \frac12(n_K^- - n_K^+) \mod4, \label{eq:eta4I}
\end{equation}
\begin{equation}
z_{2I,i=1,2,3} =C_{k_i=\pi}\mod2= \sum_{K, K_i=\pi} n_K^- \mod 2,\label{eq:z2I}
\end{equation}
where $K$ sums over the eight inversion-invariant momenta, and $n^{\pm}_K$ means the number of occupied even/odd states at the momentum $K$.
$z_{2I,i}$ is the parity of Chern number in the plane $k_i=\pi$.
$\eta_{4I}\mod 2$ is the parity of the Chern number difference between $k_z=0$ and $k_z=\pi$ planes.
Thus $\eta_{4I}=1,3$ correspond to Weyl semimetal (WSM) phase.
For $\eta_{4I}=2$ corresponds to an axion insulator phase/3D QAHI, which can not be distinguished from SI but from Wilson loop \cite{PhysRevB.84.075119}/surface state calculations. 

\paragraph{Layer constructions}
Now make use of the layer constructions of the gapped states to build a mapping from the SI to topological invariants.
Here we first give the SI of several layer constructions.
\begin{enumerate}
\item A Chern layer with $C=\pm1$ at $x=0$ gives the SI $(2100)$.
\item A Chern layer with $C=\pm1$ at $x=\frac12$ gives the SI $(0100)$.
\item A Chern layer with $C=\pm1$ at $y=0$ gives the SI $(2010)$.
\item A Chern layer with $C=\pm1$ at $y=\frac12$ gives the SI $(0010)$.
\item A Chern layer with $C=\pm1$ at $z=0$ gives the SI $(2001)$.
\item A Chern layer with $C=\pm1$ at $z=\frac12$ gives the SI $(0001)$.
\end{enumerate}

We take the Chern layers with $C=1$ at $z=0$ and $\frac12$ as two examples to show how to calculate the SI of layer-constructed states.
Since the SI indicate stable topological invariants (rather than fragile topological invariants),  states having the same stable topological invariants must have the same SI.
For layer-constructed states, the stable topological invariants are completely determined by the positions and directions of the layers \cite{song_quantitative_2018} and the Chern numbers of layers.
Other details about the layer constructions are irrelevant to determine the SI.
We first consider a single layer with $C=1$ at $z=0$. 
The $z=0$ plane is inversion-invariant, and there are four momenta in the 2D Brillouin zone, \ie $(k_x,k_y)=(0,0),\ (0,\pi),\ (\pi,0),\ (\pi,\pi)$. 
The Bloch states at these four momenta are either even/odd under the inversion.
According to the Fu-Kane formula 
\beq
(-1)^C = \prod_{K} \prod_{n\in\rm occ} \lambda_{n}^\pr(K), 
\eeq
where $K$ goes over the four inversion-invariant momenta and $\lambda^\pr_{K,n}$ is the parity of the $n$th occupied band at $K$, in a $C=1$ state, the total parity of the Bloch states at the four momenta must be odd.
Here we assume the Chern layer has a single occupied band and the parities at $(0,0),\ (0,\pi),\ (\pi,0),\ (\pi,\pi)$ are $-,+,+,+$, respectively.
Then we copy the layer to all the integer $z$ positions, \ie $z=0,\pm1,\pm2\cdots$, to construct the 3D state.
Supposing the 2D Bloch state of a single layer at $z$ is $\ket{\psi_{k_x,k_y,z}}$, then the 3D state Bloch state is given by
\beq
\ket{ \psi_{\kk} } = \frac1{\sqrt{N_z}}\sum_{z=0,\pm1\cdots} e^{i z k_z} \ket{ \psi_{k_x,k_y,z} },
\eeq
where $N_z$ is the period in the $z$-direction.
Let us calculate the inversion eigenvalues of the 3D state.
For $(k_x,k_y)=(0,0),\ (0,\pi),\ (\pi,0),\ (\pi,\pi)$, under the inversion operator $\hat{P}$, the 2D state at $z$ first gains an factor $\lambda^\pr(k_x,k_y)$ and then transforms to $-z$.
Thus we obtain
\beq
\hat{P} \ket{ \psi_{\kk} } = \lambda^\pr(k_x,k_y) \frac1{\sqrt{N_z}}\sum_{z=0,\pm1\cdots} e^{-i z k_z} \ket{ \psi_{k_x,k_y,z} }.
\eeq
For $k_z=0,\pi$, we further obtain
\beq
\hat{P} \ket{ \psi_{\kk} } = \lambda^\pr(k_x,k_y) \frac1{\sqrt{N_z}}\sum_{z=0,\pm1\cdots} e^{i z k_z} \ket{ \psi_{k_x,k_y,z} } =  \lambda^\pr(k_x,k_y) \ket{ \psi_{\kk} }.
\eeq
Thus the parities of the 3D state are given by $\lambda(k_x,k_y,k_z) = \lambda^\pr(k_x,k_y)$.
We obtain the parities of the 3D state at $(k_x,k_y,k_z)=(0,0,0),\ (\pi,0,0),\ (0,\pi,0),\ (\pi,\pi,0), 
(0,0,\pi),\ (\pi,0,\pi),\ (0,\pi,\pi),\ (\pi,\pi,\pi)$ as $-,+,+,+,-,+,+,+$, respectively.
Substituting the parities into \cref{eq:eta4I,eq:z2I}, we obtain the SI as (2001).

Then we consider the same 2D layers locating at half-integer positions, \ie $z=\pm\frac12,\pm\frac32\cdots$.
The 3D Bloch state can be written as 
\beq
\ket{ \psi_{\kk} } = \frac1{\sqrt{N_z}}\sum_{z=\pm\frac12,\pm\frac32\cdots} e^{i z k_z} \ket{ \psi_{k_x,k_y,z} }.
\eeq
For $(k_x,k_y)=(0,0),\ (0,\pi),\ (\pi,0),\ (\pi,\pi)$, we have the inversion action as
\beq
\hat{P} \ket{ \psi_{\kk} } = \lambda^\pr(k_x,k_y) \frac1{\sqrt{N_z}}\sum_{z=\pm\frac12,\pm\frac32\cdots} e^{-i z k_z} \ket{ \psi_{k_x,k_y,z} }.
\eeq
For $k_z=0,\pi$, we have $e^{-iz k_z} = e^{-2izk_z} \times e^{izk_z} = e^{ik_z} \times e^{izk_z}$ and hence
\beq
\hat{P} \ket{ \psi_{\kk} } = \lambda^\pr(k_x,k_y) e^{ik_z} \frac1{\sqrt{N_z}}\sum_{z=\pm\frac12,\pm\frac32\cdots} e^{i z k_z} \ket{ \psi_{k_x,k_y,z} } 
=  \lambda^\pr(k_x,k_y) e^{ik_z} \ket{ \psi_{\kk} }.
\eeq
Thus the parities of the 3D state are given by $\lambda(k_x,k_y,k_z) = \lambda^\pr(k_x,k_y) e^{ik_z} $.
We obtain the parities of the 3D state at $(k_x,k_y,k_z)=(0,0,0),\ (\pi,0,0),\ (0,\pi,0),\ (\pi,\pi,0), 
(0,0,\pi),\ (\pi,0,\pi),\ (0,\pi,\pi),\ (\pi,\pi,\pi)$ as $-,+,+,+,+,-,-,-$, respectively.
Substituting the parities into \cref{eq:eta4I,eq:z2I}, we obtain the SI as (0001).

The SI of other layer constructions can be similarly calculated.

\paragraph{The inversion $\ZZ_2$ invariant and axion insulator}
We find that $\eta_{4I}=2$ iff the origin point $(000)$ is occupied by odd Chern layers. 
Hence, for gapped state, we define the inversion-$\ZZ_2$ invariant as
\begin{equation}
\eta_{2I}^\pr = \frac12 \eta_{4I} \mod2.\label{eq:eta2Ip}
\end{equation}
When the total Chern number is zero, the axion $\theta$-angle is given by $\theta = \eta_{2I} \pi$.
For example, the state consists of $C=1$ layer at the $z=0$ plane and $C=-1$ Chern layer at the $z=\frac12$ plane, which has the SI (2000), has zero total Chern number and $\eta_{2I}^\pr=1$.
This state is axion insulator. 
One can see this from the boundary state: for a finite centrosymmetric sample centered at the origin, the Chern layer at $z=0$ contributes to a chiral hinge mode, whereas the chiral modes from all the other layers cancel each pairwise.
On the other hand, the state consists of $C=1$ Chern layers at both the $z=0,\frac12$ planes, which also has the SI (2000), has total Chern number 2 and also $\eta_{2I}^\pr=1$.
This state is a 3D QAH, for which the $\theta$-angle is ill-defined.
($\theta$ can still be constructed if the Chern numbers are non-zero, but will be origin-dependent \cite{varnava2019axion} and loses the physical meaning of magnetoelectric response.)

In the rest, we will directly give the SI and the corresponding interpretations.
Readers might refer to \cite{MTQC} for more details.

\subsection{Definitions for the stable indices of MSG 47.249}\label{app:F2}

\textbf{MSG 47.249 ($Pmmm$)} has the SI group $\ZZ_4\times \ZZ_2^3$.
Because of the anti-commuting mirror symmetries, all states at the inversion-invariant momenta are doubly degenerate.
The SI can be chosen as same as the MSG 47.250 $Pmmm1^\prime$ \cite{song_quantitative_2018} because for this group TRS does not change the irreps.
The $\ZZ_4$ factor is
\begin{equation}
z_4 = \sum_{K} \frac14 (n_K^- - n_K^+) \mod 4,\label{eq:z4}
\end{equation}
where $K$ sums over all inversion-invariant momenta and $n_K^\pm$ is the number of occupied even/odd states at $K$.
$z_4$ can be thought as \cref{eq:eta4I} except that $n^-_K$ are replaced by the number of odd doublets. 
Odd $z_4$ corresponds to axion insulator and $z_4=2$ corresponds to a higher-order TI (HOTI) jointly protected by mirrors and $C_2$ rotations \cite{MTQC}.
The three $\mathbb{Z}_2$ factors are the mirror Chern number parities in the $k_{1,2,3}=\pi$ planes
\begin{equation}
z_{2w,i=1,2,3} = C_{m,k_i=\pi} \mod2 = \sum_{K, K_i=\pi} \frac12 n_{K}^{-} \mod 2.\label{eq:z2w}
\end{equation}
Because of the anti-commuting mirror symmetries, net Chern numbers in all the directions vanishes.

\subsection{Definitions for the stable indices of MSG 81.33}\label{app:F3}
\textbf{MSG 81.33($P\bar4$)} has the SI group $\ZZ_4\times\ZZ_2^2$.
We choose the $\ZZ_4$ factor as the Chern number in the $k_z=\pi$ plane mod 4
\begin{equation}
z_{4S} = C_\pi \mod 4 = 2N_\mrm{occ} - \frac12 n^{\frac12}_Z + \frac12 n^{-\frac12}_Z
-\frac32 n^{\frac32}_Z + \frac32 n^{-\frac32}_Z
-\frac12 n^{\frac12}_A + \frac12 n^{-\frac12}_A
-\frac32 n^{\frac32}_A + \frac32 n^{-\frac32}_A
- n^{\frac12}_R + n^{-\frac12}_R \mod4,\label{eq:z4S}
\end{equation}
where $n^{\frac12,-\frac12,\frac32,-\frac32}_{Z,A}$ are the numbers of occupied states with $S_4$ eigenvalues $e^{-i\frac{\pi}4}$, $e^{i\frac{\pi}4}$, $e^{-i\frac{3\pi}4}$, $e^{i\frac{3\pi}4}$, and $n^{\frac12,-\frac12}_R$ are the numbers of occupied states with $C_2$ eigenvalues $e^{-i\frac{\pi}4}$, $e^{i\frac{\pi}4}$.
We choose the first $\ZZ_2$ factor as the difference of Chern numbers in the $k_z=\pi$ plane and the $k_z=0$ plane over 2 mod 2

\begin{equation}
\delta_{2S}=
- n^{\frac32}_Z +  n^{-\frac32}_Z
- n^{\frac32}_A +  n^{-\frac32}_A 
+ n^{\frac32}_\Gamma - n^{-\frac32}_\Gamma
+ n^{\frac32}_M - n^{-\frac32}_M \mod 2. \label{eq:d2S}
\end{equation}

The $\delta_{2S}=1$ phase is a WSM with 2 Weyl points between $k_z=0$ and $k_z=\pi$.
One may be curious why $C_\pi-C_0$ is an even number.
The answer is that it is enforced by the compatibility relations: the $k_z=0$ and $k_z=\pi$ planes must have the same $C_2$ eigenvalues and hence the same parity of Chern numbers.
The second $\ZZ_2$ factor is 
\begin{equation}
z_2 = \sum_{K=\Gamma, M, Z, A}\frac{n^{\frac12}_K-n^{-\frac32}_K}2  \mod2.\label{eq:z2}
\end{equation}
For $z_{4S}=0$ and $\delta_{2S}=0$, $z_2$ corresponds to axion insulator/3D QAHIstate \cite{MTQC}.

\subsection{Definitions for the stable indices of MSG 83.43}\label{app:F4}

\textbf{MSG 83.43 ($P4/m$)} has the SI group $\ZZ_4^3$.
We choose the three SI as
\begin{align}
\delta_{4m} = C_\pi^+-C_0^-\mod4 
&=\sum_{K=Z,A} \pare{-\frac12 n^{\frac12,+i}_K + \frac12 n^{-\frac12,+i}_K
-\frac32 n^{\frac32,+i}_K + \frac32 n^{-\frac32,+i}_K}
- n^{\frac12,+i}_R + n^{-\frac12,+i}_R \nono\\
&-\sum_{K=\Gamma,M} \pare{-\frac12 n^{\frac12,-i}_K + \frac12 n^{-\frac12,-i}_K
-\frac32 n^{\frac32,-i}_K + \frac32 n^{-\frac32,-i}_K}
+ n^{\frac12,-i}_X + n^{-\frac12,-i}_X \mod 4,\label{eq:d4m}
\end{align}
\begin{equation}
z_{4m,\pi}^+ = C_\pi^+\mod4
=N_{occ} + \sum_{K=Z,A} \pare{-\frac12 n^{\frac12,+i}_K + \frac12 n^{-\frac12,+i}_K
-\frac32 n^{\frac32,+i}_K + \frac32 n^{-\frac32,+i}_K}
- n^{\frac12,+i}_R + n^{-\frac12,+i}_R \mod 4,\label{eq:z4mp+}
\end{equation}
\begin{equation}
z_{4m,\pi}^- = C_\pi^-\mod4=
N_{occ} +\sum_{K=Z,A} \pare{-\frac12 n^{\frac12,-i}_K + \frac12 n^{-\frac12,-i}_K
-\frac32 n^{\frac32,-i}_K + \frac32 n^{-\frac32,-i}_K}
- n^{\frac12,-i}_R + n^{-\frac12,-i}_R \mod 4.\label{eq:z4mp-}
\end{equation}
Here $C^{\pm i}_{0,\pi}$ represents Chern number in the $\pm i$ mirror sector in the $k_z=0,\pi$ plane.

\subsection{Definitions for the stable indices of MSG 143.1}\label{app:F5}
\textbf{MSG 143.1 ($P3$)} has the SI group $\ZZ_3$.
According to the Chern number formula, the Chern number $C$ is related to the $C_3$ eigenvalues
\begin{equation}
e^{i\frac{2\pi}3 C} = (-1)^{N_\mrm{occ}} \prod_{n\in\mrm{occ}} \theta_n(\Gamma) \theta_n(K) \theta_n(KA),
\end{equation}
where $\theta_n(\Gamma,K,K^\pr)$ is the $C_3$ eigenvalue of the $n$th occupied state at the corresponding momentum.
(One should not confuse the $C_3$ eigenvalues $\theta_n$ with the axion theta angle $\theta$). 
We define the SI as,
\begin{equation}
z_{3R} = \sum_{K=A,H,HA} \pare{ 
 n^{-\frac12}_K - n^{\frac32}_K } \mod 3.\label{eq:z3R}
\end{equation}

\subsection{Stable indices of the magnetic TIs}\label{app:F6}
Using the SI defined above, we have explained the physical meaning for all of the TIs obtained from MTQC.
We tabulate all of the TIs diagnosed by MTQC in Table~\ref{tab:indice}. For each material, we list the identification number in BCSMD (BCSID), chemical formula (Formula), Coulomb interaction strength in LDA+U calculations (U),  magnetic space group (MSG), SI of the MSG calculated by the machinery of MTQC (Indices(MTQC)), whose physical meaning is unclear, the SI and the corresponding subgroup (Min-sMSG), and the physical interpretation of the Indices (Interpretations). 

\LTcapwidth=1.0\textwidth
\renewcommand\arraystretch{1.2}
\begin{longtable*}{c|c|c|c|p{0.14\columnwidth}|c|p{0.09\columnwidth}|c}
\caption{Topological indices and the physical interpretations of the TIs predicted by MTQC.} \label{tab:indice} \\
\hline
\hline
BCSID & Formula & U(eV) & MSG & Indices(MTQC) & Min-sMSG & Indices & Interpretations \\
\hline\hline
1.206 & Dy2Fe2Si2C & 0 & 2.7 ($P_S\bar 1$) & $c2=1$ & 2.4 ($P\bar 1$) & $z_{2I,1}=0$, $z_{2I,2}=0$, $z_{2I,3}=0$, $\eta_{4I}=2$ & AXI/3D QAHI\\ 
\hline
1.206 & Dy2Fe2Si2C & 2 & 2.7 ($P_S\bar 1$) & $c2=1$ & 2.4 ($P\bar 1$) & $z_{2I,1}=0$, $z_{2I,2}=0$, $z_{2I,3}=0$, $\eta_{4I}=2$ & AXI/3D QAHI\\ 
\hline
1.206 & Dy2Fe2Si2C & 4 & 2.7 ($P_S\bar 1$) & $c2=1$ & 2.4 ($P\bar 1$) & $z_{2I,1}=0$, $z_{2I,2}=0$, $z_{2I,3}=0$, $\eta_{4I}=2$ & AXI/3D QAHI\\ 
\hline
1.206 & Dy2Fe2Si2C & 6 & 2.7 ($P_S\bar 1$) & $c2=1$ & 2.4 ($P\bar 1$) & $z_{2I,1}=0$, $z_{2I,2}=0$, $z_{2I,3}=0$, $\eta_{4I}=2$ & AXI/3D QAHI\\ 
\hline
0.104 & ErVO3 & 2 & 11.54 ($P2_1'/m'$) & $c2_1=0$, $c2_2=0$, $c4=2$ & 2.4 ($P\bar 1$) & $z_{2I,1}=0$, $z_{2I,2}=0$, $z_{2I,3}=0$, $\eta_{4I}=2$ & AXI/3D QAHI\\ 
\hline
0.106 & DyVO3 & 0 & 11.54 ($P2_1'/m'$) & $c2_1=1$, $c2_2=1$, $c4=1$ & 2.4 ($P\bar 1$) & $z_{2I,1}=1$, $z_{2I,2}=0$, $z_{2I,3}=1$, $\eta_{4I}=1$ & SISM \\
\hline
\end{longtable*}
\footnotesize{$^*$ Compatible with SISM}\\

\section{Compatibility-relations along high-symmetry paths of the symmetry enforced semimetals}\label{app:G}
In TABLE~\ref{tab:compt_es}, we check all the compatibility relations between maximal $k$ vectors for the magnetic ESs in TABLE~\ref{tabledtopo}-\ref{tableftopo}. Once band structures break the compatibility relations between two maximal $k$ vectors, band crossings occur and form nodal points/nodal-lines.
\renewcommand\arraystretch{1.2}
\begin{longtable*}{c|c|c|c|c}
\caption{Compatibility-relations of the magnetic ESs. 
For each material, the first line lists BCSID, chemical formula, MSG and the value of U in LDA+U calculations.
The 1st-3rd coloumns are two maximal $k$ vectors ($k_1$ and $k_2$) 
and the intermediate path between them. 
The 4th coloumn identifies whether compatibility-relations 
between $k_1$ and $k_2$ are satisfied. 
If the answer is 'no', there have symmetry enforced band crossings between $k_1$ and $k_2$. 
The 5th coloumn gives the location of band crossings.
'Line' stands for the band crossings are protected by (screw-)rotational symmetries and there are isolated degenerate points 
on the line. 'Plane' stands for the band crossings are protected by (glide-)mirror symmetries and the crossing points form nodal-lines on the plane. The coordinates of $k$ vectors are written in the reciprocal conventional lattice, which are provided in the $CoRepresentation$ subsection of BCS website.}
\label{tab:compt_es}\\
\hline
\multicolumn{5}{c}{BCSID: 0.165;  Formula: SrMn(VO4)(OH);  MSG: 4.7 ($P2_1$);  U= 0} \\
\hline
maximal $k_1$ & intermediate path & maximal $k_2$ & satisfied? & Line/Plane \\
\hline
$\Gamma$:(0,0,0) & LD:(0,v,0) & Z:(0,1/2,0) & no & Line \\
\hline
A:(1/2,0,1/2) & U:(1/2,v,1/2) & E:(1/2,1/2,1/2) & yes & Line \\
\hline
B:(0,0,1/2) & V:(0,v,1/2) & D:(0,1/2,1/2) & yes & Line \\
\hline
C:(1/2,1/2,0) & W:(1/2,v,0) & Y:(1/2,0,0) & yes & Line \\
\hline
\multicolumn{5}{c}{} \\
\hline
\hline
\end{longtable*}

\section{Detailed discussion of the ideal magnetic TI and SMs  }\label{app:H}

\subsection{Higher-order topology of the ideal Axion insulator NpBi}\label{app:H1}

\begin{figure}[htbp] 
\centering\includegraphics[width=7.1in]{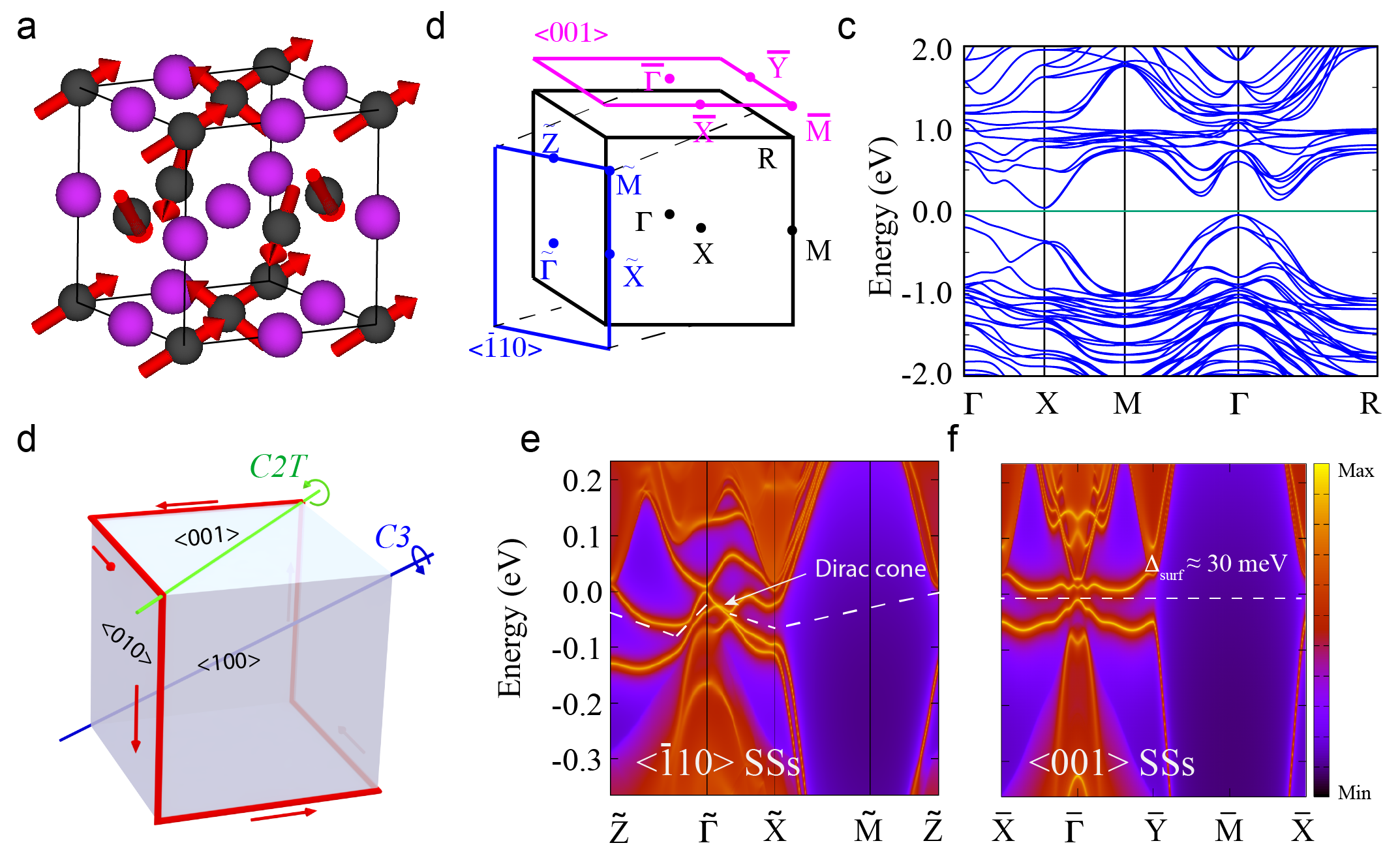} 
\caption{(a) The crystal and magnetic structures of NpBi with MSG 224.113($Pn\bar3m'$). (b) 3D bulk BZ and the projected 2D surface BZ on the $\braket{001}$(blue) and $\braket{\bar{1}10}$(pink) surfaces. 
(c) Electronic band structures of NpBi along the high symmetry paths in BZ with the interaction parameter $U=2eV$. (d) Schematic of the chiral flow of charge on the hinge of NpBi. The 1D chiral modes (red) on the six hinges are related to each other by $I$ and $C_{3xyz}$ symmetries. (e) Surface states calculation on the 
$C_{2\bar{x}y}\cdot T$ symmetric $\braket{\bar{1}10}$ surface plane. The Dirac cone surface states is along the path $\tilde{\Gamma}\tilde{X}$ constrained by the glide mirror symmetry $\tilde{M}_z=\{M_z|(0.5,0.5,0)\}$ and the dashed line indicates the "curved Fermi level", the band number below which equals that of electrons. (f) The surface states of $\braket{001}$  surface plane have a full gap in the whole BZ.   }\label{appfig1} 
\end{figure}

The crystal structure of NpBi adopts a face-centered cubic lattice with space group $Fm\bar{3}m$(No. 225) in the high temperature paramagnetic phase. 
Below 192.5 K, it undergoes a phase transition to the antiferromagnetic phase and the magnetic unit cell adopts simple cubic lattice with MSG 224.113($Pn\bar{3}m'$), as shown in FIG.~\ref{appfig1}(a). The generators of this MSG include two-fold screw rotation along the [100] direction $\tilde{C}_{2x}=\{C_{2x}|(0,0.5,0.5)\}$, three-fold rotation along the [111] direction $C_{3xyz}$, inversion $I$ and an anti-unitary symmetry $C_{2\bar xz}\cdot T$($C_{2\bar xz}$ is the two-fold rotation along the [$\bar1$10] direction and $T$ is the time reversal symmetry).
We use the experimental lattice constant $a=6.37$ {\AA} and the magnetic momentum of Np is initialized as $2.42 \mu_B$ along the (111) direction in our first-principle calculations.
In order to consider the interaction of the 5f electron on Np, we apply the LDA+U calculation with U=0,2,4,6 eV. We find that the band structures at the four Us have the same topology.  In the main text, we have calculated the parity-based stable topological indices and find $Z_4=2$, which corresponds to the axion insulator phase. Here we only chose $U=2 eV$ to analyze the material's topological surface states. 
It has been proved in Ref.~\cite{wieder2018axion} that the $C_2\cdot T$ symmetry protects the gapless surface states of axion insulators, on the plane perpendicular  to the two-fold rotation axis. Such an axion insulator also exhibits chiral hinge states~\cite{benalcazar_quantized_2017,benalcazar_electric_2017,schindler_higher-order_2018,schindler_higher-order_2018-1,song_d-2-dimensional_2017,langbehn_reflection-symmetric_2017} between the two gapped surfaces related by $C_2\cdot T$ symmetry.

There are three equivalent first Miller planes, $\braket{\bar{1}10}$, $\braket{0\bar{1}1}$ and $\braket{10\bar{1}}$ that preserve $C_2\cdot T$ symmetry. We perform the surface states calculations on the $\braket{\bar{1}10}$ plane where the $C_2\cdot T$ symmetry is preserved. As shown in FIG.~\ref{appfig1}(e), the surface Dirac cone emerges and strictly on the $\tilde{\Gamma}\tilde{Z}$ path constrained by the glide mirror symmetry $\tilde{M}_z=\{M_z|(0.5,0.5,0)\}$. For the $\braket{001}$ surface plane, it has no $C_2\cdot T$ symmetry and the surface states are fully gaped as shown in FIG.~\ref{appfig1}(f).  Considering the $I$ and $C_{3xyz}$ symmetries, the chiral flow of charge on the hinges of a cubic NpBi sample is schematically shown in FIG.~\ref{appfig1}(d).

\subsection{Topological phase diagram of the ideal antiferromagnetic nodal-line semimetal CeCo$_2$P$_2$}\label{app:H2}

\begin{figure}[htbp] 
\centering\includegraphics[width=7.1in]{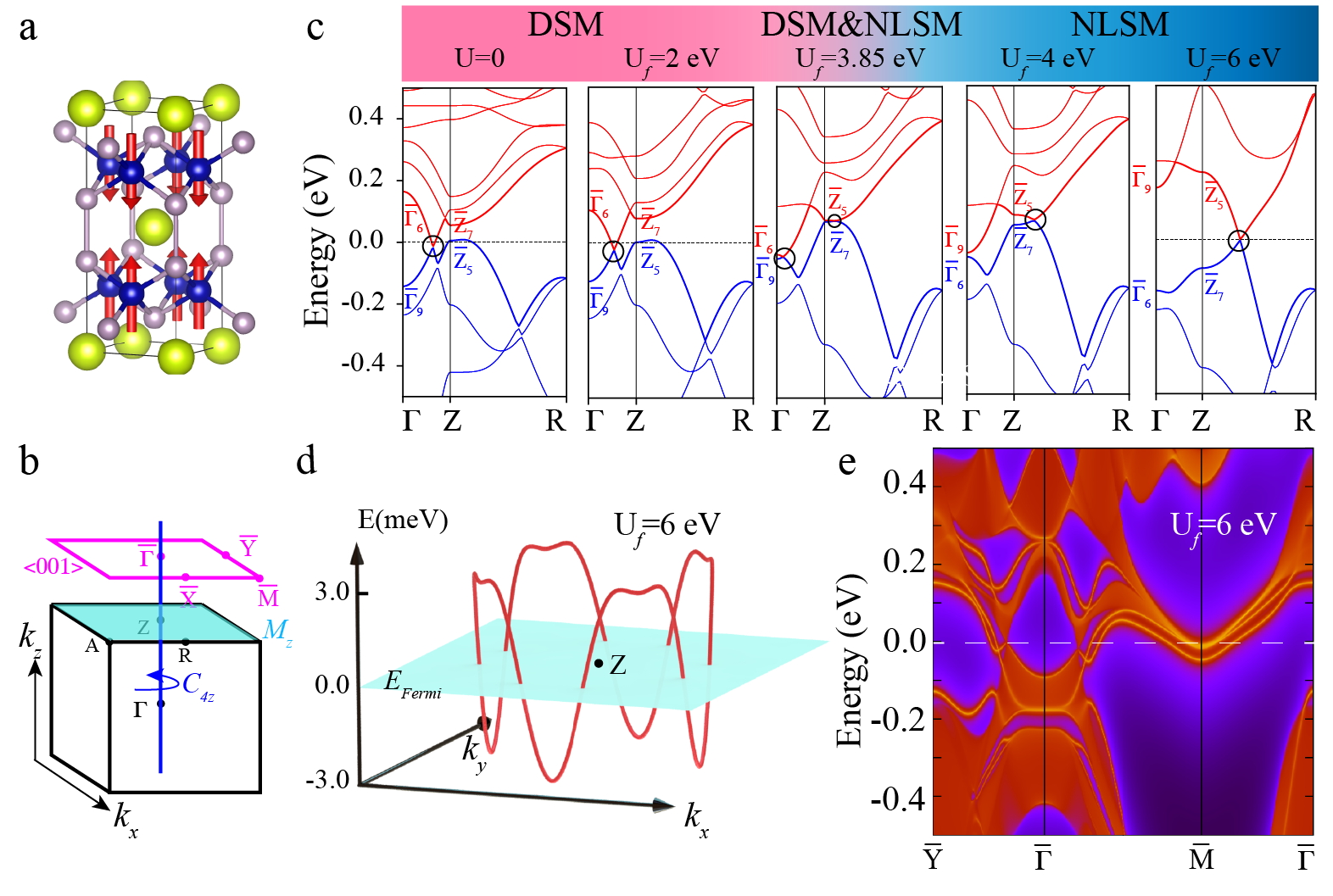} 
\caption{(a) The crystal and magnetic structures of the antiferromagnetic CeCo$_2$P$_2$ with MSG 126.386($P_I4/nnc$). (b) The 3D BZ with high symmetry momenta and the surface BZ projected on the $\braket{001}$ surface plane. The $C_{4z}$ rotation axis along $\Gamma Z$ and the $M_z$ plane $k_z=\pi$ are represented by the blue line and green plane, respectively. (c) The topological phase diagrams of CeCo$_2$P$_2$ with different Coulomb interaction strength applied. The irreps of the HOVB and LUCB are blue-colored and red-colored at both $\Gamma$ and $Z$ points. The topology with $U\le3$ eV belongs to DSM and is transformed to NLSM once $U\geq4$ eV. All of the nodal points along the high symmetry path are indicated by the black circles.
 (d) The energy dispersion of the nodal line on the $k_z=\pi$ plane, which is protected by $M_z$ symmetry. (e) The  drumhead-like surface states that connect the nodal line on the $\braket{001}$ surface. The Hubbard U of $f$ electron on Ce is equal to 6 eV in (d) and (e).}\label{appfig2} 
\end{figure}

 The space group(SG) of CeCo$_2$P$_2$ in the paramagnetic phase is $I4/mmm$. The antiferromagnetic phase transition occurs at 440 K below which the magnetic structure, as shown in FIG.~\ref{appfig2}(a), is characterized by the MSG 126.386($P_I4/nnc$). The generators of this MSG include inversion $I$, four-fold rotation $\tilde{C}_{4z}=\{C_{4z}|(0.5,0,0)\}$,  two-fold rotation $\tilde{C}_{2x}=\{C_{2x}|(0,0.5,0.5)\}$ and $\tilde{C}_{2xy}=\{C_{2xy}|(0,0,0.5)\}$, and the anti-unitary translation $T\tau(\tau=(0.5,0.5,0.5))$.
 
 Considering that the correlation effect of  the $f$ electron on Ce is very strong, we take the Coulomb interaction strength  of the $4f$ electron in the range $0\sim6$ eV. For convenience, we set U=2 eV for the $3d$ electron of the Co atom. As shown in the phase diagram Table \ref{tableftopo}, CeCo$_2$P$_2$ is ES for all U values. But we find that there is a topological phase transition between DSM and NLSM around $U=3.85$ eV. 

The band structures along $\Gamma-Z-R$ path have been plotted in the FIG.~\ref{appfig2}(c), with the Coulomb interaction $U=$ 0, 2, 3.5, 4 and 6 eV. When $U<3.85$ eV, both the band inversion between the highest occupied valence bands(HOVB) and the lowest unoccupied conduction bands(LUCB) occurs around $\Gamma$ point. The irreps of HOVB and LUCB are $\bar\Gamma_6$ and $\bar\Gamma_9$, which have different $C_{4z}$ eigenvalues (See the characters in Table~\ref{app:irreps}). 
Thus the band inversion between them creates two crossing points protected by $C_{4z}$ on the $k_x=k_y=0$ line.
As the Dirac points are symmetry protected
on the $k_x=k_y=0$ line with co-little group $4/mmm1^{\prime}$, CeCo$_2$P$_2$ is also
a higher-order topological SM.~\cite{wieder2019strong}
With $U>3.85$ eV, the band inversion around $\Gamma$ point is removed and the two Dirac points annihilate each other. However,  a new band inversion between HOVB and LUCB appears around $Z$ point.
As the irreps of the two bands at $Z$ point are $\bar Z_5$ and $\bar Z_7$, which have different $M_z$ eigenvalues, the band crossing between them form a nodal line on the $M_z$ invariant $k_z=\pi$ plane. 
We also find that for U equal to the critical point of 3.85 eV, band inversions occur both around $\Gamma$ and $Z$ points. So the Dirac points and nodal line coexist at the critical point $U=3.85$ eV.

When we  take the Coulomb interaction of $f$ electron on Ce to be 6 eV and $d$ electron interaction on Co to be 2 eV, the calculated magnetic moments on Co is 0.94$\mu_B$, which matches the experimental magnetic moments. Then the band structure in FIG.~\ref{appfig2}(d) has a clean Fermi surface and forms an ideal antiferromagnetic NLSM. The surface states on the $\braket{001}$ surface plane are calculated with the Wannier tight-binding Hamiltonian. As shown in the FIG.~\ref{appfig2}(b), there are two branches of surface states connecting to the nodal points near the Fermi level; These are distinguishable from the bulk states and can be observed by the ARPES/STM experiments. 

\begin{table}[t]
\begin{centering}
\begin{tabular}{c|c|c|c|c|c|c}
\hline \hline
$\Lambda_\alpha$ & MLCG & Irreps & $\{C^+_{4z}|(0.5,0,0)\}$ & $I$ & $\{M_z|(0.5,0.5,0)\}$ & $\{M_y|(0.5,0,0.5)\}$
\tabularnewline
\hline
\multirow{2}{*}{$\Gamma$} & \multirow{2}{*}{$4/mmm1'$} & $\bar\Gamma_6$(2) &  $-\sqrt{2}$ & 2 & 0 & 0\tabularnewline
 \cline{3-7}
& & $\bar\Gamma_9$(2) &  $\sqrt{2}$ & -2 & 0 & 0\tabularnewline
\hline
\multirow{2}{*}{Z} & \multirow{2}{*}{$4/mmm1'$} & $\bar Z_5$(2) &  $-\sqrt{2}$ & 0 & -2$i$ & 0 \tabularnewline
 \cline{3-7}
& & $\bar Z_7$(2) &  $-\sqrt{2}$ & 0 & 2$i$ & 0 \tabularnewline
\hline \hline
\end{tabular}
\end{centering}
\caption{Character table of the irreducible co-representations $\bar\Gamma_6$, $\bar\Gamma_9$, $\bar Z_5$ and $\bar Z_7$ of MSG 126.386($P_I4/nnc$) at $\Gamma$ and Z points, respectively. The first two columns are the momenta and their magnetic little co-group(MLCG); the third column is the irreps of the MLCG; the 4th-7th columns are the characters of the symmetry generators of the unitary subgroup. The $\bar\Gamma_6$ and $\bar\Gamma_9$ irreps have different eigenvalues of $C_{4z}$ and $I$, and $\bar Z_5$ and $\bar Z_7$ irreps have different eigenvalues of $M_{z}$.  }\label{app:irreps}
\end{table}

\subsection{Topological phase diagram of the antiferromagnetic Dirac semimetal MnGeO$_3$}\label{app:H3}

MnGeO$_3$ with ilmenite structure was reported to be an antiferromagnet with N\'{e}el temperature $T_N=120K$. The magnetic moment on the divalent cation Mn$^{2+}$ is about 5$\mu_B$, which is realized by the fully polarized high-spin configuration. It has a hexagonal lattice with SG $R\bar3$ and MSG 148.19($R\bar3'$) in the paramagnetic and antiferromagnetic phase, respectively. The MSG have two symmetry generators, three-fold rotation $C_{3z}$ and the combination of inversion and time reversal symmetries $P\cdot T$. 
The experimental lattice constants $a=5.012$ {\AA} and $b=14.2986$ {\AA} are used in the first principle calculations. 
Because of the absence of time reversal symmetry, a new situation not possible in the time reversal invariant space groups arises in the MSG 148.19($R\bar3'$). In this situation, all the k points on the $C_{3z}$ rotational axis are maximal k points. If the eigenvalues of $C_{3z}$ between any two points on the axis are changed, the system is predicted to be an ES.

\begin{figure}[htbp] 
\centering\includegraphics[width=7.1in]{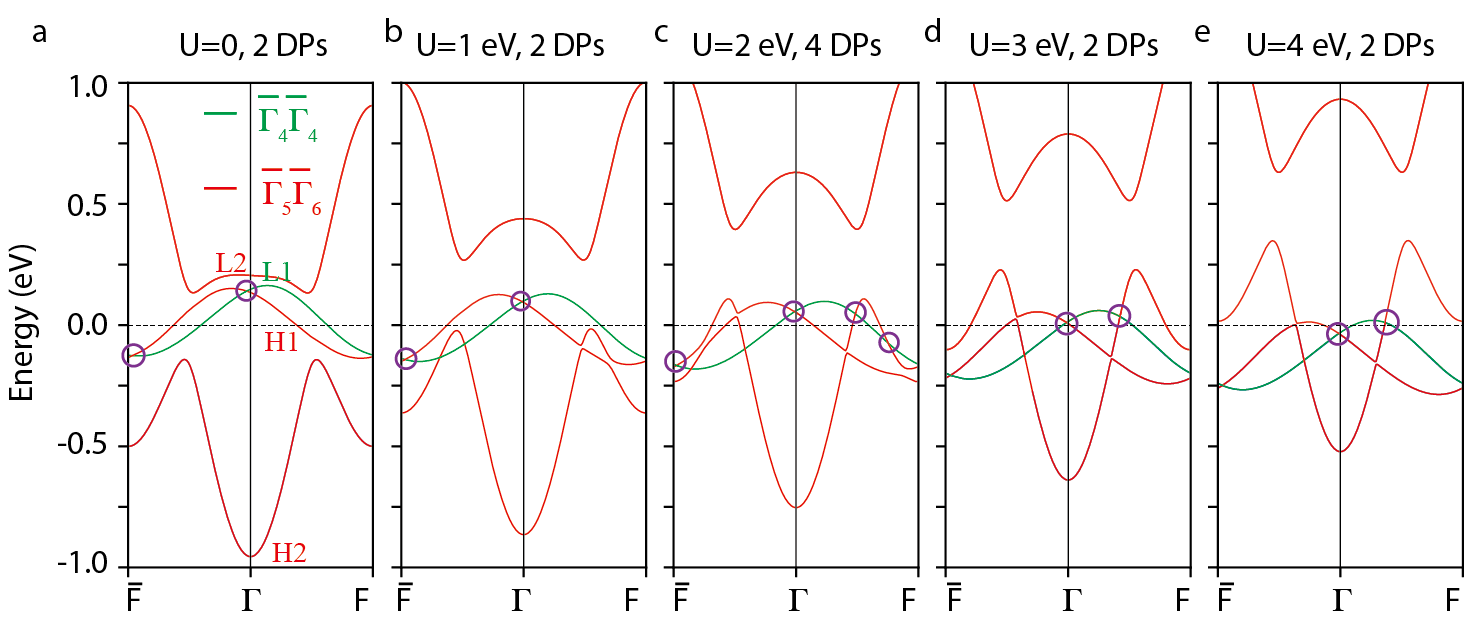} 
\caption{The band structures and topologies evolution of MnGeO$_3$ (MSG $R\bar3'$) with the Hubbard $U$ parameter increasing from 0 to 4 eV. 
The $(N_e-1)$th to $(N_e+2)$th bands are plotted, with $N_e$ being the number of electrons per unit cell.
All of the points on the $\bar{F}-\Gamma-F$ path respect the same LCG $\bar3'$, 
which have two different 2-dimensional irreps named as $\bar\Gamma_4\bar\Gamma_4$ and 
$\bar\Gamma_5\bar\Gamma_6$. The three lines H1, H2 and L2 have the same $\bar\Gamma_5\bar\Gamma_6$ representation and are red-colored; the line L1 is the $\bar\Gamma_4\bar\Gamma_4$ representation and is blue-colored. (a) For $U=0$, L1 and H1 have an anti-crossing along the $\bar F\Gamma$ path and generate two DPs. (b)-(e) With increasing the Coulomb interaction, the lines L2 and H2 move to higher energy and much more crossing generate between H2 and the branch of (L1, H1). The DPs formed by the HOVB and LUCB are indicated by the purple circles. 
  }\label{appfig3} 
\end{figure}

We take the Coulomb interaction strength of $3d$ electron of Mn from 0 to 4 eV and plot the band structures along the $C_{3z}$ symmetric path $\bar F-\Gamma-F$, as shown in the FIG.~\ref{appfig3}, where the green and red lines represent the bands that have different eigenvalues of $C_{3z}$ symmetry.
The little co-group(LCG) of the momenta along $\Gamma$F path is $\bar3'$, which have two irreducible co-representations $\bar\Gamma_5\bar\Gamma_6$ with $C_3=e^{\pm i\pi/3}$ doublet and $\bar\Gamma_4\bar\Gamma_4$ with $C_3=-1,-1$ doublet, where the two doublets are stabilized by the $P\cdot T$ symmetry.
Without taking into account the correlation effect, the bands H1 with irrep $\bar\Gamma_5\bar\Gamma_6$ and L1   with irrep $\bar\Gamma_4\bar\Gamma_4$ have an anti-crossing between the $\bar F$ and $\Gamma$ points, which  generate two Dirac points (DPs). 
Upon adding U, the second highest occupied band H2 moves up and there exist 2 pairs of DPs with $U=2$ eV, as shown in FIG.~\ref{appfig3}(c). With increasing $U$ to 3 and 4 eV, one out of the two pairs DPs annihilates, leaving one pair of DPs locating on the two sides of $\Gamma$ point.
In the FIG. 2(c)-(d) of the main text, we have calculated the surface states with $U=4$eV, on the $\braket{010}$ surface plane. The surface states connecting to the Dirac cone on $\Gamma$F path can be distinguished from the bulk states.

In symmorphic groups, Dirac nodes always appear in pairs. Since the irrep of one band is changed when it traverses though a symmetry protected DP, there must be even number of DPs for the irrep to go back to itself after a period of the BZ. It indicates that the crystalline Nielsen-Ninomiya theorem, which guarantees the doubling of the number of Dirac nodes in a crystal is still valid for the Dirac fermions in magnetic crystals.

\subsection{Weyl nodes, Nodal-lines and Anomalous Hall effect in Mn$_3$ZnC}\label{app:H4}

\begin{figure}[htbp] 
\centering\includegraphics[width=7.1in]{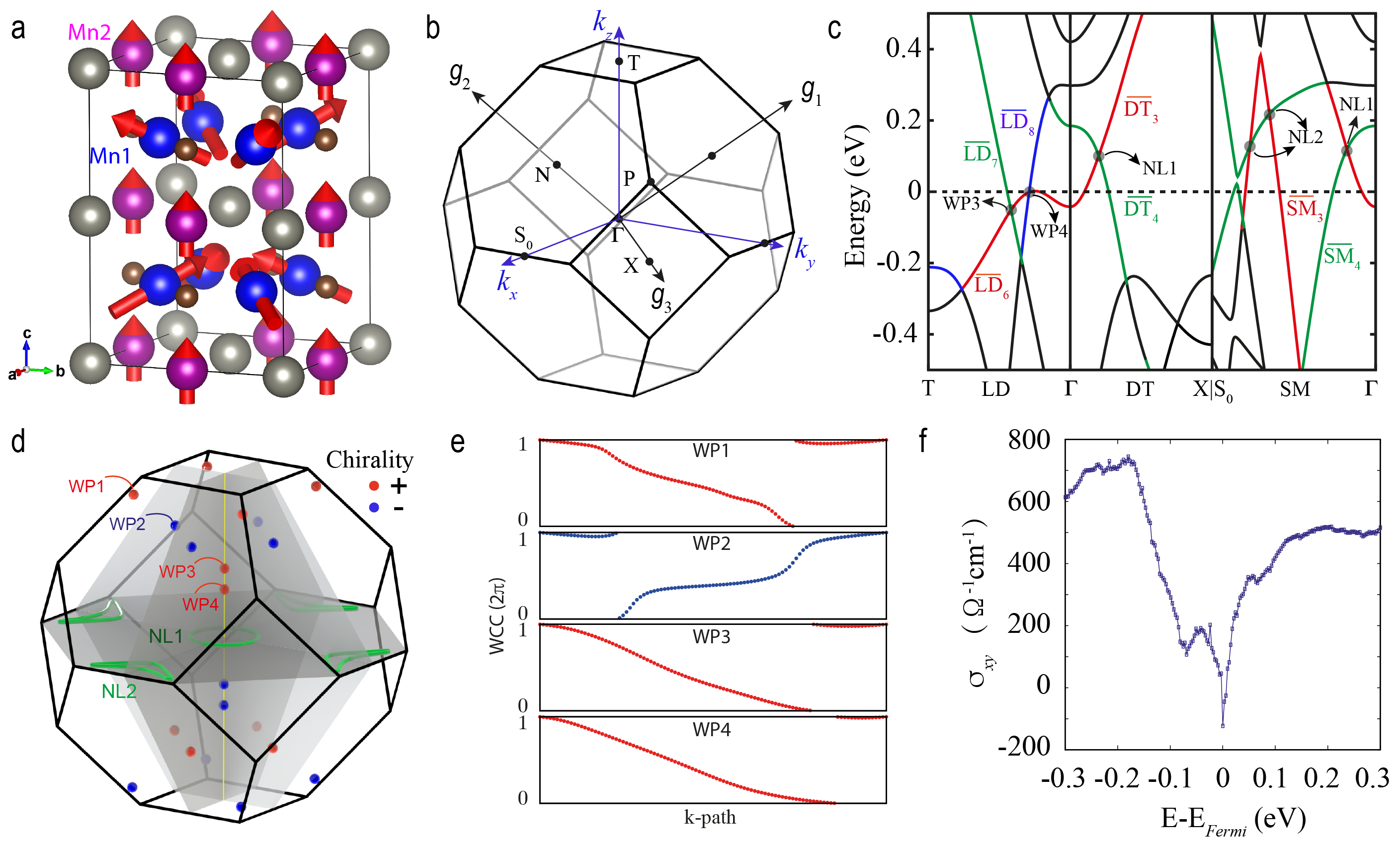} 
\caption{Weyl nodes and Nodal-lines in ferrimagnetic Mn$_3$ZnC (MSG $I_4/mm'm'$) with $U=4$eV. (a) The magnetic crystal structure and (b) the corresponding Brillouin zone (BZ) of non-collinear ferrimagnetic Mn$_3$ZnC. (c) Band structures along the high-symmetry paths $T-\Gamma-X$ and $S_0\Gamma$. The irreps of HOVB and LUCB are tagged by different colors.
Band crossings on $T\Gamma$ path form the two WPs, WP3 and WP4. 
Band crossings on the $k_z=0$ plane generate 5 NLs, which are protected by $M_z$ symmetry.
(d) Locations of WPs and NLs in the first BZ. The red(blue) balls stand for WPs with chirality +1(-1). (e) The evolution of Wannier charge centers (WCCs) on the manifold that enclosing WP1, WP2, WP3 and WP4, respectively. (f) Energy dependence of the intrinsic AHC in the non-collinear ferrimagnetic Mn$_3$ZnC. The Fermi energy is set to 0.
}\label{appfig4} 
\end{figure}

Mn$_3$ZnC adopts a cubic lattice with anti-perovskite structure in the paramagnetic phase and undergoes 
a phase transition to ferromagnetic phase with Currie temperature $T_c=470 K$. 
At 215 K, there is a second order phase transition from ferromagnetic to non-collinear ferrimagnetic phase accompanied 
with a structural transformation from cubic lattice to a body-centered tetragonal lattice, 
as shown in FIG.~\ref{appfig4}. In this article, we focus on the low-temperature non-collinear ferrimagnetic phase of Mn$_3$ZnC.

The MSG of ferrimagnetic Mn$_3$ZnC is MSG 139.537($I_4/mm'm'$), which is generated by the four-fold rotational 
symmetry along [001] direction $C_{4z}$, inversion symmetry $I$ and the anti-unitary symmetry $C_{2x}\cdot T$ ($C_{2x}$ 
is the two-fold rotational symmetry along [100] direction and $T$ is time reversal symmetry). 
There are two non-equivalent Mn atoms occupy the Wyckoff position 4$c$(0,1/2,0) and 8$f$(1/4,1/4,1/4), respectively. 
The Mn atoms on 4$c$ site form the non-collinear antiferromagnetic structure with ferromagnetic canting along [001] direction.
 While the Mn atoms on 8$f$ site are ferromagnetically polarized along [001] direction. 
In the {\it  ab initio} calculations, non-collinear ferrimagnetic Mn$_3$ZnC is diagnosed as ES for all $U$ values (i.e. $U=0,1,2,3,4$ eV for 3$d$ electron on Mn). Considering the strong correlations on 3$d$ electron, we take $U=4$eV 
and explain the band structure's topology in detail.

When $U=4$eV, compatibility-relations of the band representations are not satisfied along $\Gamma X$ and $\Gamma T$ paths. 
So there have symmetry enforced band crossings between the HOVB and LUCB along $\Gamma X$ and $\Gamma T$. In FIG.~\ref{appfig4}c, we plot the band structures along $T-\Gamma-X$ and $S_0\Gamma$ near the Fermi level. The irreps of HOVB and LUCB have been indicated by different colors. (See TABLE~\ref{app:irreps_mzc} for the characters of the irreps.)
Along $T\Gamma$ path, one can easily find two crossing points, WP3 and WP4, between the HOVB and LUCB, 
which have different $C_{4z}$ eigenvalues. So WP3 and WP4 are $C_{4z}$ protected Weyl points (WPs). 
On the $\Gamma X$ and $S_0\Gamma$ paths, there are many band crossings, which have different $M_z$ eigenvalues. 
So these crossing points on the $k_z=0$ plane form the Nodal-lines (NLs) protected by mirror symmetry, $M_z$. 
Using WannierTools package, we find 10 pairs of WPs and 5 NLs in the first BZ. The positions of them are plotted in the BZ, 
as shown in FIG.~\ref{appfig4}. The 10 pairs of WPs can be classified to four non-equivalent types, 
among which WP3 and WP4 are symmetry enforced WPs, WP1 and WP2 are generated by accidental band crossings. 
WPs in each type are related by $C_{4z}$ and $M_{z}$ symmetries. Using Wilson loop method \cite{PhysRevB.84.075119}, we have identified the 
chiralities of the WPs, as shown in FIG.~\ref{appfig4}e. We tabulated the exact positions and chiralities of them in TABLE~\ref{tab:wps}. 
The 5 NLs are classified to two types. The first type (NL1) is symmetry enforced and localized around the $\Gamma$ point. 
The other four NLs (NL2) are generated by accidental band crossings near $S_0\Gamma$ path and they are related by $C_{4z}$ symmetry.

In time reversal breaking (TRB) systems, nonzero Berry curvatures of occupied Bloch states contribute the intrinsic anomalous 
Hall conductivities (AHC). 
In non-collinear ferrimagnetic Mn$_3$ZnC, the Berry curvatures of bands near the WPs and NLs are very large, 
which will contribute giant AHC. We calculate the AHC of ferrimagnetic Mn$_3$ZnC by the sum of Berry curvatures over all occupied bands,
\begin{equation}
\sigma_{xy}=-\frac{2\pi e^{2}}{h}\int_{BZ}\frac{d^{3}\vec{k}}{(2\pi)^{3}}\sum_{n}f_{n}(\vec{k})\Omega_{n}^{z}(\vec{k})
\end{equation}
where $f_{n}(\vec{k})$ is the Fermi-Dirac distribution function, and $n$ is the index of the occupied bands. The Berry curvature is arisen from Kubo-formula derivation,
\begin{equation}
\Omega_{n}^{z}(\vec{k})=-2{\rm Im}\sum_{m\neq n}\frac{\langle\Psi_{n\vec{k}}|v_{x}|\Psi_{m\vec{k}}\rangle\langle\Psi_{m\vec{k}}|v_{y}|\Psi_{n\vec{k}}\rangle}{(E_{m}(\vec{k})-E_{n}(\vec{k}))^{2}}
\end{equation}
where $v_{x(y)}$ is the velocity operator. We calculate the intrinsic AHC with a 101$\times$101$\times$101 $k$-point grid in the first BZ based on the Wannier tight-binding Hamiltonian.

In FIG.~\ref{appfig4}f, we plot the energy dependence of intrinsic AHC near the Fermi level. On the Fermi level, the AHC is about -123 $\Omega^{-1}\cdot cm^{-1}$, which can be observed in transport experiments.

\begin{table}[ht]
\begin{centering}
\begin{tabular}{c|c|p{0.05\columnwidth}<{\centering}|p{0.05\columnwidth}<{\centering}|c|c|p{0.05\columnwidth}<{\centering}|p{0.05\columnwidth}<{\centering}} 
\hline \hline
$\Lambda_\alpha$ & MLCG & Irreps & $C^+_{4z}$ & $\Lambda_\alpha$ & MLCG & Irreps & $M_{z}$ \\
\hline
\multirow{4}{*}{LD(0,0,w)} & \multirow{4}{*}{$42'2'$} & $\overline {LD}_5$ & $\frac{-1+i}{\sqrt{2}}$ & \multirow{2}{*}{DT(u,u,0)} & \multirow{2}{*}{$m'm2'$} & $\overline {DT}_3$ & -i \\ 
 \cline{3-4}\cline{7-8}
& & $\overline {LD}_6$ & $\frac{1-i}{\sqrt{2}}$ &  &  & $\overline {DT}_4$ & i \\
\cline{3-8}
& & $\overline {LD}_7$ & $\frac{-1-i}{\sqrt{2}}$ & \multirow{2}{*}{SM(u,0,0)} & \multirow{2}{*}{$m'm2'$} & $\overline {SM}_3$ & -i \\
\cline{3-4}\cline{7-8}
& & $\overline {LD}_8$ & $\frac{1+i}{\sqrt{2}}$ &  &  & $\overline {SM}_4$ & i \\
\hline\hline
\end{tabular}
\end{centering}
\caption{Character table of the irreducible co-representations of MSG $P_I4/nnc$ at $LD(0,0,w)$, $DT(u,u,0)$ and $SM(u,0,0)$ points. The first two(5th-6th) columns are the momenta and their magnetic little co-groups(MLCG); the third(7th) column are the irreps of the MLCG; the 4th(8th) columns are the characters of the symmetry generators of the unitary subgroup. }\label{app:irreps_mzc}
\end{table}

\begin{table}
\begin{centering}
\begin{tabular}{c|p{0.3\columnwidth}<{\centering}|p{0.1\columnwidth}<{\centering}|p{0.15\columnwidth}<{\centering}}
\hline\hline
WPs & Position ($k_x$, $k_y$, $k_z$)(\AA$^{-1}$) & Chirality & $E-E_{Fermi}$(eV) \\
\hline\hline
\multirow{8}{*}{WP1} & (  0.3101,   0.3101,   0.7154) &  +1 &   -0.0767 \\
\cline{2-4}
&  ( -0.3101,  -0.3101,   0.7154) &  +1  &   -0.0767  \\ 
\cline{2-4}
&  ( -0.3101,   0.3101,   0.7154) &  +1  &   -0.0767  \\ 
\cline{2-4}
&  (  0.3101,  -0.3101,   0.7154) &  +1  &   -0.0767  \\ 
\cline{2-4}
&  (  0.3101,   0.3101,  -0.7154) &  -1  &   -0.0767  \\ 
\cline{2-4}
&  ( -0.3101,  -0.3101,  -0.7154) &  -1  &   -0.0767  \\ 
\cline{2-4}
&  ( -0.3101,   0.3101,  -0.7154) &  -1  &   -0.0767  \\ 
\cline{2-4}
&  (  0.3101,  -0.3101,  -0.7154) &  -1  &   -0.0767  \\ 
\hline
\multirow{8}{*}{WP2} &  ( 0.0,  -0.2798,   0.5009) &  -1 &   -0.2310 \\
\cline{2-4}
&  (  0.0,   0.2798,   0.5009) &  -1 &   -0.2310  \\ 
\cline{2-4}
&  (  0.2798,  0.0,   0.5009) &  -1 &   -0.2310  \\ 
\cline{2-4}
&  ( -0.2798,   0.0,   0.5009) &  -1 &   -0.2310  \\ 
\cline{2-4}
&  (  0.0,   0.2798,  -0.5009) &  +1 &   -0.2310  \\ 
\cline{2-4}
&  ( 0.0,  -0.2798,  -0.5009) &  +1 &   -0.2310  \\ 
\cline{2-4}
&  ( -0.2798,   0.0,  -0.5009) &  +1 &   -0.2310  \\ 
\cline{2-4}
&  (  0.2798,  0.0,  -0.5009) &  +1 &   -0.2310  \\ 
\hline
\multirow{2}{*}{WP3} &  (  0.0,   0.0,   0.3343) &  +1 &   -0.14 \\
\cline{2-4}
&  (  0.0,   0.0,  -0.3343) &  -1 &   -0.14 \\
\hline
\multirow{2}{*}{WP4} &  ( 0.0,  0.0,   0.2320) &  +1 &   -0.0901 \\
\cline{2-4}
&  ( 0.0,  0.0,  -0.2320) &   -1 &  -0.0901  \\
\hline
\end{tabular}
\end{centering}
\caption{The 10 pairs of WPs in the first BZ. The first column is the type of WPs. In each type, the WPs are related by $C_{4z}$ and $M_z$ symmetries. The second column is the position of each WP in the cartesian coordinate system of BZ. The 4th-5th columns are the chirality and energy related to the Fermi level of each WP.}\label{tab:wps}
\end{table}

\section{Fragile bands in the magnetic materials}\label{app:J}
In Table~\ref{tab:fragile}, we tabulate all of the magnetic materials that have fragile bands below the Fermi level. For the materials having more than one fragile brach, we provide the irreducible co-representations at high symmetry momenta of the fragile branches closest to the Fermi level.
\LTcapwidth=1.0\textwidth
\begin{longtable}{c|c|c|c|c|c|p{0.5\columnwidth}}
\caption{The fragile occupied bands in magnetic materials. In the first four columns the chemical formulae (Formula),the material identification number on BCSMD (BCSID), the magnetic space group number (MSG) and the Coulomb interaction strength (U) are tabulated. The fifth column gives the number of fragile branches (NF) in the band structure of the corresponding material. In the sixth and seventh columns, the band indices (Bands) and the irreducible representations (Irreps) at high symmetry momenta of the fragile branch closest to the Fermi level are tabulated.}
\label{tab:fragile} \\
\hline
\hline
Formula & BCSID & MSG & U(eV) & NF & Bands & Irreps \\
\hline
\hline
CeCoGe3 & 1.0.11 & 107.231 & 4 & 1 & -83$\sim$-81 & $\bar{\Gamma}_{6}+\bar{\Gamma}_{8}+\bar{M}_{6}+\bar{M}_{8}+\bar{N}_{2}+\bar{P}_{4}+\bar{X}_{4}$ \\
\hline
alpha-Mn & 1.85 & 114.282 & 0 & 5 & -43$\sim$-40 & $\bar{A}_{6}+\bar{A}_{7}+\bar{\Gamma}_{6}+\bar{M}_{7}+\bar{R}_{2}\bar{R}_{4}+\bar{R}_{3}\bar{R}_{5}+\bar{X}_{2}+\bar{X}_{3}+\bar{X}_{4}+\bar{X}_{5}+\bar{Z}_{6}+\bar{Z}_{7}$ \\
\hline
PrCo2P2 & 2.26 & 123.345 & 0 & 1 & -115$\sim$-108 & $\bar{A}_{10}+\bar{A}_{11}+\bar{A}_{12}+\bar{A}_{5}+\bar{A}_{6}+\bar{A}_{7}+\bar{A}_{8}+\bar{A}_{9}+\bar{\Gamma}_{10}+\bar{\Gamma}_{12}+\bar{\Gamma}_{6}+\bar{\Gamma}_{8}+\bar{M}_{11}+\bar{M}_{5}+\bar{M}_{6}+\bar{M}_{7}+\bar{M}_{8}+\bar{M}_{9}+\bar{R}_{3}+\bar{R}_{4}+\bar{R}_{5}+\bar{R}_{6}+\bar{X}_{3}+\bar{X}_{4}+\bar{X}_{5}+\bar{X}_{6}+\bar{Z}_{10}+\bar{Z}_{12}+\bar{Z}_{6}+\bar{Z}_{8}$ \\
\hline
NdCo2P2 & 1.251 & 124.360 & 0 & 2 & -119$\sim$-112 & $\bar{A}_{5}+\bar{A}_{6}+\bar{A}_{7}+\bar{A}_{8}+\bar{\Gamma}_{7}+\bar{\Gamma}_{9}+\bar{M}_{6}+\bar{M}_{7}+\bar{M}_{8}+\bar{R}_{3}+\bar{R}_{4}+\bar{X}_{5}+\bar{X}_{6}+\bar{Z}_{6}+\bar{Z}_{8}$ \\
\hline
\end{longtable}

\section{Magnetic moments for each materials with different Coulomb interactions}\label{app:K}
The experimental and calculated magnetic moments on the nonequivalent magnetic atoms of each material have been listed below. As it only consider the moments that contributed by spin component in the VASP, it may have an estimate for those materials that have large orbital magnetic moments. 
For all of the magnetic materials, we define the average error of the calculated magnetic moments as $\frac{1}{N}\sum_{i=1}^{N} \frac{|M^{Exp}_i - M^{T}_i|}{M^{Exp}_i} \times 100\%$, where $N$ is the number of nonequivalent magnetic atoms, $M^{Exp}$ is the experimental magnetic moments and $M^{T}$ is the calculated magnetic moments.
We label the magnetic moments that are closest to the experimental value by red color and tabulate the number  of materials with the error margin in 0~10\%, 10\%~30\%, 30\%~50\% and >50\% in Table \ref{statis_mom}.

\begin{table}
\begin{centering}
\begin{tabular}{c|c|c|c|c|c}

\hline\hline
Average error & 0$\sim$10\% & 10\%$\sim30$\% & 30\%$\sim$50\% & >50\% & Total \\
\hline
Count & 139 & 117 & 68 & 79 & 403 \\
\hline\hline

\end{tabular}
\end{centering}
\caption{Statistics of the materials with the average error of calculated magnetic moments margin in $0\sim10\%$, $10\%\sim30\%$, $30\%\sim50\%$ and >$50\%$.}\label{statis_mom}
\end{table}

\LTcapwidth=1.0\textwidth
\renewcommand\arraystretch{1.2}
\begin{longtable*}{c|p{0.16\columnwidth}|p{0.07\columnwidth}|p{0.07\columnwidth}|p{0.1\columnwidth}|p{0.08\columnwidth}|p{0.08\columnwidth}|p{0.08\columnwidth}|p{0.08\columnwidth}|p{0.08\columnwidth}|p{0.08\columnwidth}}
\caption{Magnetic moments of the magnetic materials with $3d/4d$ electrons. The first four column give the chemical formulae (Formula), BCSID, MSG for each materials. The 5th to 11th column give the magnetic moments of each nonequivalent magnetic atom, including the experimental (Exp.) magnetic moments and the calculated magnetic moments with Coulomb interaction $U=0,1,2,3,4$ eV. The calculated magnetic moments closest to the experimental value are tagged by red color.}\label{tablemagd} \\
\hline \hline
\multirow{2}{*}{No.} & \multirow{2}{*}{Formula} & \multirow{2}{*}{BCSID} & \multirow{2}{*}{MSG} & \multicolumn{7}{c}{Magnetic moments ($\mu_{B}$) }  \\
\cline{5-11}
& & & & Mag. Ele. & Exp.  & U=0 & U=1 eV & U=2 eV & U=3 eV & U=4 eV \\
\hline
\hline
\multirow{2}{*}{1} & \multirow{2}{*}{BiCrO3} & \multirow{2}{*}{0.139} & \multirow{2}{*}{2.4} & {Cr} & \textbf{2.62} & 2.66 & 2.73 & 2.79 & 2.84 & 2.88  \\
\cline{5-11}
 & & & & Error($\%$) & & \textcolor{red}{1.527} & 4.198 & 6.489 & 8.397 & 9.924  \\ 
\hline\multirow{2}{*}{2} & \multirow{2}{*}{Cr2S3} & \multirow{2}{*}{0.5} & \multirow{2}{*}{2.4} & {Cr} & \textbf{1.2} & 2.65 & 2.81 & 2.9 & 2.99 & 3.07  \\
\cline{5-11}
 & & & & Error($\%$) & & \textcolor{red}{120.833} & 134.167 & 141.667 & 149.167 & 155.833  \\ 
\hline\multirow{2}{*}{3} & \multirow{2}{*}{Fe2O3-alpha} & \multirow{2}{*}{0.66} & \multirow{2}{*}{2.4} & {Fe} & \textbf{4.22} & 3.56 & 3.8 & 3.94 & 4.05 & 4.13  \\
\hline\hline
\end{longtable*}

\begin{longtable*}{c|p{0.17\columnwidth}|p{0.08\columnwidth}|p{0.08\columnwidth}|p{0.11\columnwidth}|p{0.09\columnwidth}|p{0.09\columnwidth}|p{0.09\columnwidth}|p{0.09\columnwidth}|p{0.09\columnwidth}}
%\begin{longtable*}{c|c|c|c|c|c|c|c|c|c}
\caption{Magnetic moments of the magnetic materials with $4f/5f$ electron. The first four column give the chemical formulae (Formula), BCSID, MSG for each materials. The 5th to 10th column give the magnetic moments of each nonequivalent magnetic atom, including the experimental (Exp.) magnetic moments and the calculated magnetic moments with Coulomb interaction $U=0,2,4,6$ eV for $4f/5f$ electron. If the materials have transition metal elements at the same time, we take the Coulomb interactions of $3d/4d$ electron as 2 eV. The calculated magnetic moments closest to the experimental value are labeled by red color.}\label{tablemagd} \\
\hline \hline
\multirow{2}{*}{No.} & \multirow{2}{*}{Formula} & \multirow{2}{*}{BCSID} & \multirow{2}{*}{MSG} & \multicolumn{6}{c}{Magnetic moments ($\mu_{B}$) }  \\
\cline{5-10}
& & & & Mag. Ele. & Exp.  & U=0 & U=2 eV & U=4 eV & U=6 eV \\
\hline
\hline
\multirow{1}{*}{277} & \multirow{1}{*}{HoCr(BO3)2} & \multirow{1}{*}{1.191} & \multirow{1}{*}{2.7} & {Cr} & \textbf{3.07} & 2.84 &  - & 2.82 &  -  \\
\cline{5-10}
 & & & & Error($\%$) & & \textcolor{red}{7.492} &  - & 8.143 &  -  \\ 
\hline\multirow{1}{*}{278} & \multirow{1}{*}{Dy2Fe2Si2C} & \multirow{1}{*}{1.206} & \multirow{1}{*}{2.7} & {Dy} & \textbf{12.11} & 4.81 & 4.87 & 4.99 & 5.14  \\
\cline{5-10}
 & & & & Error($\%$) & & 60.281 & 59.785 & 58.794 & \textcolor{red}{57.556}  \\ 
\hline\multirow{2}{*}{279} & \multirow{2}{*}{Tm2BaNiO5} & \multirow{2}{*}{1.218} & \multirow{2}{*}{2.7} & {Tm} & \textbf{3.33} & 1.45 & 1.73 & 1.8 & 1.83  \\
\hline\hline
\end{longtable*}

\clearpage

\subsection{Ferro(Ferri)magnetic materials}
Among the 403 materials, there have 51 magnetic materials with net moments/ferromagnetic canting.  All of them are tabulated in Table \ref{tab:ferr} with the BCSID, chemical formula and the MSG. All of the MSGs of them are Type-I/Type-III MSG, which are compatible with ferro(ferri)magnets.
\renewcommand\arraystretch{1.2}
\begin{longtable}{|c|p{0.18\columnwidth}<{\centering}|c||c|p{0.18\columnwidth}<{\centering}|c||c|p{0.18\columnwidth}<{\centering}|c|}
\caption{Magnetic materials with non-zero net moments or ferromagnetic canting.}
\label{tab:ferr}\\
\hline
\hline
BCSID & Formula & MSG & BCSID & Formula & MSG &BCSID & Formula & MSG \\
\hline
0.26 & TmAgGe & 38.191 &0.36 & NiF2 & 58.398 &0.37 & U3Al2Si3 & 5.15  \\
\hline
0.48 & Tb2Sn2O7 & 141.557 &0.49 & Ho2Ru2O7 & 141.557 &0.64 & MnV2O4 & 88.81  \\
\hline
0.77 & Tb2Ti2O7 & 166.101 &0.78 & NiN2O6 & 148.17 &0.85 & KCo4(PO4)3 & 58.398  \\
\hline
0.91 & Rb2Fe2O(AsO4)2 & 62.448 &0.121 & Li2Co(SO4)2 & 14.79 &0.122 & Li2Mn(SO4)2 & 14.75  \\
\hline
0.149 & Nd3Ru4Al12 & 63.462 &0.151 & Tm2Mn2O7 & 141.557 &0.157 & Yb2Sn2O7 & 141.557  \\
\hline
0.165 & SrMn(VO4)(OH) & 4.7 &0.169 & U3As4 & 161.71 &0.170 & U3P4 & 161.71  \\
\hline
0.173 & Pr3Ru4Al12 & 63.462 &0.174 & Pr3Ru4Al12 & 15.89 &0.176 & Mn3Ti2Te6 & 15.89  \\
\hline
0.184 & Nd5Si4 & 92.114 &0.185 & Nd5Ge4 & 62.447 &0.191 & BaCuF4 & 36.176  \\
\hline
0.203 & Mn3Ge & 12.62 &0.220 & Mn2SiO4 & 62.446 &0.226 & NdCo2 & 15.89  \\
\hline
0.227 & NdCo2 & 141.557 &0.228 & TbCo2 & 166.101 &1.0.11 & CeCoGe3 & 107.231  \\
\hline
1.0.12 & UAu2Si2 & 71.536 &1.0.13 & FeI2 & 12.62 &2.5 & Mn3CuN & 85.59  \\
\hline
2.10 & HoP & 15.89 &2.11 & TbMg & 51.295 &2.12 & TbMg & 49.270  \\
\hline
2.19 & Mn3ZnC & 139.537 &2.28 & NpNiGa5 & 74.559 &3.3 & Ho2RhIn8 & 63.464  \\
\hline
\end{longtable}

\section{Band structures and detailed information}\label{app:I}
We list the band structures for all of the 403 magnetic materials in the following tables, with the energy range set between $-1.0\sim1.0$ eV relative to the Fermi level.
Apart from the crystal and band structures for each material, the tables also include the material identification number on BCSMD (BCSID), chemical formula (Formula), the ICSD number if available, the magnetic space group (MSG) and stable topological classification of the MSG (T.C.). For the magnetic TIs diagnosed by MTQC, see ~\ref{app:F} for the stable topological indices and the physical interpretations of them. For the magnetic ESs diagnosed by MTQC, see \ref{app:G} for the high-symmetry $k$ paths that break compatibility relations.

\LTcapwidth=1.0\textwidth
\renewcommand\arraystretch{1.0}
\input{./subtex/band}
%\clearpage

\clearpage
\bibliography{main}

\end{document}